\documentclass[12pt, english]{article}

\usepackage{changes}

\usepackage{verbatim}
\usepackage{url}
\usepackage{amsmath}
\usepackage{amsfonts}
\usepackage{amssymb}

\usepackage{todonotes}

\usepackage{footmisc}
\definecolor{sangre}{rgb}{0.6,0.18,0.19}
\definecolor{dullmagenta}{rgb}{0.4,0,0.4}
\definecolor{darkblue}{rgb}{0,0,0.6}

\ifx\pdfoutput\undefined
\usepackage[backref=none,hypertex,colorlinks,urlcolor = darkblue,citecolor = sangre,linkcolor=blue]{hyperref}
\else
\usepackage[backref=none,hypertexnames=false,colorlinks,urlcolor = darkblue,citecolor =  sangre,linkcolor=blue]{hyperref}
\fi

\usepackage{relsize,etoolbox}

\usepackage{bbm}

\usepackage{color, colortbl}
\definecolor{Gray}{gray}{0.95}
\usepackage{booktabs}
\usepackage{pdflscape}
\usepackage{amsfonts}
\usepackage{graphicx, graphics}
\usepackage{multirow}
\usepackage{epstopdf}
\usepackage{adjustbox}
\usepackage[titletoc,title]{appendix}

\usepackage[capitalize]{cleveref}
\definecolor{lavander}{cmyk}{0,0.48,0,0}
\definecolor{violet}{cmyk}{0.79,0.88,0,0}
\definecolor{burntorange}{cmyk}{0,0.52,1,0}

\def\lav{lavander!90}
\def\oran{orange!30}

\tikzstyle{peers}=[draw,circle,violet,bottom color=\lav,
                  top color= white, text=violet,minimum width=10pt]
\tikzstyle{superpeers}=[draw,circle,burntorange, left color=\oran,
                       text=violet,minimum width=20pt]
\tikzstyle{legendsp}=[rectangle, draw, burntorange, rounded corners,
                     thin,bottom color=\oran, top color=white,
                     text=burntorange, minimum width=2.5cm]
\tikzstyle{legendp}=[rectangle, draw, violet, rounded corners, thin,
                     bottom color=\lav, top color= white,
                     text= violet, minimum width= 2.5cm]
\tikzstyle{legend_general}=[rectangle, rounded corners, thin,
                           burntorange, fill= white, draw, text=violet,
                           minimum width=2.5cm, minimum height=0.8cm]

\usepackage{natbib}

\usepackage{setspace}
\usepackage[margin=1in]{geometry}
\usepackage[latin9]{inputenc}
\usepackage{array}
\usepackage{float}
\usepackage{booktabs}
\usepackage{enumitem}
\usepackage{bm}
\usepackage{amsmath}
\usepackage{amssymb}
\usepackage{graphicx}
\usepackage{setspace}
\usepackage{enumitem}

\makeatletter

\usepackage{lineno}

\usepackage{bm}
\usepackage{caption}
\usepackage{enumitem}
\usepackage{epstopdf}
\usepackage{epsfig}% to use graphics...
\usepackage{dsfont}% needed for indicator function symbol 
\usepackage{amsfonts}
\usepackage{amsthm}
\newtheorem{ex}{Example}

\newcommand{\q}{\mathfrak{q}}		%% FZ: let's use boldface for vectors
\newcommand{\E}{\mathbb{E}}

\newtheorem{assumption}{Assumption}

\newtheorem{lemma}{Lemma}

\newtheorem{theorem}{Theorem}

\usepackage{soul}

\DeclareMathAlphabet{\mathpzc}{OT1}{pzc}{m}{it}
\usepackage{lscape}
 \usepackage{babel}
\usepackage{mathrsfs}

\newcommand\sbullet[1][.4]{\mathbin{\vcenter{\hbox{\scalebox{#1}{$\bullet$}}}}}

\begin{document}

\baselineskip=1.2\baselineskip

\title{Empirical Framework for Cournot Oligopoly with Private Information\thanks{\protect\linespread{1}\protect\selectfont We thank the editor Allan Collard-Wexler and three referees for numerous helpful suggestions that substantially improved the paper. We also thank Victor Aguirregabiria, Mar\'ia F. Gabrielli, Mitsuru Igami, Sung Jae Jun, Lidia Kosenkova, and Nicholas Vreugdenhil for their thoughtful comments. We are also thankful to seminar and conference participants, at the 14th GNYMA Econometrics Colloquium, 7th Alumni Conference at Universidad de San Andr\'{e}s, 2018 NASM, Universidad Torcuato Di Tella, 2019 Triangle Econometrics Conference at Duke University, 2019 SEA, DC-MD-VA Econometrics Workshop 2020, UBC Econometrics Brownbag, and 2022 IIOC, for their comments.}
}

\author{Gaurab Aryal\thanks{ Department of Economics, Washington University in St. Louis, \href{mailto:aryalg@wustl.edu}{ aryalg@wustl.edu}.} \qquad
%} \qquad
 Federico Zincenko\thanks{College of Business, Department of Economics, University of Nebraska--Lincoln, \href{mailto:fzincenko2@unl.edu}{fzincenko2@unl.edu}.}
}

\date{\today}

\maketitle

\begin{abstract}
We propose an empirical framework for asymmetric Cournot oligopoly with private information about variable costs. First, considering a linear demand for a homogenous product with a random intercept, we characterize the Bayesian Cournot-Nash equilibrium. Then we establish the identification of the joint distribution of demand and firm-specific cost distributions. Following the identification steps, we propose a likelihood-based estimation method and apply it to the global market for crude-oil and quantify the welfare effect of private information. 
We also consider extensions of the model to include product differentiation, conduct parameters, nonlinear demand, or selective entry.  
\medskip\\
\textbf{JEL classification:} C57, D22, D43, L13. 
\smallskip\\
\textbf{Keywords:} Cournot Oligopoly, Private Information, Variable Costs, Identification, Crude-oil.
\end{abstract}

\newpage
%\linenumbers

\section{Introduction}

Competition among firms is necessary for a vibrant economy, but several factors may afford market power to firms that lower competition. One such factor is their private information \citep[e.g.,][]{BergemannHeumannMorris2019} about their production costs. Private information is also central for limit pricing and predation \citep{MilgromRoberts1982a, MilgromRoberts1982b}, collusion \citep{Roberts1985}, coordination \citep{AryalCilibertoLeyden2021}.  
Most empirical articles that study market power and estimate the associated welfare assume complete information and focus on \emph{getting the strategic aspect right}. 
However, \cite{viv02} shows that ignoring private information can generate a more significant error in our welfare calculation than if we had modeled the private information correctly but \emph{gotten the strategic aspect wrong}. His results suggest that firms' mutual information about each others' costs has a more fundamental effect on welfare estimates than is typically appreciated. 

Several important articles, e.g., \cite{Seim2006, AradillasLopez2010, PaulaTang2012} and \cite{Greico2014}, study different aspects of oligopolistic competition with private information. They, however, focus on environments with discrete actions where the source of private information is an additive ``error term" in the profit function. 
Instead, we consider a continuous game where private information is about firms' (possibly correlated) total variable costs. Modeling private information from the ``ground up" allows us to capture the nonlinear effect of private information on firms' profits and the resulting market efficiency and to provide an economic interpretation for the source of inefficiencies. For instance, in the Cournot oligopoly with homogenous goods that we consider here, complete cost information increases efficiency because only the most efficient firms produce, but the markup may rise with fewer firms. Using our method, one can determine which factor dominates.  

Our main contribution is to develop an empirical framework for asymmetric Cournot competition with private information about their costs.
To this end, we build on \cite{viv02} and consider a market for a homogenous product with linear and stochastic demand, where firms are asymmetric and have private information about their marginal costs. Also, we allow for a common but unobserved (to the econometrician) market-level technology shock that shifts and induces correlation across firms' costs. 

We characterize the Bayesian Cournot-Nash equilibrium for this game and propose a constructive strategy to identify the model parameters assuming that the observed quantities and prices are equilibrium outcomes of the game. 
Our identification strategy uses the results that the equilibrium strategies are linear and strictly decreasing in their own cost and that demand and costs shocks are exogenous (``shifters") independent and identically distributed across markets.

In particular, we show that the variation in observed prices and outputs identifies the demand parameters and that the variation in firms' outputs and the monotonicity of the equilibrium strategies identify the cost distributions. 
 Then we show that the joint variation in firms' outputs identifies the unobserved (common) technology shock distribution. 
 
Our identification strategy borrows some insights from the empirical literature on Bayesian games. For instance, in empirical auctions with independent private values, strict monotonicity of bidding strategies plays a central role in the identification; see, for example, \cite{gpv00}. See \cite{EinavNevo2006}, who provide the link between the classic demand and pricing literature and empirical auctions.\footnote{We also discuss how we can view our identification problem as classic identification of simultaneous equations system that determines demand and supply.} Similarly, our idea of using joint variation in firms' outputs to identify the distribution of the common technology shock is akin to the identification strategy in \cite{Krasnokutskaya2011} for auctions with unobserved heterogeneity. 

To illustrate our method, we study the monthly global market for crude-oil. 
We consider 20 major crude-oil-producing countries from January 1992 to December 2019. 
In this environment, variable costs comprise rental rates for drilling rigs, prices for steel, site preparation costs, construction costs, capital costs, and general equipment rental costs averaged across all oil fields.\footnote{While some of these costs  (e.g., steel prices) may be commonly known, others (e.g., rental rates, equipment rental, and capital costs) are likely to be private information. Thus, treating the variable costs as countries' private information is reasonable.} Although we propose a semi-nonparametric identification strategy, given our small sample size (of 336 months), we make distributional assumptions and propose a maximum likelihood estimation procedure to ensure good finite-sample performance. 

In our empirical exercise, we treat each oil-producing country as if it is a competitive firm in our model. This assumption is consistent with the fact that in most oil-producing countries, production decisions are centralized, and state-run companies exploit the reserves.\footnote{An exception is the U.S., where production is decentralized. 
We also consider an extension with conduct parameters that allows price-taking firms as a special case. However, given that we only observe the total U.S. production and our focus is on methodology, we treat the U.S. as one firm.} In such cases, the estimated variable costs are an aggregate measure of costs from several oil reserves within each country. 
In that regard, our application is closer to \cite{cdfq13} than to \cite{AskerWexlerLoecker2019}, where the latter provides a detailed empirical analysis of the effect of heterogeneity across oil fields, within and across countries, on total efficiency. Using counterfactual exercise, we quantify the welfare effect of private information. In particular, we estimate the deadweight loss under private information at 16.3\% higher than under complete information.

We also consider extensions of our model in four directions and study their identification. First, we consider differentiated products, and second, the possibility that firms do not play the (static) Bayesian Nash-Cournot equilibrium; instead, they play a conjectural variation equilibrium by allowing the firms to have different conduct parameters. Third, we consider a nonlinear demand function. Fourth, we consider Cournot oligopoly with a selective entry, where firms are \emph{symmetric} and observe a signal about their cost, make a costly entry decision, and then choose their outputs after entering.

Our article contributes to several strands of research in industrial organization. First, it is related to the literature \citep{Vives1984, viv02} that studies the role of private information in Cournot competition. Second, in terms of our empirical application, we complement \cite{Rosen2006} who also studies the identification of marginal costs under incomplete information. Third, our article is also related to the literature that estimates games with private information, such as \cite{Seim2006}, \cite{Sweeting2009}, and \cite{Greico2014}. We complement this research, but in contrast, we model the source of private information (about cost) and use it to determine the expected payoff structure, resulting in a nonseparable model that requires a new approach to identify the cost parameters. Our empirical approach is similar in spirit to \cite{SweetingRobertsGedge2020}, where costs are firms' private information. 

We also contribute to a large and varied literature on the crude-oil industry \cite[see, e.g.,][]{LasservePierru2021} by introducing private information. In so far as the oil extraction decisions involve inter-temporal tradeoffs  \citep{Hotelling1931, CremerWeitzman1976, Loury1986}, our estimate of the size of private information misses these tradeoffs. Consequently, our estimate also does not incorporate any adverse effects of future oil price uncertainty on oil production \citep{Kellogg2014}. 

While our empirical application considers the crude-oil market, our framework applies more broadly and can be used to study other industries characterized by asymmetric Cournot competition. Some of these industries may include the lysine market \citep{Roos2006}, the Portland cement industry \citep{Ryan2012}, the ready-mix concrete industry \citep{HortacsuSyverson2007, Wexler2013}, and the coffee bean market \citep{Igami2015}.

The rest of our paper proceeds as follows. Sections \ref{section:model} and \ref{section:identification} describe our model and the identification strategies, respectively. Section \ref{section:data} describes the data. Sections \ref{subsec:est} provides the estimation procedure and Monte Carlo simulations. Section \ref{section:empirical} reports our empirical findings followed by a discussion of the model and the estimates in Section \ref{section:discussion}. Section \ref{section:conclusion} concludes. The proofs of all the results stated in the main text are relegated to Appendix A, and additional estimation results to Appendix B. 

In Supplementary Appendix S, we consider four extensions of our baseline model: (i) differentiated Cournot competition; (ii) possibility that firms do not play Bayesian Cournot-Nash equilibrium by allowing them to have different conduct parameters; (iii) homogenous Cournot competition with nonlinear demand; and (iv) homogenous Cournot competition with a selective entry. For each, we discuss how our empirical strategy extends to that case. 

\vspace{.1in}

{\bf Notation.} All vectors and their concatenation with a comma are column vectors. We use boldface to denote vectors (or random vectors) and regular letters for scalars (or random variables). 
For generic random variables $(Y,X)$, $F_{Y,X}$ and $f_{Y,X}$ denote their joint cumulative distribution function (CDF) and probability density function (PDF), respectively. Further, $F_{Y|X}(\cdot| x)$, $\mu_{Y|X}(x)$ and $F^{-1}_{Y|X}(\cdot|x)$ denote the conditional CDF, conditional mean and conditional quantile function of $Y$ given $X=x$, respectively. We use $\mu_X$ for the unconditional mean and write $X\perp Y$ when $X$ and $Y$ are independent. We also employ the same notation for random vectors. For example, if $\mathbf Y$ and $\mathbf X$ are random vectors each with dimension $2 \times 1$, $F_{\mathbf Y, \mathbf X}$ denotes the joint CDF of the $4\times1$ random vector $(\mathbf Y, \mathbf X)$. Finally, for a given vector $\mathbf x = (x_1,\dots, x_N)$, we write $x^+ = \sum_{i=1}^N x_i$, $\mathbf x_{-i} = (x_1,\dots, x_{i-1},x_{i+1},\dots,x_N)$, and $\|\mathbf x\| = \sqrt{\sum_{i=1}^N x_i^2 }$.
We use $\boldsymbol{\iota}_\mathcal{I}$ for a $\mathcal I \times 1$ vector of ones, and $\mathds{I}_{\mathcal I}$ for the $\mathcal I\times \mathcal I$- identity matrix.

\section{Model} \label{section:model}

In this section, we present our model of Cournot oligopoly with homogeneous goods where asymmetric firms have private information about their variable costs. To this end, we extend \cite{viv02} to allow for stochastic demand and a common technology shock, and then we characterize the equilibrium strategies.

Let there be $\mathcal T$ markets, and in each market $t=1,\dots,\mathcal T$, let there be $\mathcal M \in \mathbb N$ consumers. 
Each consumer $m=1,\dots,\mathcal M$, has quasi-linear preferences for an homogeneous good and in market $t$, maximizes the net benefit function \begin{equation}
\mathfrak u ( c_{mt} ; p_t , U_{t} ) = U_{t} \times c_{mt} - \frac{\tilde\beta}{2} c_{mt}^2- p_t \times c_{mt}, 
\label{eq:ut}
\end{equation}
where $c_{mt}$ is the quantity consumed by $m$, $p_t $ is the per-unit price of the product, $ U_{t} \geq0$ is a (one-dimensional) demand shock that affects the consumer's willingness to pay, and $\tilde\beta>0$ is a common utility parameter. 
A consumer in market $t$ takes the market price $p_t$ as given and chooses quantity consumed according to $\mathfrak c ( p_t, U_{t} ) =\arg\max_{c\geq0} \mathfrak u ( c ; p_t , U_{t})$. Then summing the demand over consumers in market $t$ gives the inverse demand function
\begin{equation}
\mathfrak p (c_t^+, U_t) = U_t - \frac{\tilde\beta}{\mathcal M} c_t^+ = U_t -\beta c_t^+, 
\label{eq:demand}
\end{equation}where $c_t^+= \sum_{m =1}^{\mathcal M} c_{mt}$ is the total consumption and $\beta = \tilde\beta /\mathcal M$ is the demand parameter. 

On the supply side, let there be $\mathcal I \geq 2 $ firms in each market $t$ that compete in quantities. And let $\mathscr I = \{ 1,\dots, \mathcal I\}$ denote the set of firms.
We begin by assuming that firms are heterogeneous in their production costs. For $i=1,\dots,\mathcal I$, let $V_{it}\geq 0$ denote firm $i$'s inefficiency parameter (or simply, firm $i$'s private cost) in market $t$, and we assume that $V_{it}$ is firm $i$'s private information. Furthermore, we allow $i$'s variable cost in market $t$ to depend on $i$'s private cost $V_{it}$ and a cost shock $W_t\in\mathbb{R}$ common across all firms.
 
In particular, let $i$'s total variable cost of producing $q_{it}$ in market $t$ be
\begin{equation}
\mathfrak{v c} (q_{it} ; V_{it}, W_t) =( V_{it} + W_t)\times q_{it} + \frac{\lambda}{2} q_{it}^2 = {V}_{it}^\ast q_{it} + \frac{\lambda}{2} q_{it}^2, \label{eq:vc}
\end{equation}
where $\lambda\geq0$ is a cost parameter and $V_{it}^\ast : = V_{it} + W_t$ is $i$'s total variable cost. Thus, firms with higher $V_{it}^\ast$ are less efficient, and have higher marginal costs, than firms with lower $V_{it}^\ast$.

Letting $\mathbf V_t = (V_{1t},\dots,V_{\mathcal I t}) $, hereafter, we assume that $\{ (\mathbf V_t , W_t, U_t) : t = 1,\dots, \mathcal T \}$ are random vectors that satisfy the following assumption. Let $\mathbf V = ( V_{1},\dots,V_{\mathcal I} )$ and $( U ,W )$ be random vectors representing the private cost shocks and the common demand and technology shocks, respectively. We begin with the following modeling assumptions. 

\begin{assumption}

\label{assump:distributions}

The random vectors $\{ (\mathbf V_t , W_t, U_t) : t = 1,\dots, \mathcal T \}$ are IID as $(\mathbf{V} , W, U )$. Further, the distribution $(\mathbf{V} , W, U )$ satisfies the next conditions.
 
\begin{enumerate}[label=(\roman*)]

\item The firms' types $\mathbf{V}$ and the common shocks $(U,W)$ are independent, i.e., $(U,W) \perp \mathbf{V}$. Also, the firm-specific cost shocks $\{V_{1},\dots,V_{\mathcal I}\}$ are mutually independent.

\item For each $i\in\mathscr I$, $V_i$ has support given by $[\underline{v}_i,\bar{v}_i]$ with $0 \leq \underline{v}_i < \bar{v}_i < \infty$. It also admits a PDF $f_{V_i}$ that is strictly positive and continuously differentiable on $(\underline{v}_i,\bar{v}_i)$.

\item The random vector $(U,W)$ has rectangular support given by $[\underline u, \infty)\times [\underline{w},\bar{w}] \subset\mathbb{R}_+\times \mathbb{R} $ with $\underline w < \bar w$. It also admits a joint PDF $f_{U,W}$ that is strictly positive and continuously differentiable on $ (\underline{u},\infty)\times (\underline w, \bar w) $.

%\item Both $W$ and $W\times U$ have zero mean, i.e., $\mu_W = 0$ and $\mu_{W\times U} = 0$.

\end{enumerate}

\end{assumption}

 We remark that even though Assumption \ref{assump:distributions}-(i) implies that $V_{it} \perp V_{jt}$ for any two firms $i \neq j$, the total variable costs $V_{it}^\ast = V_{it} + W_t$ can be correlated across firms because of $W_t$. We can interpret $W_t$ as an unobserved technology shock that shifts production costs for all the firms. Henceforth, we refer to $V_{it}$ as firm $i$'s private cost shock and $W_t$ as the common cost shock observed by all the firms. Furthermore, throughout this section, we allow the common cost shock $W_t$ and the demand shock $U_t$ to be correlated. 

We have made several assumptions about the supports in light of our empirical application and model tractability. First, we assume that the demand shock has a positive lower bound, $\underline{u}>0$, a reasonable assumption because $\underline{u}\leq 0$ in Equation (\ref{eq:ut}) would imply a zero demand with positive probability. Second, we follow the extant literature on games with incomplete information and assume that $V$ and $W$ have bounded support. These assumptions, together with Assumption \ref{assump:distributions2} defined shortly below, ensure that firms' ex-ante expected profit is finite and that private costs, equilibrium outputs, and market-clearing prices are nonnegative. 

Thus, we can allow the upper bounds $\bar{v}_i$ to be unbounded, as long as the private cost distribution $F_{V_i}$ is such that the ex-ante expected profit is finite. However, if firms' costs are unbounded from above, firms probably will not produce anything; however, we do not observe zero production in our sample.

In the rest of this section, we present the timing of the game and derive the equilibrium strategies for which we assume that (i) the market-clearing condition holds in each market, i.e., aggregate consumption equals total output, (ii) there is no fixed cost of production, and (iii) the joint distribution $F_{\mathbf{V}, W, U}$ is common knowledge among all firms. 

Specifically, in market $t$, nature draws $(\mathbf V_t , W_t, U_t)\sim F_{\mathbf{V}, W, U}$ and each firm $i$ observes its private cost $V_{it}$, as well as $(W_t , U_t)$. Then all firms simultaneously choose their outputs, and the market clears. 
We consider static Bayesian Cournot-Nash equilibria in pure strategies for each market. The common shocks $(W_t, U_t)$ are observed by all the firms, so they can be treated as commonly known constants when choosing the (expected) profit-maximizing output. For a given $(V_{it}, W_t , U_t) = ( v , w , u)$ and given strategies of the opponents $ \q_j (\cdot, w , u):[\underline{v}_j,\bar{v}_j] \rightarrow \mathbb{R}_+$, where $j\neq i$, firm $i$ chooses its quantity that maximizes its expected profit:
\begin{equation}
\max_{q \geq 0 } \ \underbrace{ q \times \E\left\{ \mathfrak p( q + {\q}^+_{-i}(\mathbf{V}_{-i , t} , w , u ) , u) | V_{it}= v, W_t = w, U_t = u \right\}}_{\texttt{revenue at interim expected market-clearing price}}- \underbrace{ (( v + w ) q + \frac{\lambda}{2} q^2)}_{\texttt {variable cost}} ,
\label{eq:interprofit}
\end{equation}
where $\mathfrak{\q}^+_{-i}(\mathbf{V}_{-i,t} , w , u ) = \sum_{\substack{j\neq i} }{\q}_j (V_{jt}, w , u)$ is the total quantities produced by $i$'s opponents, and the expectation is with respect to $i$'s interim belief about its opponents' costs $\mathbf{V}_{-i,t} :=(V_{1t},\cdots, V_{i-1,t}, V_{i+1,t}, \cdots, V_{\mathcal I,t})$ is distributed as $\mathbf{V}_{-i} : =(V_{1},\cdots, V_{i-1}, V_{i+1}, \cdots, V_{\mathcal I}) \sim \prod_{\substack{j\neq i}} F_{V_{j}}(\cdot) $. Then the equilibrium strategy must satisfy the following first-order condition:
\begin{eqnarray}
\q_i ( v , w , u ) & = & \frac{u -\beta \E[{\q}^+_{-i}(\mathbf{V}_{-i,t}, w , u ) | V_{it}= v, W_t = w, U_t = u ] - w - v}{\lambda + 2\beta} \nonumber \\
& = & \frac{u -\beta \E[{\q}^+_{-i}(\mathbf{V}_{-i,t},w,u) ] - w - v}{\lambda + 2\beta},
\label{strategyFOC} 
\end{eqnarray}
where the second equality follows from Assumption \ref{assump:distributions}. Thus, the equilibrium strategies are linear in private costs. We impose additional assumptions on the parameters and their supports to guarantee a unique solution with nonnegative quantities and a market-clearing price.

\begin{assumption} \label{assump:distributions2}
We have that $\beta>0$, $\lambda\geq0$, and {\footnotesize\begin{equation*}
 \frac{(\lambda + \beta) \underline u }{\lambda + (\mathcal I + 1)\beta} + \beta \sum_{i=1}^\mathcal I \left[\frac{\underline v_i - \mu_{V_i} }{\lambda + 2\beta} +\frac{1}{\lambda + (\mathcal I+1)\beta} \left\{ \underline w + \frac{1}{\lambda + \beta} \left[ (\lambda +\mathcal I \beta) \mu_{V_i} - \beta \sum_{j \neq i} \mu_{V_j} \right]\right\} \right] \geq 0 .
\end{equation*}}Furthermore, \begin{equation*}
 \frac{1}{\lambda + (\mathcal I+1)\beta} \left\{ \underline u - \bar w - \frac{1}{\lambda + \beta} \left[ (\lambda +\mathcal I \beta) \mu_{V_i} - \beta \sum_{j \neq i} \mu_{V_j} \right]\right\} - \frac{\bar{v}_i - \mu_{V_i} }{\lambda + 2\beta}\geq 0 \quad \forall \ i \in \mathscr I.
\end{equation*}
\end{assumption}

The first part of Assumption \ref{assump:distributions2} is a technical requirement that ensures a nonnegative market-clearing price; see \cite{EinyHaimankoMorenoShitovitz2010} and \cite{Hurkens2014} for a detailed discussion on this topic. The second part ensures that it is always profitable for every firm to produce. In particular, it implies that even when firm $i$ realizes the highest cost and demand is the lowest, it is still profitable for such a firm to choose nonnegative output.
 Note that Assumption \ref{assump:distributions2} is automatically satisfied, e.g., when $\underline u $ is sufficiently large in comparison with the upper boundaries $\{ \bar v_1,\dots, \bar v_\mathcal{I}, \bar w \}$. Alternatively, these boundaries can be arbitrarily large if we allow $\underline{u}$ to be sufficiently large.

The following lemma, which builds on \citet[Proposition 1]{viv02}, establishes the existence and uniqueness of the Bayesian Cournot-Nash equilibrium in strictly increasing strategies. The proof of the lemma is in Appendix \ref{section:proof}. 
\begin{lemma}	\label{lem:strategies}
If Assumptions \ref{assump:distributions} and \ref{assump:distributions2} hold, there exists a unique Bayesian Cournot-Nash equilibrium. Specifically, $i$'s equilibrium strategy $\mathfrak \q_i (\cdot , w, u):[\underline{v}_i,\bar{v}_i] \rightarrow \mathbb{R}_+$ is 
\begin{equation*}
\mathfrak{\q}_i (v_i , w, u) = \frac{1}{\lambda + (\mathcal I+1)\beta} \left\{ u - w - \frac{1}{\lambda + \beta} \left[ (\lambda + \mathcal I\beta) \mu_{V_i} - \beta \sum_{j \neq i} \mu_{V_j} \right]\right\}- \frac{v_i - \mu_{V_i} }{\lambda + 2\beta} . 
\end{equation*} 
\end{lemma}

This lemma states that the equilibrium strategy for a firm is linear and strictly decreasing in its private cost. Each firm responds to the average cost type of its opponent, so the equilibrium may not be Pareto efficient because some firms may produce more than the socially optimal quantities. Finally, we remark that if the parameters $(\beta, \lambda )$ and the shocks $(V_{it}, W_t , U_t )$ are scaled by some constant $c >0$, then the equilibrium quantities will not be affected, but the new equilibrium price will be $c \times P_t$. 
Thus the observed price and quantities cannot be rationalized by two sets of structural parameters if one of them is a scaled version of the other, aiding in the identification as we study next.

%%%%%%%%%%%%%%%%%%%%%%%%%%%%%%%%%%%%%%%%%%%%%%%%%%%%
%%%%%%%%%%%%%%%%%%%%%%%%%%%%%%%%%%%%%%%%%%%%%%%%%%%%
%%%%%%%%%%%%%%%%%%%%%%%%%%%%%%%%%%%%%%%%%%%%%%%%%%%%

\section{Identification} \label{section:identification}

In this section, we study the identification of our model and propose a constructive multi-step identification strategy. More specifically, we determine conditions on our model and the data under which we can use the joint CDF of the equilibrium prices and quantities $F_{P, \mathbf Q}$, where $P = \mathfrak p (Q^+,U) $, $Q^+ = \sum_{i \in \mathscr I} Q_i$, $Q_i = \mathfrak q_i (V_i , W,U)$ for $i \in \mathscr I$, and $\mathbf Q = (Q_1,\dots, Q_\mathcal{I})$ to uniquely determine all the model parameters. Recall that our model parameters are (i) the slope of the demand function, $\beta$, (ii) the marginal CDF of the demand shock $F_U$, (ii) the parameter of the cost function, $\lambda$, (iii) marginal distributions of private costs, $\{F_{V_i}: i \in \mathscr I\}$, and (iv) the conditional CDF of the technological shock $W$ given $U$, $F_{W|U}$. Even though we do not know $F_{P, \mathbf Q}$, in practice, we can consistently estimate it from the observables $\{ (P_t, \mathbf Q_{t}) : t =1,\dots, \mathcal T \}$ as $\mathcal T \rightarrow \infty$, where $P_{t} =\mathfrak p (Q_t^+,U_t) $ is the market-clearing price in market $t$, $Q_t^+ = \sum_{i \in \mathscr I} Q_{it}$, $Q_{it} = \mathfrak q_i (V_{it} , W_t,U_t)$, and $\mathbf Q_{t}= (Q_{1t}, \ldots, Q_{\mathcal I t})$ is the output produced by the firms in market $t$. Thus, the data can be interpreted as realizations of our Bayesian Cournot-Nash model over many markets.

To simplify the exposition hereafter, we do not consider additional exogenous features that can affect the costs or the demand, even though we can accommodate such features as follows. Let $\mathbf{X}_t = ( X_{1t},\dots,   X_{\mathcal{I}t})$ be observed firms' characteristics that affect private costs and that are common knowledge among firms. We can then model $V_{it} \big|_{X_{it}}  \sim F_{V_i | X_i}(\cdot|X_{it})$ for each $i \in \mathscr{I}$, as well as $W_t\big|_{\mathbf{X}_t} \sim F_{W | \mathbf{X}} (\cdot , \cdot | \mathbf{X}_t)$, and build our identification strategy from the conditional distribution $F_{P, \mathbf Q | \mathbf{X}}$. Similarly, we can accommodate demand shifters such as income and demographic characteristics. To determine the limits of our identification strategy from relying solely on the game-theoretic structure, we do not consider observed firms' characteristics in the remainder of this paper.

Before presenting the formal identification results, for intuition, we sketch the idea and discuss the identifying variations for a simplified case when there are two symmetric firms with $\mu_{V_1} = \mu_{V_2} =: \mu_V$. 
The market demand in period $t$, denoted as $Q_t^d $ in (\ref{eq:demand}), is 
\begin{eqnarray}
 Q_t^d = \frac{U_t}{\beta}- \frac{1}{\beta} P_t,
 \label{eq:demand0} 
\end{eqnarray}
where $U_t$ can be interpreted as an exogenous demand shifter. On the supply side, firm $i$'s first-order conditions can be written as 
\begin{equation}
U_t - \beta [Q_{it}+ \E_t (Q_{jt}) ] - Q_{it} (\beta + \lambda)= V_{it} + W_t , \ \ j\neq i, i, j\in\{1,2\}, 
\label{eq:supply2}
\end{equation}
where from Lemma \ref{lem:strategies} we know $\E_t (Q_{jt}) = (U_t - W_t - \mu_V)/(3\beta + \lambda)$. 
Thus, the average $\E_t (Q_{jt})$ depends on both $U_t$ and $W_t$ because the firms observe them before they choose their productions. 
To express the market supply $Q_t^s := Q_{1t} + Q_{2t}$, where the superscript $s$ denotes total supply, as a function of the price $P_t$ and the supply shocks $(V_{1t} , V_{2t} , W_t )$, we substitute $U_t = P_t + \beta Q_t$ in (\ref{eq:supply2}) and sum over the two firms, which gives us 
\begin{eqnarray}
 Q_t^s = \frac{2}{\lambda + \beta} P_t + \frac{-2}{\lambda + \beta} (W_t + \mu_V) + \frac{-(\lambda+3\beta)}{(\lambda +\beta )(\lambda + 2\beta )}(V_{1t} + V_{2t} - 2 \mu_V).
 \label{eq:supply0}
\end{eqnarray}

Equations (\ref{eq:demand0}) and (\ref{eq:supply0}) simultaneously determine the equilibrium price and quantity $(P_t, Q_t)$ under the equilibrium condition $Q_t^d = Q_t^s$. If it is written this way, we can interpret $U_t$ as an exogenous demand shifter and $(W_t, V_{1t}, V_{2t})$ as exogenous supply shifters, which are crucial for the identification. To wit, if we condition on $Q_{1t}$ being at the lowest, $Q_{1t} = \underline q_1$, the strict monotonicity of the equilibrium strategy implies that $(U_t,W_t ,V_{1t}) = (\underline u, \bar w, \bar v_1)$. Thus conditioning on $Q_{1t} = \underline q_1$, we can see that in Equations (\ref{eq:demand0})-(\ref{eq:supply0}) under the equilibrium condition $Q_t^d = Q_t^s$, $P_t$ and $Q_{2t}$ are pinned down by $V_{2t}$. Thus, as $V_2$ varies across $t$, $Q_{2t}$ and $P_{2t}$ vary, which in turn identifies the demand slope $\beta$. Once we know $\beta$, we can identify the demand shock (i.e., the intercept) $U_t$ from (\ref{eq:demand0}). 

Next, focusing on each firm separately, we can use the fact that output decreases with its costs. If we ignore $W_t$, this monotonicity of the equilibrium allows us to use the distribution of production to identify the distribution of private costs. But with $W_t$, we identify the distribution of $V_{it}^\ast= V_{it}+W_t$. To separately identify the distribution of $V_{i}$ from $W$, we use a deconvolution method exploiting the fact that $W_t$ is common across firms.

Before we proceed, we make the following normalization assumption, important for the identification. 
\begin{assumption}

\label{assump:distributions_new}
 Both $W$ and $W\times U$ have zero mean, i.e., $\mu_W = 0$ and $\mu_{W\times U} = 0$.
\end{assumption}

Assumption \ref{assump:distributions_new} is a technical assumption that is helpful in the identification as it is a location normalization. Essentially, $\mu_W = 0$ is a normalization that allows us to identify the means of the private cost shocks $\mu_{\mathbf{V}}$, while $\mu_{W\times U} = 0$ is similar to the exogeneity assumption used in nonlinear models. Clearly, $\mu_W = 0$ and $\mu_{W\times U} = 0$ imply that $W$ and $U$ must be uncorrelated.

\subsection*{Demand Parameters}

We begin with identifying $\beta$ and the marginal CDF $F_{U}$. For this purpose, consider firm $i\in \mathscr I$ and the range of its output $Q_i$, which is given by $[\underline q_i , \infty)$ with $\underline q_i \geq 0 $ (Lemma \ref{lem:testableimplications}). This interval can be identified from $F_{P,\mathbf Q}$, as the support of $F_{Q_i}$. Strict monotonicity of the equilibrium strategy (Lemma \ref{lem:strategies}) implies that $i$'s smallest output is associated with its highest cost and the smallest demand shock, i.e., $\underline q_i = \mathfrak{\q}_i (\bar v_i , \bar w, \underline u)$. In other words, the event $\{Q_i = \underline q_i \}$ is equivalent to $\{(V_i, W, U) = (\bar v_i, \bar w, \underline u)\}$.

Now consider the total output of firm $i$'s competitors, $Q^+_{-i} = \sum_{j\neq i} Q_j$. From Lemma \ref{lem:testableimplications}, we know that $F_{Q^+_{-i} | Q_i} (\cdot| \underline q_i)$ is continuous and is supported on $[\underline q_{-i} , \infty)$. Moreover, its density $f_{Q^+_{-i} | Q_i} (\cdot| \underline q_i)$ is strictly positive in the interior of this set, which implies that the conditional quantile function $F_{Q^+_{-i} | Q_i}^{-1}(\cdot|\underline q_i)$ is a well-defined and strictly increasing function. Thus, for any two distinct $\alpha, \alpha^{\prime}\in[0,1]$, from the inverse demand function (\ref{eq:demand}) we obtain 
\begin{eqnarray*}
F^{-1}_{P | Q_i} (\alpha| \underline q_i) &=& \underline u - \beta \times \left[ \underline q_i + F^{-1}_{Q^+_{-i} | Q_i} \left(1- \alpha| \underline q_i \right) \right] , \\
F^{-1}_{P | Q_i} (\alpha^\prime| \underline q_i) &=& \underline u - \beta \times \left[ \underline q_i + F^{-1}_{Q^+_{-i} | Q_i} \left(1- \alpha^\prime| \underline q_i \right) \right].
\end{eqnarray*}
So the slope parameter can be identified by subtracting the first equation from the second:
\begin{equation}
\beta = \frac{F^{-1}_{P | Q_i} (\alpha^\prime | \underline q_i) - F^{-1}_{P | Q_i} (\alpha| \underline q_i)}{ F^{-1}_{Q^+_{-i} | Q_i} (1 - \alpha| \underline q_i ) - F^{-1}_{Q^+_{-i} | Q_i} (1 - \alpha^\prime| \underline q_i ) },
\label{eq:idb}
\end{equation}
Note that $\alpha\neq \alpha^{\prime}$ implies $F^{-1}_{P | Q_i} (\alpha| \underline q_i)\neq F^{-1}_{P | Q_i} (\alpha^{\prime}| \underline q_i)$ and $F^{-1}_{Q^+_{-i} | Q_i} (1 - \alpha| \underline q_i ) \neq F^{-1}_{Q^+_{-i} | Q_i} (1 - \alpha^\prime| \underline q_i )$, so the denominator on the RHS of (\ref{eq:idb}) is nonzero and $\beta$ is well defined. Heuristically, the slope of the demand function is identified by the ``derivative" of the inverse demand function with respect to the equilibrium quantities produced by the other firms while holding $Q_i$ at $\underline{q}_i$. 
The choice of $i$ and the quantiles were arbitrary, suggesting that $\beta$ is over-identified. Once $\beta$ is identified, we can recover the demand shock as $U =P + \beta Q^+$ and identify its CDF as $F_U(u) =F_{ P + \beta Q^+} (u)$ for $u\in\mathbb{R}$.

\subsection*{Cost Parameter}

Next, we consider identifying the common cost parameter $\lambda>0$. 
In particular, we can use the variation in the output produced $Q_i$ that can be explained by variation in $U$ across markets to identify $\lambda$. A high value of $\lambda$ means the marginal cost is increasing, so even if the demand increases because $U$ increases, in equilibrium firm $i$'s output $Q_i$ does respond, and vice versa. To formalize this intuition, for $i \in \mathscr I$, let
 \begin{eqnarray*}
\gamma_{0,i} & = & \frac{ - 1}{[\lambda + ( \mathcal I+1)\beta](\lambda + \beta)} \left[ (\lambda + \mathcal I \beta) \mu_{V_i} - \beta \sum_{j \neq i} \mu_{V_j} \right], \\ 
\gamma_{1} & = & \frac{1}{\lambda + (\mathcal I+1)\beta},\quad \text{and,}\quad 
\tilde{V}_{i} = \frac{\mu_{V_i} - V_{i} }{\lambda + 2\beta} - \frac{W}{\lambda + (\mathcal I+1)\beta}.
\end{eqnarray*}
Then, after substituting $U =P + \beta Q^+$ and $(\gamma_{0,i}, \gamma_{1}, \tilde{V}_{i} )$ in firm $i$'s equilibrium strategy (Lemma \ref{lem:strategies}), we obtain the following linear expression:
\begin{equation}
Q_{i} = \gamma_{0,i} + \gamma_{1} \left( P + \beta Q^+ \right) + \tilde{V}_{i} .
\label{eq:strategyOLS} 
\end{equation}Assumption \ref{assump:distributions} implies that $\mu_{U \times \tilde{V}_{i} } = 0$, which in turn implies that the ``regressor" $ P + \beta Q^+$ and the ``error" $\tilde{V}_{i}$ satisfy the orthogonality condition that allows us to identify the slope as \begin{equation}
\gamma_1 = \frac{\mathrm{cov}(Q_i,P + \beta Q^+)}{\mathrm{var}(P + \beta Q^+)}, \label{eq:estimationm} 
\end{equation}
which in turn identifies the $ \lambda = \frac{1}{\gamma_1} - (\mathcal I + 1) \beta$.

\subsection*{Distributions of Cost Shocks}

In this subsection, we focus on identifying the marginal CDFs of the private cost shocks, i.e., $\{F_{V_1},\dots,F_{V_\mathcal{I}} \}$, and the joint distribution of common shocks $F_{W,U}$. Here, our identification strategy relies on the variation in firms' output, and the equilibrium strategies are linear and strictly decrease private costs.

The intuition behind our identification approach is that all else equal, firms with higher costs choose lower quantities than firms with lower costs. So, for any two firms $i\neq j$, 
if we hold firm $j$'s output fixed at its lowest level, $\underline q_j$, the conditional quantile of $i$'s output, $Q_i$, can be expressed as a linear function of the quantile function of $V_i$ because $Q_j = \underline q_j$ implies $(W,U) = (\bar w, \underline u)$, while the distribution of $V_i$ is unaffected by independence. Then the variation in the conditional quantiles of $Q_i$ identifies $F^{-1}_{V_i}$ and hence $F_{V_i}$. 

Next, if we keep $U$ fixed, $Q_i$ is a linear combination of the firm-specific shock $V_i$ and the common-cost shock $W$. Then, once we identify $F_{V_i}$, we can identify the characteristic function of $W$ using a deconvolution method, which uniquely identifies the CDF $F_{W}$ because there is a one-to-one correspondence between a CDF and a characteristic function. In the remainder of this section, we formalize these arguments.

We begin by identifying the means $\mu_{\mathbf V } = (\mu_{V_1},\dots, \mu_{V_\mathcal{I}} )$. After applying the law of iterated expectations to the equilibrium first-order condition (\ref{strategyFOC}), for any $i\in \mathscr I$, we obtain
\begin{equation}
\mu_{V_i} = \mu_{P + \beta Q^+} - \beta \mu_{Q^+_{-i}} - (\lambda + 2\beta)\mu_{Q_i},
\label{eq:idmuthetai}
\end{equation}
where all the parameters on the right-hand side are known or have been identified.

We are ready to state the following result that shows how we can use $F_{P, \mathbf{Q}}$ to identify the distributions mentioned above nonparametrically. Let $\varphi_{W | U} (\cdot | u ) $ denote the conditional characteristic function of $W$ given $U = u$, i.e., $\varphi_{W | U} (z |u) = \E \left[\exp\left( {\bf i}z \times W \right) \middle| U = u \right] $, where $z \in \mathbb R$ and ${\bf i}=\sqrt{-1}$ denotes the imaginary unit.

\begin{theorem} \label{thm:identification}

Suppose that $F_{P, \mathbf{Q}}$ is known and that Assumptions \ref{assump:distributions}, \ref{assump:distributions2} and \ref{assump:distributions_new} hold.

\begin{enumerate}

\item Then, for any $i\in \mathscr I$, $F_{V_i}$ is identified from the conditional distribution $F_{Q_i | Q_j}$ as
 \begin{equation}
F_{V_i}( v) = 1 - F_{Q_i | Q_j} \left[ - \frac{v - \mu_{V_i}}{\lambda + 2\beta} + \mu_{Q_i | Q_j } \left( \underline q_j \right) \middle| \underline q_j \right], \quad v \in \mathbb R, j\neq i.
\label{eq:idFthetai}
\end{equation}
\item For any $(w,u) \in \mathbb R \times [\underline u , \infty )$, we can identify the conditional characteristic function of $W$ given $U =u$ as 
 \begin{eqnarray}
\varphi_{W | U} (w | u) & = & \exp\left[ {\bf i} w \left\{ u - \frac{1}{\lambda + \beta} \left[ (\lambda + \mathcal{I} \beta ) \mu_{V_i} - \beta \sum_{j \neq i} \mu_{V_j} \right]\right\} \right] \nonumber \\
& & \qquad \quad \times \ \frac{ \E \left[ \exp\left\{ - {\bf i} w \left[ \lambda + (\mathcal{I}+1)\beta \right] Q_i \right\} \middle| P + \beta Q^+ = u \right] }{ \E \left[ \exp\left\{ {\bf i} w \left[ \lambda + (\mathcal{I}+1)\beta \right] \frac{V_i - \mu_{V_i} }{\lambda + 2\beta} \right\} \right] },
\label{eq:idFupsiloneta}
\end{eqnarray}
which in turn identifies $F_{W|U}(\cdot|u)$ by uniqueness of the characteristic function.
\end{enumerate}

\end{theorem}

The first part of this theorem provides a closed-form expression for $F_{V_i}$ in terms of $F_{Q_i | Q_j}, \mu_{Q_i | Q_j } $, and $\{ \beta,\lambda, \mu_{V_i}\}$. Our identification strategy relies essentially on the linearity and strict monotonicity of the equilibrium strategies and the independence between $\mathbf{V}$ and $(W, U)$. Here, note that the distribution $F_{V_i}$ is over-identified because when $\mathcal I>2$, for each $i$, there is more than one $j\neq i$.

The second part of Theorem \ref{thm:identification} provides an expression for the conditional characteristic function associated with $F_{W|U}(\cdot|\cdot)$. We derive this expression from the equality
\begin{eqnarray}
 - [ \lambda + (\mathcal I+1)\beta ] Q_i + U + \frac{1}{\lambda + \beta} \left[ (\lambda + \mathcal I \beta) \mu_{V_i} - \beta \sum_{j \neq i} \mu_{V_j} \right] 
 = \left[ \lambda + (\mathcal I+1) \beta \right] \frac{V_i - \mu_{V_i} }{\lambda + 2\beta} + W   \   \ \  \label{eq:deconvo}  
\end{eqnarray}
that follows by Lemma \ref{lem:strategies}. Observe that the left-hand side is observable at this step of the identification process, while the first term on the right hand is unobservable, but its distribution is known. Thus, the distribution of $W$ can be recovered by applying deconvolution techniques usually employed in panel data, and error components models \cite[see, e.g.,][]{hm96}. 
We note that if firms were symmetric, (\ref{eq:deconvo}) could also be used to identify $F_V$ as an alternative to the first part of this theorem.

The deconvolution method identifies conditional characteristic function $\varphi_{W | U} (\cdot | \cdot)$ as a function of the data $F_{P,\mathbf Q}$ and $F_{V_i}$. 
There is a one-to-one mapping between the conditional characteristic function and the conditional CDF, identifying $F_{W|U}(\cdot| u ), \forall u \in [\underline u ,\infty)$; indeed, $F_{W|U}$ is overidentified as $i\in \mathscr I$ in (\ref{eq:idFupsiloneta}) is arbitrary. Formally, if the conditional PDF $f_{W|U}(\cdot|u)$ satisfies the regularity conditions in \citet[Theorem 3]{Shephard1991}, then $\forall w \in \mathbb R$, we obtain \begin{equation*}
F_{W|U}(w | u )=\frac{1}{2}-\frac{1}{2\pi} \int_0^{\infty}\left[\frac{\varphi_{W | U} (z | u )\exp(-{\bf i} z w)}{\imath z}+ \frac{\varphi_{W | U} (-z | u) \exp({\bf i} z w)}{-\imath z}\right]dz.
\end{equation*}Heuristically, given $U=u$, the unobserved common technological shock generates dependence between quantities produced in the same market. We can use this dependence to recover $F_{W|U}$. In particular, the marginal distributions of $Q_i$'s are insufficient to identify the conditional CDF because there is no unique decomposition of the sum $V_i^*$ into its common and individual components $V_i$ and $W$. Finally, we can identify the joint CDF of $(W,U)$ at $(w, u) \in \mathbb R^2$ as \begin{equation*}
F_{W,U}( w, u ) = \left\{ \begin{array}{ll}
0 & \text{ if } u < \underline u , \\
\int_{\underline u}^u F_{W| U}(w| z) f_U (z) dz& \text{ if } \underline u \leq u .
\end{array}\right.
\end{equation*}

\section{Data} \label{section:data}

For our application, we consider the global market for crude-oil from January 1992 to December 2019. Besides being an important industry globally, crude-oil is a homogeneous product where producers compete in quantities, making it an appropriate application. The production data are available from the \emph{Monthly Energy Review} published by the U.S. Energy Information Administration.\footnote{The \emph{Monthly Energy Review} is available from this website \url{https://rb.gy/rygmcz}.} We observe the monthly productions of 20 major oil-producing countries. We treat each country as a competing firm. 

\begin{figure}[t!]
\caption{Monthly Crude-Oil Production by Countries (mm barrels) \label{fig:rawoutput}}
\hspace{-0.5in}
 \includegraphics[scale=.55]{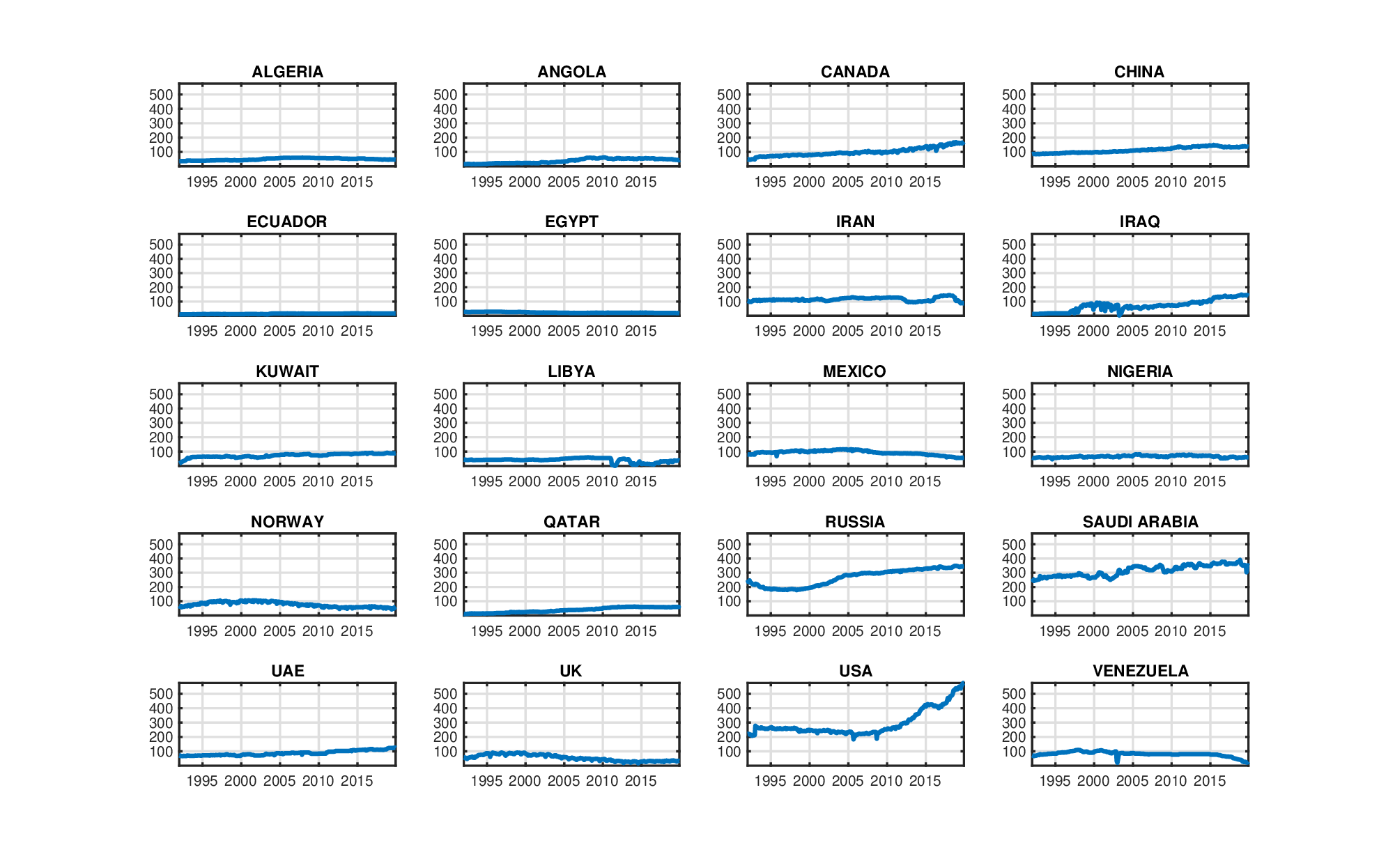}
 \caption*{\footnotesize {\bf Notes.} Monthly production of crude oil (in millions of barrels), by countries. Source of the data: \emph{Monthly Energy Review} published by the U.S. Energy Information Administration.}
\end{figure}

Figure \ref{fig:rawoutput} displays time series of monthly output data (measured in millions of barrels), while Table \ref{tab:summary} reports summary statistics. As we can observe, all countries produce strictly positive output each month, which is consistent with the second part of Assumption \ref{assump:distributions2}. We may observe zero production if we consider production over a shorter period. We also see that countries differ in their productions, from Ecuador, Egypt, and Libya at the lowest end of production to Russia, Saudi Arabia, and the U.S. at the highest. These differences suggest cost asymmetry. Furthermore, the productions have a time trend, so we de-trend them first. 

\begin{figure}[ht!]
\caption{Price of Crude Oil }\label{fig:prices}
 \begin{center}
 \includegraphics[scale=0.45]{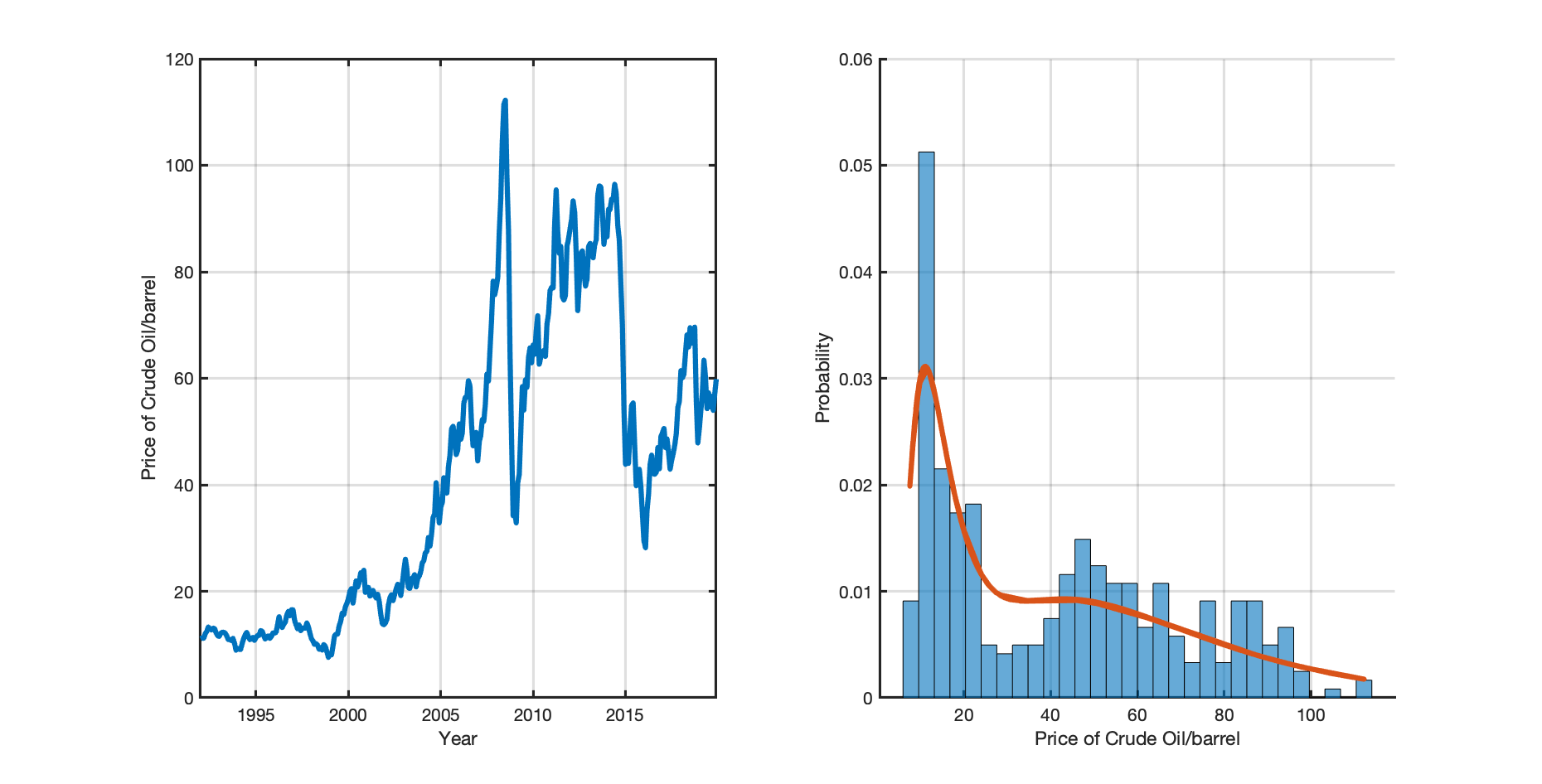}
 \end{center}
 \caption*{\footnotesize {\bf Notes.} Left: time series of the monthly price of crude-oil per barrel, in 2019:Q4 U.S. dollars. Right: histogram and estimated density of the price in our sample.}
\end{figure}

We also observe oil prices (per barrel) published by the St. Louis Federal Reserve. Figure \ref{fig:prices} shows the time series and histogram of these prices, expressed in 2019:Q4 U.S.\ dollars. The prices range from \$7.57 to \$112.24, with a mean of \$41.43 and a standard deviation of \$27.71.

\begin{table}[ht!]
\caption{Summary Statistics of Crude-Oil Production by Countries \label{tab:summary}}
\begin{center}
\begin{tabular}{lllll}
\toprule
\textbf{Countries} & \textbf{Min} & \textbf{Mean} & \textbf{Std.\ Dev.} & \textbf{Max} \\ \midrule
Algeria & 35.85 & 49.32 & 7.2 & 59.85 \\
Angola & 14.49 & 37.8 & 15.83 & 61.5 \\
Canada & 45.51 & 99.3 & 28.69 & 171.3 \\
China & 83.76 & 113.78 & 19.25 & 149 \\
Ecuador & 8.91 & 13.963 & 2.3 & 17.1 \\
Egypt & 18.6 & 22.79 &2.9&28.53 \\
Iran & 87.39&115.51&12.11&144.84 \\
Iraq & 1.89&71.43&38.86&148.98 \\
Kuwait & 17.43&72.88&12.91&91.26 \\
Libya & 0.6&40.5&13.5&59.91 \\
Mexico & 55.23&92.69&15.46&116.88 \\
Nigeria & 46.35&65.99&6.53&80.85 \\
Norway &41.9&77&17.72&107.37 \\
Qatar & 9.75&38.4&17.59&61.92 \\
Russia & 177.8&266.84&58.62&349.83 \\
Saudi Arabia & 243.27&314.94&38.65&390.5 \\
U.A.E & 66.51&87.66&16.22&126.27 \\
U.K. & 16.02&54.64&21.64&91.8 \\
U.S.A. &183.39&293.82&91.23&575.91 \\
Venezuela & 20.4&80.6&16.8&110.13\\ \bottomrule  
\end{tabular}
\caption*{\footnotesize {\bf Notes.} Table displays summary statistics of monthly output (in millions of barrels) by country. Our sample period is from January 1992 to December 2019.} 
\end{center}
\end{table}

\section{Estimation\label{subsec:est}}

In this section, we propose a MLE procedure based on observed prices and quantities $\{(P_t,\mathbf Q_t): t=1,\dots, \mathcal T=336\}$. Although our identification is semi-nonparametric, we make parametric assumptions due to our sample size. In particular, we assume that the distributions of $(U, W)$ and the private costs belong to certain parametric families. Moreover, to capture the time-series nature of our data, we also allow time trends in $U_t$ and $V_{i,t}$, $i \in \mathscr I$. Finally, in Subsection \ref{sec:mc} below, we provide a Monte Carlo experiment to evaluate the finite-sample performance of the proposed estimators.

We begin with the following assumption that will allow us to de-trend $\{ \mathbf Q_t \}$. 
\begin{assumption} \label{assump:timetrend}
\begin{enumerate}[label=(\roman*)]
\item For each $i \in \mathscr I$, $V_{it}$ can be expressed as
$
V_{it} = \tau_{i}^s e^{ - \tau t} + V_{it}^{\mathrm{dt}},
$
where $\tau_{i}^s, \tau \geq 0$ are parameters and $\{ (V_{1t}^{\mathrm{dt}},\dots ,V_{\mathcal I t}^{\mathrm{dt}})\}$ are IID.

\item The demand shock can be expressed as $U_t = \tau^d e^{- \tau t} + U_t^{\mathrm{dt}}$, where $\tau^d , \tau \geq 0$ are parameters and $\{ U_t^{\mathrm{dt}} \}$ is strictly stationary and ergodic.

\item The parameters $(\tau^d, \tau^s, \lambda, \beta)$ satisfy the relationship $\tau^d = \frac{-\beta}{\lambda+\beta} \sum_{i \in \mathscr I} {\tau_{i}^s}$.

\item The technology shock process $\{W_t\}$ is strictly stationary and ergodic. 

\end{enumerate}
\end{assumption}
\noindent Assumptions \ref{assump:timetrend}-(i) and \ref{assump:timetrend}-(ii) impose an additively separable time trend. Additive separability can be restrictive, but these assumptions enable us to keep the model tractable. In particular, they will allow us to express the equilibrium outputs as an additively separable function of time-trend; see Equation (\ref{eq:apptrans}) below. Assumption \ref{assump:timetrend}-(iii) implies that the prices do not have a time trend; more specifically, that $\{ P_t\}$ is strictly stationary and ergodic. In our empirical application, $V_{it}^{\mathrm{dt}}$ captures the temporary or transient cost shock facing $i$, and the time-series components capture some persistent shocks. 

By applying Assumption \ref{assump:timetrend} to demand equation (\ref{eq:demand}) and to the equilibrium outputs from Lemma \ref{lem:strategies}, the equilibrium price and quantities can be written as
\begin{eqnarray}
\left( \begin{array}{c}
P_t \\
\mathbf Q_t 
\end{array} \right)  
& = & \mathds H_1 ( \lambda, \beta ) \left( \begin{array}{c}
U_t \\
\frac{W_t}{\lambda + (\mathcal I+1)\beta} + \frac{\mathbf V_t}{ \lambda + 2\beta}
\end{array} \right) + \mathds H_2 (\lambda, \beta, \mu_{\mathbf V_t}) 	 \nonumber \\
& = & e^{- \tau t} \left( \begin{array}{c} 0 \\
\frac{-1}{\lambda + \beta} \boldsymbol{\tau}^s \end{array} \right) + \mathds H_1 ( \lambda, \beta ) \left( \begin{array}{c}
U_t^{\mathrm{dt}} \\
\frac{W_t}{\lambda + (\mathcal I+1)\beta} + \frac{\mathbf V_t^{\mathrm{dt}} }{ \lambda + 2\beta}
\end{array} \right) + \mathds H_2 (\lambda, \beta, \mu_{\mathbf V}^\mathrm{dt}) , \label{eq:apptrans}
\end{eqnarray}
where $ \boldsymbol{\tau}^s = (\tau_1^s , \dots, \tau_{\mathcal{I}}^s )$, $\mu_{\mathbf V}^{\mathrm{dt}} = (\mu_{V_1}^{\mathrm{dt}} ,\dots, \mu_{V_{\mathcal I}}^{\mathrm{dt}})$ is a $\mathcal I \times 1$ vector containing the means of the de-trended private-costs, $\mathds H_1 ( \lambda, \beta )$ is a $({\mathcal I}+1)\times ({\mathcal I}+1)$ matrix of the form
\begin{equation*}
\mathds H_1 ( \lambda, \beta ) = \left( \begin{array}{cc}
\frac{\lambda + \beta}{\lambda + (\mathcal I+1)\beta} & \beta \boldsymbol{\iota}_\mathcal{I}^\prime \\
 \frac{1}{\lambda + (\mathcal I+1)\beta} \boldsymbol{\iota}_\mathcal{I} & -\mathds{I}_{\mathcal I}
\end{array}\right) , 
\end{equation*}
where $\boldsymbol{\iota}_\mathcal{I}$ is a $\mathcal I \times 1$ vector of ones, $\mathds{I}_{\mathcal I}$ denotes the identity matrix of dimension $\mathcal I$, and $\mathds H_2 (\lambda, \beta, \mu_{\mathbf V}) $ is a $(\mathcal I+1) \times 1$ vector whose $(i+1)^{th}$ element is given by 
\begin{eqnarray*}
\mathds H_{2,i+1} (\lambda, \beta, \mu_{\mathbf V}) = \frac{-1}{[\lambda + (\mathcal I+1)\beta](\lambda + \beta)} \left[ (\lambda + \mathcal I\beta) \mu_{V_i} - \beta \sum_{j \neq i}\mu_{V_j}\right]+ \frac{ \mu_{V_i}}{\lambda + 2\beta}
\end{eqnarray*}
and the first element is given by $\mathds H_{2,1} (\lambda, \beta, \mu_{\mathbf V}) = - \beta \sum_{i \in \mathscr I} \mathds H_{2,i+1} (\lambda, \beta, \mu_{\mathbf V})$.

Next, we make the following assumptions about the distribution of $ (U_t^{\mathrm{dt}}, W_t, \mathbf V_t^{\mathrm{dt}} )$.
However, before that, we introduce a few new notations. Let $\mathcal{T}r{\mathcal N}$ denote a truncated normal random variable. With a slight abuse of notation, for given constants $(\zeta_0,\zeta_1)$, we write $Z\sim \zeta_0+ \zeta_1\times \mathrm{Beta}(a,b)$ to denote that $Z$ has the same distribution as $\zeta_0+ \zeta_1\times \mathrm{Beta}(a,b)$, where $\mathrm{Beta}^*(a,b)$ stands for a Beta random variable with parameters $(a,b)$.

\begin{assumption}\label{assumption:densities}
\begin{enumerate}[label=(\roman*)]
\item Demand Shock: Let $U_t^{\mathrm{dt}} \sim \mathcal{T}r{\mathcal N}( \mu_U , \sigma_U^2)$ on $[ \underline u , \infty )$ with $\underline u>0$. 
\item Technology Shock: Let $W_t \mid_{U_t^{\mathrm{dt}}=u} \sim \bar{w} \times \left( 2\mathrm{Beta}\left( a_w (u) , a_w (u) \right) - 1 \right)$ where $a_w (u) := \exp(\tilde a_1 + \tilde a_2 u)$ and $\bar w, \tilde a_1 , \tilde a_2>0$. 
\item Private Costs: For $i \in \mathscr I$, let $V_{it}^{\mathrm{dt}} \sim \bar w\times ( \mathrm{Beta}(a_i ,b_i) + 1)$ where $a_i>0$ and  $b_i>0$.
\end{enumerate}
\end{assumption}

Thus we assume that the cost distributions belong to the Beta family. Beta densities are versatile and are widely used to model many types of uncertainties, because it can be unimodal, increasing, decreasing, or constant depending on the values of the parameters; see \cite{JohnsonKotzBalakrishnan1994}.
From Assumption \ref{assumption:densities}-(ii), it follows that the common cost shock $W_t$ is supported on $[- \bar w, \bar w] $, so it can be negative with range. In contrast, the total de-trended marginal cost, $V_{it}^{\mathrm{dt}} + W_t$, has support $[0,2\bar w]$ with mean $\mu_{V_i}^{\mathrm{dt}} =(2 a_i + b_i ) \bar w /(a_i + b_i) $. 
In this parametric framework, the model parameters are 
\begin{equation*}
\theta := ( \beta,\lambda , \underline u, \mu_U , \sigma_U^2 , \bar{w},\tilde a_1 , \tilde a_2, a_1 ,b_1,\dots,a_\mathcal{I} ,b_\mathcal{I}).
\end{equation*}

To construct the estimator of $\theta$, the first step is to estimate $\tau$ and $\mathbf{c}_1 := \boldsymbol{\tau}^s/(\lambda+\beta)$ by nonlinear least squares. 
Specifically, letting $\mathbf c_2$ be a $\mathcal I \times1$ vector of constants, we set $(\hat\tau, \hat{\mathbf{c}}_1, \hat{\mathbf{c}}_2 ) = \mathrm{argmin} \sum_t \| \mathbf{Q}_t + \mathbf{c}_1 e^{- \tau t}- \mathbf{c}_2 \|^2$. 
Then let $\mathbf{Q}_t^\mathrm{dt} := \mathbf{Q}_t + \hat{\mathbf{c}}_1 e^{- \hat\tau t}= \mathbf{Q}_t + \frac{e^{- \tau t}}{\lambda+\beta} \boldsymbol{\tau}^s $ be the vector of de-trended quantities. 
After applying the change of variable formula to (\ref{eq:apptrans}) and since $\mathrm{det}[ \mathds H_1 ( \lambda, \beta ) ] = 1$ for all $\lambda,\beta>0$, the joint PDF of $(P_t,\mathbf{Q}_t^\mathrm{dt})$ can be written as 
\begin{equation*}
f_{P,\mathbf{Q}^\mathrm{dt}} (p , \mathbf q ; \theta) = f_{U^\mathrm{dt}, \tilde{\mathbf V}^\ast} \left( \mathds H_1 ( \lambda, \beta ) ^{-1} \left[ \left(\begin{array}{c}
p \\
\mathbf q 
\end{array}\right) - \mathds H_2 (\lambda, \beta, \mu_{\mathbf V}^\mathrm{dt}(\theta) ) \right] ; \theta \right) ,
\end{equation*}
where $ f_{U^\mathrm{dt}, \tilde{\mathbf V}^\ast}(\cdot; \theta)$ is the joint density of $
\left( U^\mathrm{dt}_t , \frac{W_t}{\lambda + (\mathcal I+1)\beta} + \frac{\mathbf V_t^\mathrm{dt}}{ \lambda + 2\beta} \right).
$
Then, the estimator of $\theta$ is obtained by maximizing the log-likelihood function over some compact set $\Theta$:
\begin{equation*}
\hat\theta = \underset{\theta \in \Theta}\arg\max\quad\!\! \sum_{t=1}^{\mathcal T} \log \left[ f_{U^\mathrm{dt}, \tilde{\mathbf V}^\ast} \left( \mathds H_1 ( \lambda, \beta ) ^{-1} \left[ \left(\begin{array}{c}
P_t \\
\mathbf Q_t^\mathrm{dt} 
\end{array}\right) - \mathds H_2 (\lambda, \beta, \mu_{\mathbf V}^\mathrm{dt}(\theta) ) \right];\theta \right) \right].
\end{equation*}

Next, we can use Assumption \ref{assumption:densities} to determine the closed-form expression for $ f_{U^\mathrm{dt}, \tilde{\mathbf V}^\ast}(\cdot; \theta)$. 
For notational simplicity, we suppress the dependence on $\theta$ and begin with $
 f_{U^\mathrm{dt}, \tilde{\mathbf V}^\ast} ( u , \tilde{\mathbf v}^\ast ) = f_{U^\mathrm{dt}} ( u ) \times f_{ \tilde{\mathbf V}^\ast | U^\mathrm{dt}} ( \tilde{\mathbf v}^\ast | u),$
where Assumption \ref{assumption:densities}-(i) implies that $ f_{U^\mathrm{dt}} $ is truncated-normal, and 
\begin{equation*}
f_{\tilde{\mathbf V}^\ast| U^\mathrm{dt}} ( \tilde{\mathbf v}^\ast | u) = \left(\frac{\lambda + 2\beta}{\bar w}\right)^{\mathcal I} \int_{\frac{-\bar w}{\lambda + (\mathcal I+1)\beta}}^{\frac{\bar w }{\lambda + (\mathcal I+1)\beta}} f_{ \tilde{\mathbf V}^\ast |\left( U^\mathrm{dt} , \frac{W}{\lambda + (\mathcal I+1)\beta} \right)} ( \tilde{\mathbf v}^\ast | u , \tilde w) \times f_{\tilde W|U^\mathrm{dt}} (\tilde w |u) d \tilde w .
\end{equation*}

From Assumption \ref{assumption:densities}-(ii) and the change of variable formula, it follows that  
\begin{equation*}
f_{\tilde W|U^\mathrm{dt}} (\tilde w | u) = \frac{\lambda + (\mathcal I+1)\beta}{2 \bar w} f_{\mathrm{Be}\left( a_w (u) , a_w (u) \right) } \left( \frac{\tilde w [\lambda + (\mathcal I+1)\beta]}{2 \bar w} + \frac{1}{2}\right) ,
\end{equation*}
where $f_{\mathrm{Be}\left( a_w (u) , a_w (u) \right) }$ denotes the Beta density with parameter $(a_w (u) , a_w (u))$. The conditional joint density of $\tilde{\mathbf V}^\ast:= (\tilde V^\ast_1,\dots, \tilde V^\ast_\mathcal{I})$ can be written as a product of re-scaled Beta densities as follows:
\begin{equation*}
f_{ \tilde{\mathbf V}^\ast |\left( U^\mathrm{dt} , \frac{W}{\lambda + (\mathcal I+1)\beta} \right)} ( \tilde{\mathbf v}^\ast | u , \tilde w) = \left(\frac{\lambda + 2\beta}{\bar w}\right)^{\mathcal I} \prod_{i \in \mathscr I} f_{\mathrm{Be}(a_i,b_i)} \left( \frac{(\tilde v^\ast_i - \tilde w )(\lambda + 2\beta )}{ \bar w } - 1 \right). 
\end{equation*}
The identification results in Section \ref{section:identification} imply that, for any two distinct parameters $\tilde\theta \neq \theta$, we have that $f_{P,\mathbf{Q}^\mathrm{dt}} (p , \mathbf q ; \tilde\theta) \neq f_{P,\mathbf{Q}^\mathrm{dt}} (p , \mathbf q ; \theta) $. If the log-likelihood function 
\begin{eqnarray}
\sum_t \log \left( f_{P,\mathbf{Q}^\mathrm{dt}} (P_t,\mathbf{Q}_t^\mathrm{dt} ; \tilde\theta) \right), \label{eq:loglikelihood}
\end{eqnarray}
is continuous at every $\tilde\theta\in \Theta$, with probability one, then the consistency of $\hat\theta$ would follow from \citet[Theorem 2.5]{NeweyMcFadden94}. However, in our setting, the supports of $U^\mathrm{dt}$, $W_t$, and $V_{it}^{\mathrm{dt}}$ depend on two unknown boundary-parameters, $\underline{u}$ and $\bar w$. So, the log-likelihood function in (\ref{eq:loglikelihood}) is discontinuous with positive probability. 

Establishing consistency of our estimator and obtaining the limiting distribution require extending \cite{chernozhukov2004likelihood} to allow for a nonseparable model given in (\ref{eq:apptrans}). We remark that \cite{chernozhukov2004likelihood}, considering an additive separable model, establish that the boundary-parameters converge at the rate of $\mathcal T$, and other regular parameters converge at the parametric rate $\sqrt{\mathcal T}$. Given our likelihood function, we conjecture that their results apply in our setting, but the formal proof is beyond the scope of this article. 

However, to evaluate the finite-sample performance of the estimator, in the following subsection, we use Monte Carlo experiments. Furthermore, to build the confidence intervals (see the application in Section \ref{section:empirical} below), we apply the subsampling procedure (to the de-trended data) described in \cite{PolitisRomanoWolf1999} Chapter 3, for stationary time series because of its robustness properties.

\subsection*{Monte Carlo Experiments}		\label{sec:mc}

In this section, we present estimation results using simulated data to assess the finite-sample performance of our estimator. 
In light of our empirical application, we consider ${\mathcal I}=20$ firms and divide them into six groups, each with a different cost distribution. In particular, for group $g=1,\ldots 6$, the private cost shocks $V \sim 5\times ( \mathrm{Beta}^{\dagger}(a_g ,b_g) + 1)$, where $\mathrm{Beta}^{\dagger}(a_g ,b_g)$ is Beta distribution with parameters $(a_g ,b_g)$, see the second column of Table \ref{table:mc}, truncated at $[0.025,0.975]$. We assume that the firms' groups are common knowledge and held fixed. 

Let the demand and cost parameters be $\beta=0.5$ and $\lambda=0.03$, respectively. 
Let the demand shock to be a truncated normal random variable, i.e., $U_t\sim {\mathcal T}r {\mathcal N}(\mu_U=300, \sigma_U^2=800)$ on $[\underline{u}=400, \infty)$, and the common technology shock given $U_t=u$ to be, $W_t|U_t=u \sim 5\times ( 2\times \mathrm{Beta}^{\dagger}(a(u), a(u)) + 1)$ with $a(u) := \exp(\tilde{a}_1+\tilde{a}_2\times u)$ with $\tilde{a}_1=\tilde{a}_2=0.001$. 

We consider two sample sizes $\mathcal T\in\{350, 700\}$, and for $t=1,\ldots, \mathcal T$, we first draw the individual costs $(V_{1t},\ldots, V_{20t})$ and the common demand and cost shocks $(U_t, W_t)$ from the distributions specified above, and then use Lemma \ref{lem:strategies} and (\ref{eq:demand}) to determine the equilibrium outputs $(Q_{1t},\ldots, Q_{{\mathcal I}t})$ and the market-clearing price $P_t$, respectively. Then we apply the estimation procedure to this sample. We repeat this procedure 500 times and, using the estimates from each round, in Table \ref{table:mc} we calculate the simulated bias, standard deviation (SD), and root mean squared error (RMSE), expressed as a fraction of the true parameter.

As we see, our estimation performs well. The average bias is small, and so are the SD and RMSE. 
Comparing the results across the two sample sizes, we see that the estimates improve with a larger sample, suggesting that our estimates will be good with $\mathcal T=336$.

\begin{table}[t!]
\caption{Estimation Results Using Simulated Data.\label{table:mc}}
\hspace{-0.65in}
\begin{tabular}{l|l|rrr | rrr }
\toprule
\multirow{2}{*}{Parameters } & True & \multicolumn{3}{c|}{$ T = 350$} & \multicolumn{3}{c}{$ T = 700$ } \\
	& Values & Bias & SD & RMSE & Bias & SD & RMSE \\
\midrule
Demand slope ($\beta$) & 0.50 & 0.0222&	0.0038&	0.0224&	0.0220&	0.0031	&0.0222 \\
Mean of demand shock ($\mu_U $) & 300 & -0.0082&	0.0264&	0.0276	&-0.0092	&0.0222	&0.0240 \\
Variance of demand shock ($\sigma^2_U$) & 800 & 0.0052&	0.0369&	0.0373	&0.0025	&0.0415	&0.0415 \\
Left Truncation of demand shock ($\underline u$) 	& 400 & 0.0205	&0.0035&	0.0208	&0.0203	&0.0029	&0.0205\\
Parameter of the cost function ($\lambda$) & 0.03 & 0.0290&	0.0274	&0.0399	&0.0325&	0.0306&	0.0446 \\
 Type 1 cost parameter: $a_1$ & 0.5 & 0.0156	&0.0135&	0.0206&	0.0184	&0.0128&	0.0224 \\
 Type 1 cost parameter: $b_1$ & 0.2 & -0.0077	&0.0132&	0.0153	&-0.0117	&0.0128&	0.0173 \\
 Type 2 cost parameter: $a_2$ & 0.6 & 0.0172	&0.0141&	0.0223	&0.0234&	0.0132&	0.0268\\
 Type 2 cost parameter: $b_2$ & 0.2 & -0.0088&0.0138&	0.0164	&-0.0114	&0.0137&	0.0178 \\
 Type 3 cost parameter: $a_3$ & 0.4 & 0.0117&	0.0154	&0.0194	&0.0162&	0.0145&	0.0217\\
 Type 3 cost parameter: $b_3$ & 0.1 & -0.0090	&0.0146&	0.0172	&-0.0147	&0.0148	&0.0209\\
 Type 4 cost parameter: $a_4$ & 0.5 & 0.0191	&0.0145	&0.0240&	0.0186	&0.0133	&0.0229\\
 Type 4 cost parameter: $b_4$ & 0.3 & 0.0038&	0.0146&	0.0151	&0.0022	&0.0137	&0.0139 \\
 Type 5 cost parameter: $a_5$ & 0.60 & 0.0024	&0.0168	&0.0169	&0.0006	&0.0165	&0.0165 \\
 Type 5 cost parameter: $b_5$ & 0.7 & -0.0189&	0.0128&	0.0228&	-0.0203	&0.0109	&0.0230 \\
 Type 6 cost parameter: $a_6$ & 0.4 & 0.0035&	0.0185	&0.0188	&0.0006&0.0175&	0.0175\\
 Type 6 cost parameter: $b_6$ & 0.3 &-0.0176	&0.0128&	0.0218&	-0.0198	&0.0109	&0.0226 \\
 Parameter of the technology shock ($W$): $\tilde{a}_1$ & 0.001 &0.0094&	0.0386&	0.0397&	0.0105&	0.0391&	0.0404 \\
 Parameter of the technology shock ($W$): $\tilde{a}_2$ & 0.001 & 0.0216	&0.0400	&0.0454	&0.0309	&0.0414	&0.0516 \\
$\overline{w}$ & 5 & -0.0268	&0.0040&	0.0271	&-0.0268	&0.0032	&0.0270 \\
\bottomrule  
\end{tabular}
\caption*{\footnotesize {\bf Notes.} The table displays estimation results from using simulated data. Column 2 shows the true parameters, and columns 3 and 4 show the results using sample sizes $T=350$ and $T=700$, respectively. Bias, standard deviation (SD), and root mean square errors (RMSE) are computed across 500 replications for $T\in\{350, 700\}$, and for ease of comparison, expressed as a proportion of the true parameter value.}
\end{table}

\section{Estimation Results} \label{section:empirical}
In this section, we present the estimation results. First, we discuss the results from k-means clustering \cite[e.g.,][]{CoatesNg2012} to group countries into similar types based on the average and the standard deviation of their production. Second, we estimate the model parameters assuming that countries that belong to the same group have the same distribution of private costs: it follows from Lemma \ref{lem:strategies} that countries with the same distribution of private shocks will have symmetric strategies. Third, in a counterfactual exercise, we determine the effect of firms sharing information about their costs on consumer surplus.

We assume that the private cost distributions are common knowledge. 
However, knowledge of these distributions is based on countries' geological information (e.g., nature and size of the reserves) and extraction technologies. 
Thus countries may have similar technologies and thus have symmetric cost distributions.\footnote{  
Countries may strategically announce their reserves or other features of their extraction technologies, say, to influence competitors' beliefs about them.
For our empirical analysis, we assume that for reasons exogenous to our model, countries know each others' cost distributions, and those cost innovations are independent and identically distributed across months.} 
To capture this feature, and in light of our sample size (Figure \ref{fig:rawoutput}), we divide countries into finite groups.
For instance, it is reasonable to assume that larger producers such as the U.S., Saudi Arabia, and Russia have different production costs than smaller producers such as Libya and Venezuela. Similarly, countries in similar geography are likely to have symmetric costs. 

\begin{figure}[th!]
\caption{Grouping using K-means Clustering \label{fig:kmeans}}
 \begin{center}
\includegraphics[scale=0.4]{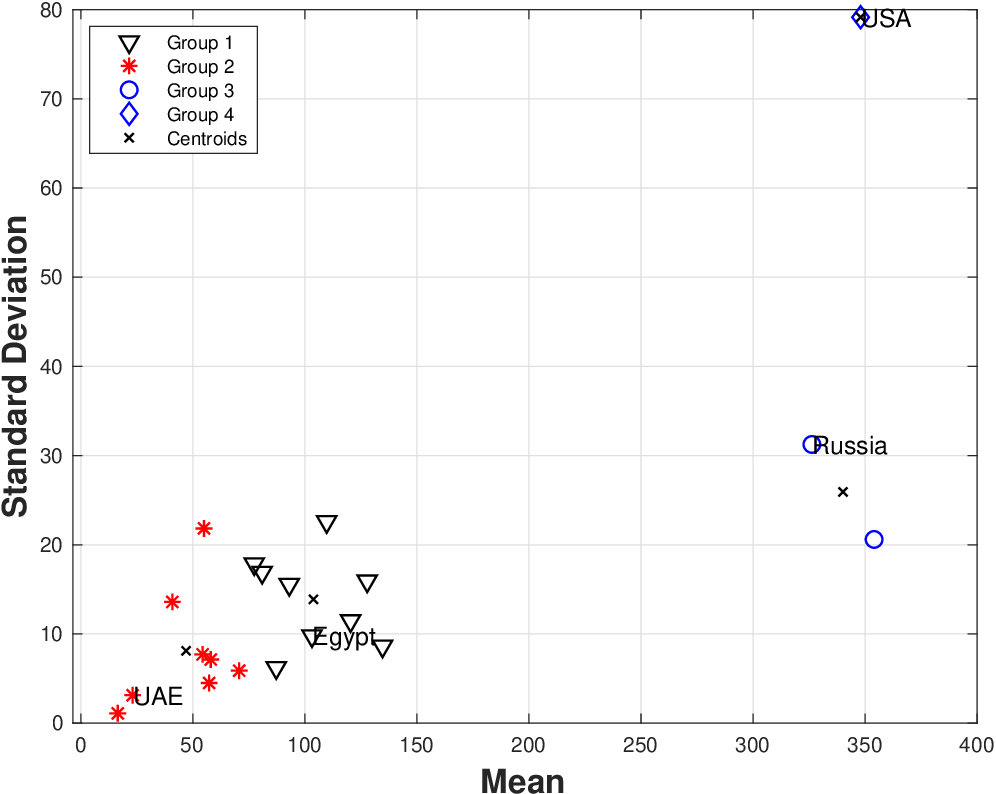}
 \caption*{\footnotesize {\bf Notes.} These figure displays the results from k-means clustering. There are 4 groups, with centroids (mean and standard deviation): $(103.8, 13.88), (46.95 , 8.11), (340.15, 25.92)$ and $(348, 79.15)$, respectively. Group 1 consists of China, Canada, Iran, Iraq, Kuwait, Mexico, Norway, U.A.E., and Venezuela; Group 2 consists of Algeria, Angola, Ecuador, Egypt, Libya, Qatar, Nigeria, and the U.K.; Group 3 consists of Russia and Saudi Arabia; and Group 4 consists of U.S.A. }
 \end{center}
 \end{figure}

To this end, we first apply the unsupervised k-means clustering based on averages and standard deviations of the de-trended productions to classify countries with similar production. Then, given that classification, we further classify countries to belong to similar geographic areas.  

\begin{figure}[t!]
\caption{De-trended Productions, by Group\label{fig:detrendQ}}
 %\begin{center}
\hspace{-0.45in}\includegraphics[scale=0.5]{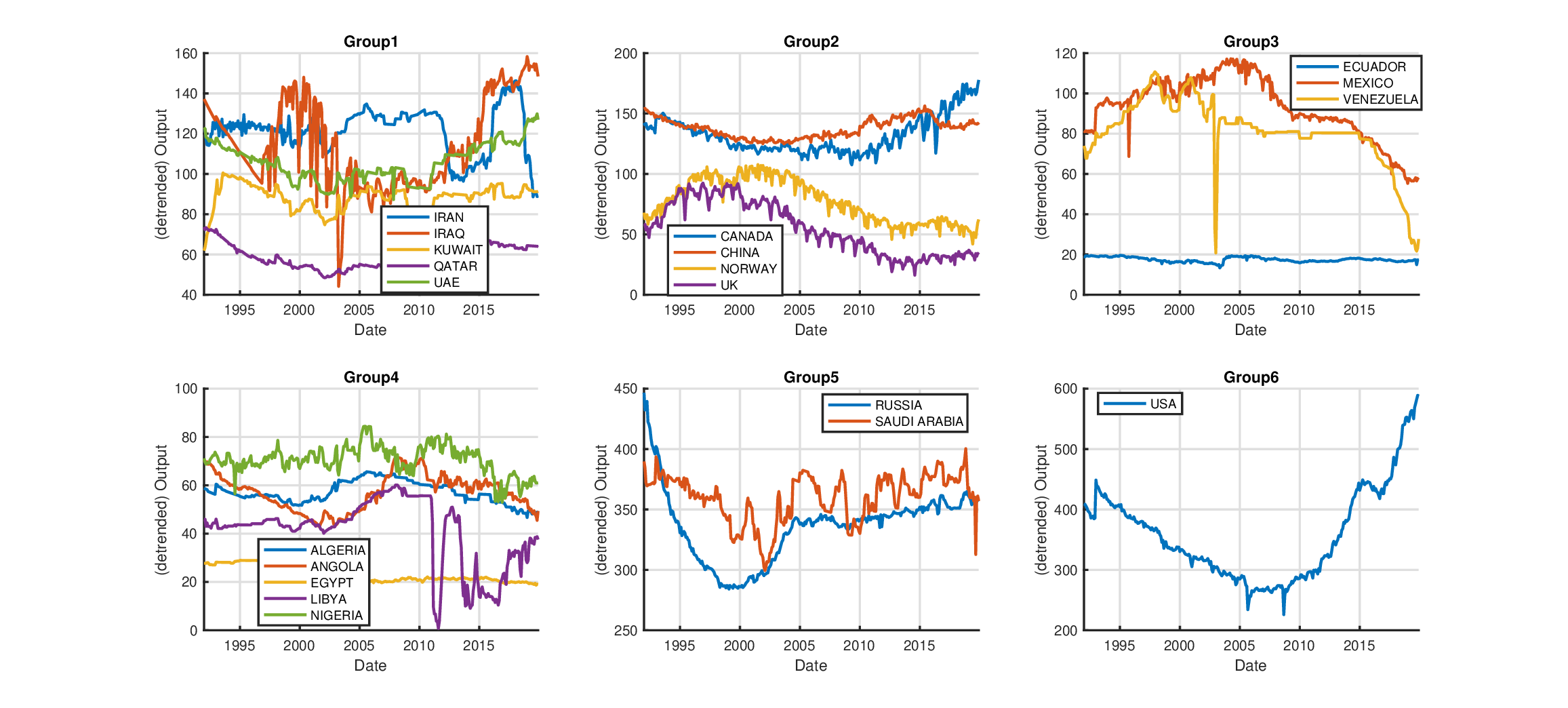}
 \caption*{\footnotesize {\bf Notes.} These figures display de-trended time-series of crude-oil production (mm of barrels) for four groups.}
 %\end{center}
 \end{figure}

We display the k-means clustering exercise results in Figure \ref{fig:kmeans}, which shows that we can classify countries into four groups. The classification is consistent with what we would expect from ``eyeballing" Figure \ref{fig:rawoutput} that countries that belong to a group have similar production patterns. Given these four groups, we classify a few into smaller groups based on locations. In the end, we get six groups with the following memberships: Group 1 (Iran, Iraq, Kuwait, Qatar, and U.A.E.), Group 2 (Canada, China, Norway, U.K.), Group 3 (Mexico, Venezuela, Ecuador), Group 4 (Algeria, Angola, Egypt, Libya, and Nigeria), Group 5 (Russia and Saudi Arabia) and Group 6 (U.S.). We display the production pattern, by group, in Figure \ref{fig:detrendQ}. Hence, countries use the type-symmetric equilibrium for estimation.

\begin{figure}[ht!]
\caption{Estimated PDFs of Costs and Demand Shocks\label{fig:CDF_theta}}
\centering
\includegraphics[scale=0.45]{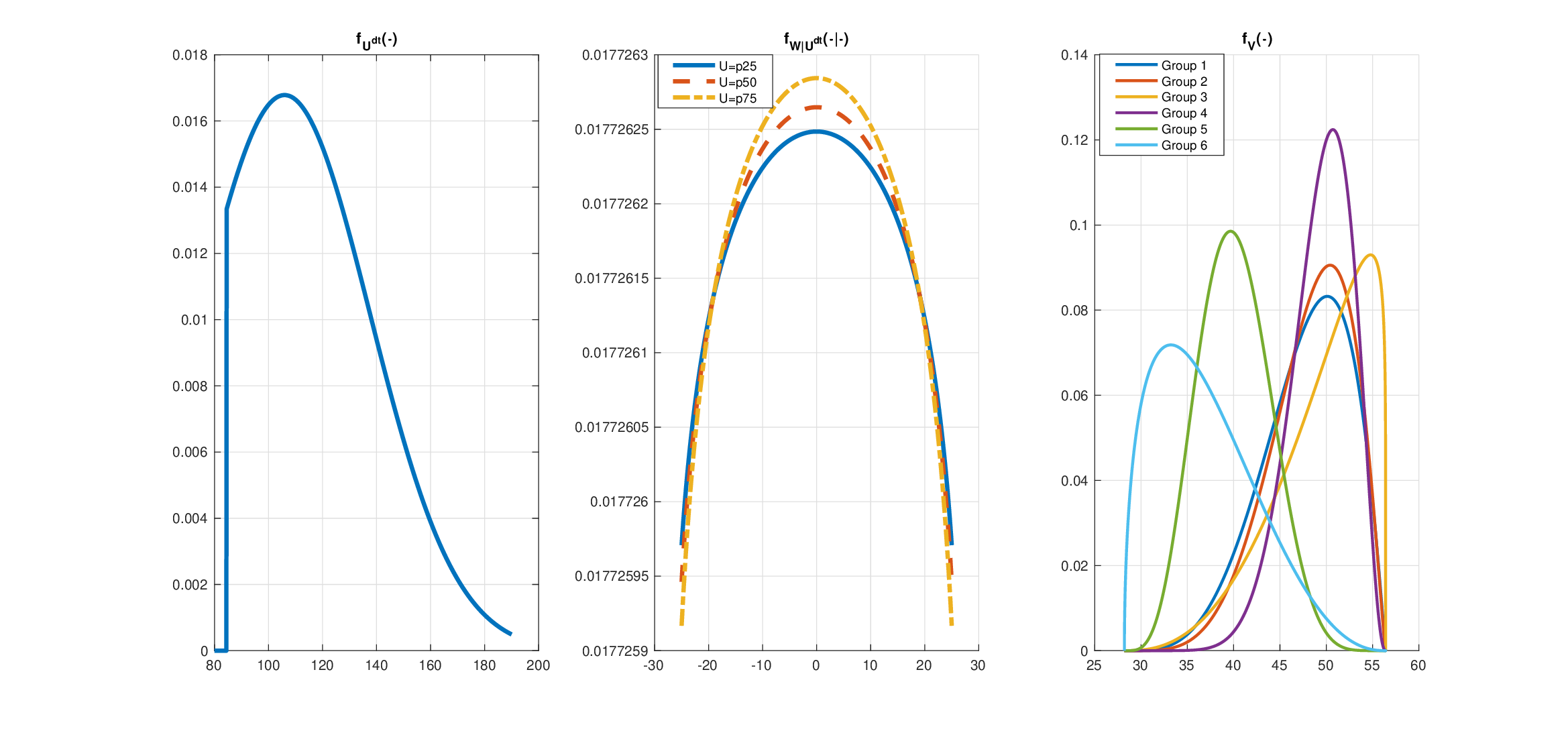}
 \caption*{\footnotesize {\bf Notes.} These figures display (from left to right) the estimated density of (a) (de-trended) demand shock $U^{\mathrm{dt}}$; (b) technology shock $W$ given $U^{\mathrm{dt}}\in\{p25, p50, p75\}$; and (c) private costs, by group, with means and variances ($47.83, 
48.32, 
49.49, 
49.42, 
40.01, 
36.98$) and ($21.46, 
18.63, 
25.79, 
10.61, 
14.67, 
29.25$), respectively.}

 \end{figure}

\paragraph{Estimates.} We apply our estimation method to our sample. 
The estimates together with their 95\% confidence intervals are displayed in Table \ref{tab:estimations}. To estimate the confidence interval we use subsampling method for stationary time series; see \cite{PolitisRomanoWolf1999}, Chapter 3.\footnote{In particular, we use a block size of $b_T=\lfloor T^{0.9}\rfloor=187$, which gives us a total of 150 unique subsamples, and use Theorem 3.2.1 in the book to determine the confidence intervals.}
As we can see, the estimated slope parameter is $\hat{\beta}=0.025$, which means that the demand is downward sloping (as expected) and inelastic, and the estimated parameter of the cost function is $\hat{\lambda}=0.016$. 
We estimate the mean demand shock, or the demand intercept (or the choke point) to be $\hat\mu_U=92.9$, and its estimated variance is $\hat{\sigma}_U^2=673.36$, with left truncation at $\hat{\underline{u}}=66.99$. Figure \ref{fig:CDF_theta} displays the estimated densities of the demand shock, conditional density of the technology shock, and group-specific cost densities. 

\begin{table}[t!]
\caption{Estimation Results \label{tab:estimations}}
\begin{center}
\begin{tabular}{l|c|ll}
\toprule
Parameters & Estimates & \multicolumn{2}{c}{95\% Confidence Intervals} \\
\midrule
Demand slope ($\beta$) & 0.033 & [0.03, & 0.1204] \\
Mean of demand shock $(\mu_U)$ & 105.945 & [$1\times 10^{-7}$, & 293.713] \\
Variance of demand shock $(\sigma_U^2)$ &1,000 & [377.27, & 4,350] \\
Left truncation of demand shock $(\underline{u})$ & 84.443 & [84.10, & 268.036]\\
Parameter of the cost function ($\lambda$) & $7.035\times10^{-4}$ & [$1\times 10^{-7}$,& 0.022] \\
Group 1 cost parameters: $a_1$ & 4.763 & [4.081, & 36.746] \\
Group 1 cost parameters: $b_1$ & 2.082 & [2.043, & 16.443] \\
Group 2 cost parameters: $a_2$ & 5.514 & [3.160 & 15.744] \\
Group 2 cost parameters: $b_2$ & 2.217 & [0.805, & 10.247] \\
Group 3 cost parameters: $a_3$ & 3.555 & [2.759, & 17.562] \\
Group 3 cost parameters: $b_3$ & 1.156 & [0.663, & 5.248] \\
Group 4 cost parameters: $a_4$ & 9.753 & [7.498, & 52.417] \\
Group 4 cost parameters: $b_4$ & 3.215 & [1.20, & 17.109] \\
Group 5 cost parameters: $a_5$ & 5.154 & [1.566, & 34.778] \\
Group 5 cost parameters: $b_5$ & 7.071 & [2.919, & 45.301] \\
Group 6 cost parameters: $a_6$ & 1.503 & [0.619, & 11.276] \\
Group 6 cost parameters: $b_6$ & 3.326 & [0.589, & 22.111] \\
Parameter of the technology shock ($W$): $\tilde{a}_1$ & $1\times 10^{-7}$ & [$1\times 10^{-7}$, & $0.032$] \\
Parameter of the technology shock ($W$): $\tilde{a}_2$ & $1\times 10^{-7}$ & [$1\times 10^{-7}$,& 0.054] \\
$\overline{w}$ & 28.206 & [26.247, & 77.789] \\
\bottomrule  
\end{tabular}
 \caption*{\footnotesize {\bf Notes.} The table displays maximum likelihood estimates of the parameters, with the group membership defined in Figure \ref{fig:detrendQ}. The third column displays the 95\% confidence interval estimated using the subsampling method for stationary time series.}

\end{center}
\end{table}

Regarding the individual cost parameters, the estimates suggest asymmetries across the four groups. 
As seen from the third and the fourth panels in Figure \ref{fig:CDF_theta}, Groups 5 and 6 are the most efficient. These orderings are consistent with the fact that the U.S. (Group 6), Saudi Arabia, and Russia (Group 5) are the largest and therefore the most efficient oil producers, while others have smaller production, which our model interprets as having higher costs. 

\paragraph{Welfare Cost of Private Information.} Next we determine the effect of firms sharing their cost information on output, prices, and consumer surplus. 
We use the estimated parameters and follow \cite{HarrisHowisonSircar2010} and \cite{SarkarGuptaPal1998} to determine equilibrium outputs and prices under complete cost information.
 Once we determine the outputs and price for each month $t$, we also determine the consumer surplus. 
 We repeat this exercise 1000 times and take an average across the simulation draws.  
 
We find that under complete information, in many instances, countries do not produce anything: when costs are known, for some countries producing zero is the best response, and, consequently, the market efficiency increases. The outputs under complete information can be either higher or lower than those under incomplete information, depending on the cost densities. However, we find that the mean quantity under complete information is on average 7\% more than under incomplete information, where the mean is taken across simulations and countries. We also find that, on average, there is a threefold increase in the variance in production across firms as we move from complete information to incomplete information.  
 Correspondingly, the market-clearing price decreases on average by 18\%, and the complete information decreases the deadweight loss (using the method in \cite{Daskin1991}) by 16.3\%.\footnote{ For each simulation, we calculate the deadweight loss built on the benchmark of the cost of the most efficient country. Then we average the deadweight loss across 1,000 simulations and compare the average under incomplete information with complete information.} 

\section{Discussion\label{section:discussion}}
We have analyzed the crude-oil market using a model of static Cournot competition with private cost information to illustrate the application of our method.
To this end, we have set aside several important issues about the oil industry.
In this section, we briefly discuss three issues: (i) dynamics in oil extraction, (ii) non-competitive productions by the Organization of the Petroleum Exporting Countries (OPEC), and (iii) the nature of cost innovations. 

\subsection*{Dynamics in Oil Extraction}

While we have abstracted from any dynamics in the oil market, oil producers face inter-temporal tradeoffs because oil reserves are limited, and countries may allocate the production decisions over time. In such settings, production decisions are nonseparable across time. As our static model does not capture these effects, we should exercise caution in interpreting estimation results from our static model. 

First and foremost, dynamics affect the interpretation of the private costs and, consequently, change the nature of the oligopolistic competition. Indeed, \cite{Loury1986} shows that if countries have their cumulative extraction limited by the size of their initial reserves, the resource scarcity affects the Cournot oligopoly. Similarly, \cite{CremerWeitzman1976} use a data structure similar to ours and show that a dynamic resource extraction model can rationalize the data and the role of OPEC producers. Our static model cannot capture these inter-temporal tradeoffs and their effects on welfare.

Furthermore, countries have strong asymmetry in terms of their cost. For instance, producing a barrel in the U.S. is more expensive than in Saudi Arabia, which explain Saudi Arabia's oil rent and market power. 
 Although we allow asymmetry across countries, the static model we have developed can neither capture scarcity rents nor explain the source of cost asymmetry, where scarcity rents---due to the finite reserve size and or capacity constraint---are determined by the gap between the market price and the extraction cost. 

Nonetheless, extending exhaustible resource extraction problems, i.e., the so-called Hotelling problem \citep{Hotelling1931}, to allow private information about costs is a complex problem to solve. At the same time, there is much uncertainty in the empirical literature about the applicability of the Hotelling model \citep{Gaudet2007,AndersonKelloggSalant2018}. 

Furthermore, we treat countries that own oil rigs and concessionaires as the same. In practice, however, concessionaires will likely have better information about the initial oil reserve than the owners, which affects the optimal extraction path \citep{MartimortPouyetRicci2018}. How these extraction paths change with competition and the signaling effect of productions are other important but difficult questions to address. Proposing a model that captures all of these effects is beyond the scope of this paper.

\subsection*{Non-competitive OPEC Members}

We have also assumed that all countries in our sample, including OPEC members, behave competitively.\footnote{ OPEC was founded in 1960 by Iran, Iraq, Kuwait, Saudi Arabia, and Venezuela. Later they were joined by Qatar (1961), Libya (1962), the UAE (1967), Algeria (1969), Nigeria (1971), Ecuador (1973), and Angola (2007). Although Ecuador suspended its membership in December 1992 and rejoined in October 2007, we treat it as a member in our analysis here. Thus, 12 out of 20 countries in our sample are OPEC members.} As the evidence of collusion among OPEC members is mixed \citep{Spilimbergo2001, AlmogueraDouglasHerrera2011, OkulloReynes2016}, it is desirable to assess the robustness of our estimate of the size of private information to this assumption.

To this end, we consider a variation of our model in which OPEC members choose their (possibly coordinated) productions \emph{for reasons that are exogenous to the model} before the nonmembers. In particular, we estimate a model where OPEC moves first and chooses its quantity. Then other countries compete \'a la Cournot conditional on the OPEC choice.

Two remarks are noteworthy. First, while conceptually straightforward, developing an equilibrium model of Stackelberg competition with private cost information where the ``leader" is a cartel is complex and beyond the scope of our article, not least because such a model must incorporate adverse selection \citep{Roberts1985, AtheyBagwell2001} and the signaling effect of OPEC's production on others' beliefs. Second, without an equilibrium notion to rationalize OPEC choices, we cannot estimate the cost parameters of OPEC members. Nonetheless, as our objective is to assess the robustness of our welfare estimate with respect to the OPEC behavior, our model is reasonable because, given its additive separability, we can compare the welfare using only the nonmember productions. 

In particular, relying on the linearity and additive separability of the demand, we can separate the production between OPEC members and nonmembers and express it as 
\begin{equation}
\mathfrak p (c_t^+, U_t) = U_t -\beta Q_t^+= U_t -\beta \sum_{m\in {\mathcal M}^O}Q_{mt}-\beta \sum_{i\in {\mathcal M}\backslash{\mathcal M}^O}Q_{it} = U_t^* -\beta \sum_{i\in {\mathcal M}\backslash{\mathcal M}^O}Q_{it}, \label{eq:demand_opec}
\end{equation}
where ${\mathcal M}^O$ denotes the set of OPEC members and $U_t^* = (U_t-\beta \sum_{m\in {\mathcal M}^O}Q_{mt})$ is the new random demand intercept. From Eq.\ (\ref{eq:demand_opec}) and after appropriately adapting Assumptions \ref{assump:distributions} and \ref{assump:distributions2} to $U^*$, we can show that the equilibrium characterization in Lemma \ref{lem:strategies} still applies to the nonmembers in ${\mathcal M}\backslash{\mathcal M}^O$, as long as the total OPEC output $\sum_{m\in {\mathcal M}^O}Q_{mt}$ is exogenous. Consequently, we can nonparametrically identify $\beta$, $F_{U^*, W}$, and the cost distributions for firms in ${\mathcal M}\backslash{\mathcal M}^O$. Since we observe $\sum_{i\in {\mathcal M}^O}Q_{it}$, we can also identify the conditional distribution $F_{U|W}(\cdot|\cdot)$ from $F_{U^*|W}(\cdot|\cdot)$.

\begin{table}[t!]
\caption{Estimation Results (Non-OPEC Producers) \label{tab:estimations_nonopec}}
\begin{center}
\begin{tabular}{l|c|ll}
\toprule
Parameters & Estimates & \multicolumn{2}{c}{95\% Confidence Intervals} \\
\midrule
Demand slope ($\beta$) & 0.027 & [0.02, & 0.05] \\
Mean of demand shock $(\mu_U)$ & 60.029 & [2.02, & 101.79] \\
Variance of demand shock $(\sigma_U^2)$ &532.026 & [255.477, & 1,630.167] \\
Left truncation of demand shock $(\underline{u})$ & 70.22 & [55.755, & 135.647]\\
Left truncation of demand shock $(\underline{u}^*)$ & 39.283 & [31.533, & 75.01]\\
Parameter of the cost function ($\lambda$) & 0.028 & [$5.53\times 10^{-4}$,& 0.038] \\
Group 2 cost parameters: $a_2$ & 5.798 & [4.272 & 11.389] \\
Group 2 cost parameters: $b_2$ & 2.088 & [1.824, & 4.752] \\
Group 3 cost parameters: $a_3$ & 7.851 & [3.072, & 20.74] \\
Group 3 cost parameters: $b_3$ & 2.627 & [1.705, & 5.356] \\
Group 4 cost parameters: $a_4$ & 12.048 & [5.023, & 22.164] \\
Group 4 cost parameters: $b_4$ & 2.652 & [1.212, & 4.283] \\
Group 5 cost parameters: $a_5$ & 16.22 & [3.712, & 18.602] \\
Group 5 cost parameters: $b_5$ & 24.456 & [12.412, & 24.839] \\
Group 6 cost parameters: $a_6$ & 3.173 & [1.972, & 6.965] \\
Group 6 cost parameters: $b_6$ & 4.33 & [3.015, & 12.728] \\
Parameter of the technology shock ($W$): $\tilde{a}_1$ & $2.638\times 10^{-6}$ & [$2.376\times 10^{-10}$, & $2.559\times 10^{-5}$] \\
Parameter of the technology shock ($W$): $\tilde{a}_2$ & $9.548\times 10^{-5}$ & [$2.782\times 10^{-10}$,& $7.864\times10^{-5}$] \\
$\overline{w}$ & 42.935 & [30.379, & 75.01] \\
\bottomrule  
\end{tabular}
 \caption*{\footnotesize {\bf Notes.} The table displays maximum likelihood estimates of the parameters using outputs of non-OPEC countries. Here Group 2 includes Canada, China, the U.K., and Norway, and Groups 3--5 are Mexico, Egypt, Russia, and the U.S., respectively. The third column displays the 95\% confidence interval estimated using the subsampling method for stationary time series.}

\end{center}
\end{table}

We present the estimation results in Table \ref{tab:estimations_nonopec}, which is comparable to Table \ref{tab:estimations}. To make the comparison easy, we keep the same group numbering after removing the OPEC members. For instance, all countries except Egypt in Group 4 are OPEC members. So, the ``new" Group 4 in Table \ref{tab:estimations_nonopec} includes only Egypt. Likewise, Group 1 is excluded because all the countries in that group are OPEC members. 

We find the estimates are reasonably similar even though now we have a smaller sample size. We have also obtained that the average output under complete information has a 9\% higher mean and 60\% higher variance than under incomplete information. Correspondingly, the market-clearing price decreases on average, while the deadweight loss decreases by 15.1\%. 

Thus, the estimate is similar to the one obtained previously, although here, we use the information only from non-OPEC members to quantify the size of private information and that welfare variation captures only the cost of
private information for non-OPEC countries. 
Although we should exercise caution and not interpret these estimates to mean that OPEC does not affect market efficiency \cite[]{AskerWexlerLoecker2019}.
These results only suggest that the cost of private information outside or inside OPEC are of similar magnitude since
including OPEC before or considering only non-OPEC countries here 
gives similar estimates. 

\subsection*{Nature of Cost Innovations}

So far, we have assumed that country-specific shocks $\{ V_{it}\}$ are independent and identically distributed (i.i.d.) across $t$ for every $i \in \mathscr{I}$. Regarding the independence condition, even though we allow auto-correlation in observed quantities induced by $\{ (W_t, U_t) \}$, we may still have some auto-correlation in $\{ V_{it}\}$ when studying the global crude-oil market. Some country-specific structural changes might make this i.i.d. assumption unrealistic. Below, we discuss how we can adapt our framework to capture these two features.

First, month-to-month cost innovations may be associated with common changes in input costs across the industry, for instance, when all countries hire from a common set of offshore drilling rigs or comparable pools of oil workers. As we mentioned, some of these common shocks can be captured by the time-series components (Assumption \ref{assump:timetrend}) and common cost shock $W_t$, but we may still miss some correlation left. One way to capture this dependence is to use the U.S. Bureau of Labor Statistic's PPI for oil and gas drilling (PCU213111213111P) as a deflator. The estimates using the new deflator are in Appendix \ref{section:BLS}. 
We estimate deadweight loss under incomplete information becomes 12.9\% larger than under complete information. 

Second, our framework allows for asymmetric distributions in private costs shocks, so publicly known country-specific shocks can be captured by changing the functional form of the corresponding distributions. For instance, if there is a war in Libya, we can consider Libya as a separate group with two cost distributions: $F_{V_{\mathrm{Libya}}}(\cdot)$ and another distribution $\tilde{F}_{V_{\mathrm{Libya}}}(\cdot)$ with higher mean to capture higher costs during wartime. If these distributions satisfy Assumption \ref{assump:distributions} and are common knowledge, then there will be two equilibria: one before and another after the war. Moreover, our identification strategies will still apply if we observe when country-specific shocks occur. 

To estimate the new model, we would need to introduce observed within-country heterogeneity by adapting Assumption \ref{assumption:densities}-(iii) and modifying Eq.\ (\ref{eq:loglikelihood}) accordingly. For instance, we can continue to assume Assumption \ref{assumption:densities}-(iii) holds so that the two Beta distributions can be modeled with four parameters parameters $(a_{\mathrm{Libya}} , b_{\mathrm{Libya}} , a_{\mathrm{Libya}}^\prime, b_{\mathrm{Libya}}^\prime )$. Then, we can write the log-likelihood function as
 \begin{eqnarray*}
 & & \sum_t ( 1- D_t) \log \left[ f_{P,\mathbf{Q}^\mathrm{dt}} (P_t,\mathbf{Q}_t^\mathrm{dt} ; ( \beta,\lambda , \underline u, \mu_U , \sigma_U^2 , \bar{w},\tilde a_1 , \tilde a_2, a_{\mathrm{Libya}} , b_{\mathrm{Libya}} , a_2 ,b_2,\dots,a_\mathcal{I} ,b_\mathcal{I}) ) \right] \\
 & & \ \ + \ \sum_t D_t \log \left[ f_{P,\mathbf{Q}^\mathrm{dt}} (P_t,\mathbf{Q}_t^\mathrm{dt} ; ( \beta,\lambda , \underline u, \mu_U , \sigma_U^2 , \bar{w},\tilde a_1 , \tilde a_2, a_{\mathrm{Libya}}^\prime, b_{\mathrm{Libya}}^\prime , a_2 ,b_2 ,\dots,a_\mathcal{I} ,b_\mathcal{I}) ) \right] ,
\end{eqnarray*}
where $D_t\in\{0,1\}$ is equal to one during the Libyan war periods and zero otherwise. If the war permanently affects costs, then we can set $D_t=1$ from the start of the war.

Similarly, the shale-oil revolution may be the reason behind the increase in U.S. production over the last decade; see Figure \ref{fig:detrendQ}. However, the deposits behind it (e.g., tight oil formations) are different from conventional oil deposits previously exploited, requiring different extraction technologies with different costs than before. We can follow the same procedure outlined above to capture these changes, modeling the U.S. with two different cost distributions and adapting the above log-likelihood function accordingly.

 Third, some innovations might simultaneously impact countries with similar deposits, i.e., countries with similar oil types, sizes, operators, extraction technologies, and locations. In other words, innovations may affect multiple countries, but not all of them. Our model can still be adapted to capture such changes in costs, as long as those changes are publicly observed. In particular, we can define groups based on deposits' observables (e.g., type of the oil, location) instead of productions and geographic locations, as we have done above. Although such modification does not affect the identification results because, in practice, one has a finite sample, there is a tradeoff between the number of groups and the variance of the estimators. 

\section{Conclusion} \label{section:conclusion}

We have developed a model of Cournot competition with private information about (possibly asymmetric) costs. We have specified that the inverse demand function is linear in total quantity with stochastic intercept (or choke price). We have also allowed for an unobserved market characteristic that affects the costs of all firms. In this context, we have first characterized the equilibrium strategies and established the semi-nonparametric identification of the model's parameters. The identification and estimation strategies exploited the strictly monotonic relationship between a firm's output and unobserved shocks. In sum, we rely on the optimality conditions, functional form assumptions about demand, and the assumptions that firms have correct mutual beliefs for the identification. We applied our method using crude-oil production data to quantify private information's role. Finally, we extended our analysis to consider several extensions.

There are several avenues for future research. 
First, and as we discussed earlier, we abstracted from any dynamics in the oil market. However, understanding the role of private information in dynamic Cournot competition is important. Although there has been substantial development on this topic \citep{BonattiCisternasToikka2017}, its application to an empirical setting, say, by adopting and extending the method in \cite{BBL2007}, is still open. Furthermore, an extension of such a model to a resource extraction problem with asymmetric information between countries and their concessionaire, in the spirit of \cite{MartimortPouyetRicci2018}, is yet another topic for future research. 

Second, we may consider the possibility that firms have imprecise beliefs about the cost distribution of their competitors. For example, in the context of the crude-oil market, knowledge about cost distributions is based on geological information (e.g., nature and size of the reserves), which is either private or when public, comes with its uncertainty given the incentives of each country to use these announcements for strategic purposes.
There are at least three ways to model this feature. We can follow the approach in \cite{AryalGrundlKimZhu2018} for auctions and model firms with multiple priors, or allow non-equilibrium beliefs \citep{AguirregabiriaMagesan2019}, or like \cite{MganolfiRoncoroni2021}, use \emph{Bayes Correlated Equilibrium} \citep{BergemannMorris2016}. 

%\clearpage 

\appendix
\setcounter{section}{0}
\setcounter{subsection}{0}
\setcounter{equation}{0}
\setcounter{lemma}{0}
\renewcommand{\thesection}{A}
\renewcommand{\thesubsection}{A.\arabic{subsection}}
\renewcommand{\theequation}{A.\arabic{equation}}
\renewcommand{\thelemma}{A.\arabic{lemma}}

%
%\begin{center}
%{\Large\bf Appendix: Proofs}
% \end{center}

\section{Appendix: Proofs}	\label{section:proof}

This section provides the proofs of Lemma \ref{lem:strategies} and Theorem \ref{thm:identification}. The proof of the latter relies on an auxiliary lemma provided in Section \ref{subsec:alem} together with its proof and a brief discussion on its testable implications.

\subsection{Proof of Lemma \ref{lem:strategies}} 

Existence of the equilibrium strategies follows immediately by checking that $\{\mathfrak q_i : i \in \mathscr I \}$ satisfy the first-order conditions (\ref{strategyFOC}), as well as the second-order conditions, which is trivial because $-2\beta - \lambda <0$ by Assumption \ref{assump:distributions2}. Observe also that such strategies are nonnegative due to the second part of this assumption.

To establish uniqueness, let $\{\tilde{\mathfrak q}_i : i \in \mathscr I \}$ be equilibrium strategies and fix $(w,u)$. By (\ref{strategyFOC}), the former must satisfy \begin{eqnarray*}
\tilde{\mathfrak q}_i ( v_i , w,u )  & =  &   \frac{u -\beta \E[{\tilde \q}^+_{-i}(\mathbf{V}_{-i,t},w,u) ]  - w  }{\lambda + 2\beta} + \frac{-1}{\lambda + 2\beta} \times v_i  = :  \tilde{\mathfrak g}_{i,1} ( w,u ) + \tilde{ g}_{2} \times v_i
\end{eqnarray*}for each $i\in \mathscr I$; note that $ \E[{\q}^+_{-i}(\mathbf{V}_{-i,t},w,u) ] $ depends only on $(w,u)$ and $(\beta, \lambda)$. Thus, in vector notation, we can write \begin{equation}
\tilde{\mathfrak g}_{1} ( w,u ) + \tilde{ g}_{2} \mathbf v =   \frac{-\beta}{\lambda + 2\beta } \mathds M_1 \left[ \tilde{\mathfrak g}_{1} ( w,u ) +  \tilde{g}_{2} \mu_{\mathbf V} \right] + \frac{u - w  }{\lambda + 2\beta } \boldsymbol{\iota}_\mathcal{I} + \tilde{ g}_{2} \mathbf v ,
\label{eq:help1}
\end{equation}where $\mathds M_1$ is $\mathcal I \times \mathcal I$ matrix that has zeros in the main diagonal and ones outside, $\boldsymbol{\iota}_\mathcal{I}$ is a $\mathcal I \times 1$ vector of ones, and $\tilde{\mathfrak g}_{1} = (\tilde{\mathfrak g}_{1,1},\dots, \tilde{\mathfrak g}_{\mathcal I,1})$. Expression (\ref{eq:help1}) can be rewritten as\begin{equation}
\left( \mathds{I}_{\mathcal I} +  \frac{\beta}{\lambda + 2\beta }\mathds M_1 \right)  \tilde{\mathfrak g}_{1} ( w,u ) =  \frac{-\beta \tilde{ g}_{2}}{\lambda + 2\beta } \mathds M_1 + \frac{u - w  }{\lambda + 2\beta } \boldsymbol{\iota}_\mathcal{I} ,
\label{eq:help2}
\end{equation}where $\mathds{I}_{\mathcal I}$ denotes the identity matrix of dimension $\mathcal I$. To complete the proof, it suffices to show that the matrix on the left-hand side is invertible. To do so, write \begin{equation*}
\mathds{I}_{\mathcal I} +  \frac{\beta}{\lambda + 2\beta }\mathds M_1  =  \left( 1 - \frac{\beta}{\lambda + 2\beta }\right) \mathds{I}_{\mathcal I} + \frac{\beta}{\lambda + 2\beta }  \boldsymbol{\iota}_\mathcal{I} \boldsymbol{\iota}_\mathcal{I}^\prime 
 =  \frac{\lambda + \beta}{\lambda + 2\beta} \mathds{I}_{\mathcal I} + \frac{\beta}{\lambda + 2\beta }  \boldsymbol{\iota}_\mathcal{I} \boldsymbol{\iota}_\mathcal{I}^\prime .
\end{equation*}and note that the desired result is obtained from the Sherman-Morrison formula because $1 + [\mathcal I \beta/(\lambda + \beta) ] \neq 0$; see Sec.\ 2.7.1. in \cite{press2007}	\qed

%\subsubsection{Additional Results}
\subsection{An Auxiliary Lemma}		\label{subsec:alem}

The next auxiliary lemma establishes smoothness conditions on the distributions of quantities, as well as on certain conditional distributions. These results will be employed in the proof of Theorem \ref{thm:identification} below.

\begin{lemma} \label{lem:testableimplications}
If Assumptions \ref{assump:distributions} and \ref{assump:distributions2} hold, then the following conditions are satisfied $\forall i \in \mathscr I$.

\begin{enumerate}

\item $F_{Q_i}$ has support $[\underline q_i, \infty )$. It also admits a PDF $f_{Q_i}$ that is strictly positive and continuously differentiable on the interior of this set.

\item The supports of conditional CDFs $F_{Q_{-i}^+|Q_i}(\cdot| \underline q_i )$ and $F_{P|Q_i}(\cdot| \underline q_i )$ are given by $[\underline \rho_i , \bar \rho_i ] \subset \mathbb{R}_+$ and $[\underline \varrho_i , \bar \varrho_i ] \subset \mathbb{R}_+$, respectively. Furthermore, $F_{Q_{-i}^+|Q_i}(\cdot| \underline q_i )$ and $F_{P|Q_i}(\cdot| \underline q_i )$ admit conditional PDFs $f_{Q_{-i}^+|Q_i}(\cdot | \underline q_i )$ and $f_{P|Q_i}(\cdot | \underline q_i )$ that are strictly positive and continuously differentiable on $(\underline \rho_i , \bar \rho_i )$ and $(\underline \varrho_i , \bar \varrho_i )$, respectively.

\end{enumerate}

\end{lemma}

Before proceeding to the proof of this lemma, we highlight that this lemma can be of interest by itself as it provides testable implications of our model. Specifically, the first part of Lemma \ref{lem:testableimplications} establishes that $Q_i$ is a continuous random variable supported on an interval. This result follows that $Q_i$ is a linear combination of $V_i$ and $(U - W)$. As an example of a model whose equilibrium outputs are not continuous, consider a static (or dynamic) Cournot model with entry and exit. In such a model, a potential entrant first decides whether to pay a fixed cost to enter and, upon entry, choose the optimal quantity. If the fixed entry cost is sufficiently high and we observe the set of all potential entrants, then some firms do not enter with a positive probability. As a result, we would observe that the outputs of some firms are equal to zero, and therefore the distribution of these outputs would have a mass point at zero. Similarly, an incumbent might exit, which means the outputs would have a mass point at zero.

The second part of Lemma \ref{lem:testableimplications} states that, for any $i \in \mathscr I$, $Q_{-i}^+ $ is a continuous random variable when we condition on $Q_i = \underline q_i$. This prediction does not hold, e.g., under complete information, because the equilibrium strategy for a firm $i\in \mathscr I$ is given by
 \begin{equation*}
\mathfrak q^{\mathrm c}_{i} (\mathbf{v} ,w,u) = - \frac{v_i - v^{\sbullet}}{\lambda + \beta} + \frac{u - w - v^{\sbullet}}{\lambda + (I+1)\beta},
\end{equation*}where $v^{\sbullet} = (1/\mathcal I) \sum_{i=1}^{\mathcal I} v_i$ stands for the average type \citep[see][Proposition 1]{viv02}.\footnote{Note that these quantities can be negative without further restrictions on the parameter values. See Section 2 in \cite{HarrisHowisonSircar2010} for a discussion on the complete-information Cournot equilibrium with non-identical marginal costs.} Unlike with $\mathfrak q_i$ in Lemma \ref{lem:strategies}, the firm $i$'s strategy $\mathfrak \q^{\mathrm c}_i $ depends on its type $v_i$ and also on its competitors type $\mathbf{v}_{-i}$ in a strictly monotonic way and thus 
\begin{eqnarray}
\mathfrak{q}_i^{\mathrm c} (\mathbf V , W, U) = \underline q_i^{\mathrm c} \ \rightarrow \ ( \mathbf{V}, W,U) = (\bar{\underline{\mathbf{v}}}_{-i} , \bar w ,\underline u) \ \rightarrow \ \sum_{j\neq i} \mathfrak{q}_j^{\mathrm c} (\mathbf V , W, U) = \sum_{j\neq i} {\q}_j^{\mathrm c} (\bar{\underline{\mathbf{v}}}_{-i} , \bar w ,\underline u),\quad
\label{eq:implication}
\end{eqnarray}
where $\underline q_i^{\mathrm c} = \min_{ (\mathbf{v},w,u) } \mathfrak{\q}^C_i (\mathbf{v},w,u)$ and $\bar{\underline{\mathbf{v}}}_{-i} = (\underline v_{1},\dots, \underline v_{i-1}, \bar v_i , \underline v_{i+1}, \dots, \underline v_{\mathcal I})$. In other words, under complete information, firm $i$ produces at the lowest level when its costs are the highest and its competitors' costs are at the lowest. As a result, letting $Q_i^{\mathrm c} = \mathfrak{q}_i^{\mathrm c} (\mathbf V, W,U) $ be firm $i$'s equilibrium output under complete information, we have that $Q_i^{\mathrm c} =\underline q_i^{\mathrm c} $ implies $\sum_{j\neq i} Q_j^{\mathrm c} = \sum_{j\neq i} {\q}_j^{\mathrm c} (\bar{\underline{\mathbf{v}}}_{-i} , \bar w ,\underline u) $ with probability one. Thus, $\sum_{j\neq i} Q_j^{\mathrm c} $ becomes a degenerated random variable after conditioning on $Q_i^\mathrm{c} = \underline q_i^\mathrm{c} $, so $F_{Q_{-i}^+|Q_i}(\cdot| \underline q_i)$ cannot be rationalized by a Cournot model with complete information.

\paragraph{Proof of Lemma \ref{lem:testableimplications}.} For the first part, by the transformation formula and $V_i \perp (W, U)$, observe that  $f_{Q_i | W, U} (q_i | w,u ) = f_{V_i} \left(\frac{q_i - {\mathfrak g}_{i,1} ( w,u ) }{g_2} \right) \left| \frac{1}{g_2} \right|$ for $q_i \in \mathbb R$ and $(w,u) \in \mathscr S_{WU} : = [\underline{w},\bar{w}] \times [\underline u, \infty) $, where $g_2 = -1/ (\lambda + 2\beta )$ and \begin{equation*}
{\mathfrak g}_{i,1} ( w,u ) = \frac{1}{\lambda + (\mathcal I+1)\beta} \left\{ u - w - \frac{1}{\lambda + \beta} \left[  (\lambda + \mathcal I\beta) \mu_{V_i} - \beta \sum_{j \neq i} \mu_{V_j} \right]\right\} + \frac{\mu_{V_i} }{\lambda + 2\beta} .
\end{equation*}
Thus, 
\begin{equation}
f_{Q_i } (q_i ) = \left| \frac{1}{g_2} \right| \underset{\mathscr S_{WU}}{\int \int}  f_{V_i} \left(\frac{q_i - {\mathfrak g}_{i,1} ( w,u ) }{g_2} \right) f_{W,U} (w,u) dw du .
\label{eq:fQi}
\end{equation}
Now pick $q_i \in (\underline q_i , \infty)$. Observe that there exists a nonempty open neighborhood $\mathscr N_2 \subset \mathscr S_{WU}$ and $\underline v_i < v_i^{(1)} < v_i^{(2)} < \bar v_i$ such that $
 v_i^{(1)}   \leq  \frac{q_i - {\mathfrak g}_{i,1} ( w,u )}{g_2} \leq  v_i^{(2)}, \forall \ (w,u ) \in \mathscr N_2 .$
Thus, since $f_{V_i}$ is bounded away from zero on $[ v_i^{(1)}, v_i^{(2)} ]$, we obtain \begin{equation*}
f_{Q_i } (q_i ) \geq \left| \frac{1}{g_2} \right| \underset{\mathscr N_2}{\int \int}  f_{V_i} \left(\frac{q_i - {\mathfrak g}_{i,1} ( w,u ) }{g_2} \right) f_{W,U} (w,u) dw du > 0 .
\end{equation*}As $q_i \in (\underline q_i , \infty)$ has been arbitrarily chosen, from this inequality we can conclude that the support of $Q_i$ is $[\underline q_i , \infty )$ and that $f_{Q_i}$ is strictly positive on $(\underline q_i , \infty)$. Moreover, by Assumption \ref{assump:distributions} and expression (\ref{eq:fQi}), it follows that $f_{Q_i}$ is continuously differentiable on this set.

For the second part, to simplify the exposition and without loss of generality, we prove the statement for $i=1$ and $q = \underline q_1$. The conditional CDF $F_{Q^+_{-1}| Q_1 }$ can be expressed as $
F_{Q^+_{-1}| Q_1 } (\cdot | \underline q_1 ) = F_{Q^+_{-1}| V_1,W,U } (\cdot | \bar v_1, \bar w, \underline u ) =  F_{Q^+_{-1}| W,U } (\cdot | \bar w, \underline u ) $ and that the boundaries of its (conditional) support are given by \begin{eqnarray*}
 \underline \rho ( \underline q_1) & = & \sum_{i \neq 1} \left[ \frac{ 1}{\lambda + (\mathcal I+1)\beta} \left\{ \underline u - \bar w - \frac{1}{\lambda + \beta} \left[  (\lambda +\mathcal I \beta) \mu_{V_i} - \beta \sum_{j \neq i} \mu_{V_j} \right]\right\}  - \frac{\bar v_i - \mu_{V_i} }{\lambda + 2\beta} \right] , \\ 
 \bar \rho ( \underline q_1)  & = &  \sum_{i \neq 1} \left[ \frac{1}{\lambda + (\mathcal I+1)\beta} \left\{ \bar u - \underline w - \frac{1}{\lambda + \beta} \left[  (\lambda + \mathcal I \beta) \mu_{V_i} - \beta \sum_{j \neq i} \mu_{V_j} \right]\right\} - \frac{\underline v_i - \mu_{V_i} }{\lambda + 2\beta} \right] .
\end{eqnarray*}
Now let $\mathds T$ be a $(\mathcal I - 1)\times (\mathcal I -1)$ matrix of the form 
%\begin{equation*}
$\mathds T = \left( \begin{array}{cccccc}
1 & 0 & 0 & \dots & 0 & 0 \\
0 & 1& 0 & \dots & 0 & 0 \\
\vdots & \vdots & \vdots & \vdots & \vdots & \vdots \\
0 & 0 & 0 & \dots & 1 & 0 \\
1 & 1 & 1 & \dots & 1 & 1
\end{array}\right)  
$%\end{equation*}
and write 
%\begin{equation*}
$\mathbf Q^\ast : = \left( \begin{array}{c}
Q_2 \\
\vdots \\
Q_{\mathcal I-1} \\
Q_{-1}^+
\end{array}\right) 
=
\mathds T \left[ {\mathfrak g}_{1}^\ast ( W, U ) + g_2 \mathbf V_{-1} \right] ,
$%\end{equation*}
where ${\mathfrak g}^\ast_{1} = ({\mathfrak g}_{2,1},\dots, {\mathfrak g}_{\mathcal I,1} )$, $\mathbf V_{-1} = (V_2,\dots,V_{\mathcal I})$, and each ${\mathfrak g}_{i,1}$ has been defined in the previous subsection. Since $\mathds T$ is nonsingular and $\{V_1, \mathbf V_{-1} , (W,U)\}$ are mutually independent, by the transformation formula and monotonicity of the equilibrium strategies (Lemma \ref{lem:strategies}), we have \begin{equation*}
f_{\mathbf Q^\ast | Q_1 } ( \mathbf q^\ast  | \underline q_1 ) = f_{\mathbf Q^\ast | V_1, W, U } ( \mathbf q^\ast  | \bar v_1, \bar w, \underline u ) = f_{\mathbf Q^\ast | W, U } \left( \mathbf q^\ast  | \bar w, \underline u \right) = f_{ \mathbf V_{-1} } \left( \frac{ \mathds T^{-1} \mathbf q^\ast - \mathfrak{g}_{1}^\ast (\bar w, \underline u ) }{g_2} \right)  . 
\end{equation*}for $\mathbf q^\ast \in \mathbb R^{\mathcal I-1}$. Thus, the conditional PDF $f_{Q_{-1}^+ | Q_1 } ( \cdot | \underline q_1 )$ can be obtained by integrating out the utmost right-hand side with respect to the first $\mathcal I-2$ elements of $\mathbf q^\ast$. Then it follows by standard arguments that $f_{Q_{-1}^+ | Q_1 } ( \cdot | \underline q_1 )$ is strictly positive and continuously differentiable on $( \underline \rho ( \underline q_1) , \bar \rho ( \underline q_1) )$. Finally, the desired results regarding $F_{P | Q_1 } ( \cdot | \underline q_1 )$ and $f_{P | Q_1 } ( \cdot | \underline q_1 )$ can be obtained by noting that $
F_{P | Q_1 } ( p | \underline q_1 )  = 1  - F_{Q_{-1}^+ | Q_1 } \left( \frac{\underline u - p}{\beta} - \underline q_1 \middle| \underline q_1 \right).$ 	 \qed

\subsection{Proof of Theorem \ref{thm:identification}} \label{app:proof:thm:identification}

We start with the identification of $F_{V_i}(\cdot)$. Note that $\mu_{U|Q_j} ( \underline q_j) = \underline u$ for any $j\neq i$ by strict monotonicity of firm $j$'s equilibrium strategy (Lemma \ref{lem:strategies}). Further, we have that $\mu_{V_i | Q_j } (\underline q_j ) = \mu_{V_i| ( V_j, W ,U )}  ( \bar v_j , \bar w , \underline u ) = \mu_{V_i}$ by Assumption \ref{assump:distributions}, so we can write $
\mu_{Q_i | Q_j } (\underline q_j ) = \frac{1}{\lambda + (\mathcal I+1)\beta}\left\{ \underline u - \bar w - \frac{1}{\lambda + \beta} \left[  (\lambda + \mathcal I \beta) \mu_{V_i} - \beta \sum_{j \neq i} \mu_{V_j} \right]\right\}$,
where we use the fact that the events $\{ Q_j =\underline q_j \}$ and $\{( V_j, W ,U ) = ( \bar v_j , \bar w , \underline u ) \}$ are equivalent. Using Lemma \ref{lem:strategies}, we can express the conditional quantile function of $Q_i$ given $ Q_j =\underline q_j $ as 
\begin{equation}
 F^{-1}_{Q_i | Q_j} (1-\alpha |\underline q_j)  =  \frac{-1}{\lambda + 2\beta} \left[ F^{-1}_{V_i | ( V_j, W ,U )} (\alpha |\bar v_j , \bar w , \underline u  ) - \mu_{V_i} \right] + \mu_{Q_i | Q_j } (\underline q_j )  
 \label{eq:quantileQiQj}
\end{equation}
for all $\alpha \in [0,1]$, where $\beta$, $\lambda$, and $\mu_{V_i}$ have  been identified in  (\ref{eq:idb}), (\ref{eq:estimationm}), and (\ref{eq:idmuthetai}), respectively. As $V_i$ and $ ( V_j, W ,U )$ are independent by Assumption \ref{assump:distributions}, the unconditional quantile function of $V_i$ satisfies $ F^{-1}_{V_i} (\cdot ) = F^{-1}_{V_i | ( V_j, W ,U )} (\cdot|\bar v_j , \bar w , \underline u  )$. Hence, after rearranging (\ref{eq:quantileQiQj}), we get (\ref{eq:idFthetai}).

Regarding the second statement, note that Lemma \ref{lem:strategies} implies 
\begin{eqnarray*}
- [ \lambda + (\mathcal I+1)\beta ] Q_i = \  \left[ \lambda + (\mathcal I+1) \beta \right] \frac{V_i - \mu_{V_i} }{\lambda + 2\beta} -  \left\{U - W - \frac{1}{\lambda + \beta} \left[  (\lambda + \mathcal I \beta) \mu_{V_i} - \beta \sum_{j \neq i} \mu_{V_j} \right]\right\}.
\end{eqnarray*}

Now consider any $z \in \mathbb R$ and write \begin{eqnarray*}
& & \E \left[ \exp\left\{ - {\bf i} z [ \lambda + (\mathcal I+1)\beta ] Q_i \right\}  \middle| U = u \right]  = \ \exp\left[ -{\bf i} z \left\{ u  - \frac{1}{\lambda + \beta} \left[  (\lambda + \mathcal I \beta) \mu_{V_i} - \beta \sum_{j \neq i} \mu_{V_j} \right]\right\} \right] \\
& & \quad \quad \quad \quad \times \ \E \left[ \exp\left\{ {\bf i} z  [ \lambda + (\mathcal I+1)\beta ] \frac{V_i - \mu_{V_i}}{\lambda + 2\beta} + {\bf i} z W \right\}   \middle| U = u \right].
\end{eqnarray*} 
Assumption \ref{assump:distributions} implies that $(V_i\perp W)|U=u$, so, the second term on the right-hand side becomes \begin{eqnarray*}
&  & \E \left[ \exp\left\{ {\bf i} z  [ \lambda + (\mathcal I+1)\beta ] \frac{V_i - \mu_{V_i}}{\lambda + 2\beta} + {\bf i} z W \right\}   \middle| U = u \right] \\
& & \quad \quad \quad \quad =  \  \E \left[ \exp\left\{  {\bf i} z  [ \lambda + ( \mathcal I+1)\beta ] \frac{V_i - \mu_{V_i} }{\lambda + 2\beta}  \right\}   \middle| U = u \right]   \times   \E \left[ \exp\left(  {\bf i} z W \right)   \middle| U = u \right].
\end{eqnarray*}
Then, using $V_i\perp U$, we get \begin{eqnarray*}
& &  \exp\left[ {\bf i} z \left\{ u  - \frac{1}{\lambda + \beta} \left[  (\lambda + \mathcal I \beta) \mu_{V_i} - \beta \sum_{j \neq i} \mu_{V_j} \right]\right\} \right]   \\
& & \quad  \quad \times \ \frac{  \E \left[ \exp\left\{ - {\bf i} z  [ \lambda + (\mathcal I+1)\beta ] Q_i \right\}  \middle| U = u \right] }{  \E \left[ \exp\left\{ {\bf i} z  [ \lambda + (\mathcal I+1)\beta ]  \frac{V_i - \mu_{V_i} }{\lambda + 2\beta}  \right\}  \right] }  \ = \ \E \left[ \exp\left(  {\bf i} z W \right)   \middle| U = u \right].
\end{eqnarray*}The desired result follows by recalling that $U = P + \beta Q^+$, being $\beta$ an identified object. \qed

\setcounter{section}{0}
\setcounter{subsection}{0}
\setcounter{equation}{0}
\setcounter{lemma}{0}
\renewcommand{\thesection}{B}
\renewcommand{\thesubsection}{B.\arabic{subsection}}
\renewcommand{\theequation}{B.\arabic{equation}}
\renewcommand{\thelemma}{B.\arabic{lemma}}
\renewcommand{\thetable}{B.\arabic{section}}
\renewcommand{\thefigure}{B.\arabic{section}}

 \section{Cost Innovation.\label{section:BLS}} 
 Here, we present the estimation results that uses BLS's PPI for oil and gas drilling (PCU213111213111P) as the price deflator as discussed in Section \ref{section:discussion} under ``Nature of Cost Innovations."

\begin{table}[t!]
\caption{\bf Estimated Parameters (Using BLS Oil and Gas Drilling Deflator) \label{tab:estimationsBLS}}
\begin{center}
\scalebox{0.8}{\begin{tabular}{l|c|ll}
\toprule
Parameters & Estimates & \multicolumn{2}{c}{95\% Confidence Intervals}\\
\midrule
Demand slope ($\beta$) & 0.05 & [0.5, 0.19] \\
Mean of demand shock $(\mu_U)$ & 128.937 &[$1\times10^{-7},422.151$] \\
Variance of demand shock $(\sigma_U^2)$ &2,722.666&[$1,539.517,11,679.792$] \\
Left truncation of demand shock $(\underline{u})$ & 115.987 &[$112.244,408.569$]\\
Parameter of the cost function ($\lambda$) & $6.947\times10^{-8}$ &[$1\times10^{-8},0.008$] \\
Group 1 cost parameters: $a_1$ & 13.832 &[$9.856,34.888$]\\
Group 1 cost parameters: $b_1$ & 5.064 &[$4.082,17.706$] \\
Group 2 cost parameters: $a_2$ & 9.01 &[$4.6,15.64$] \\
Group 2 cost parameters: $b_2$ & 2.927 &[$2.064,10.724$]\\
Group 3 cost parameters: $a_3$ & 8.853 &[$4.097,18.541$] \\
Group 3 cost parameters: $b_3$ & 2.417&[$1.308,6.344$] \\
Group 4 cost parameters: $a_4$ & 13.888&[$13.087,52.034$] \\
Group 4 cost parameters: $b_4$ & 3.731&[$3.7,19.45$] \\
Group 5 cost parameters: $a_5$ & 3.619 &[$1.634,41.904$]\\
Group 5 cost parameters: $b_5$ & 3.967 &[$4, 52.462$]\\
Group 6 cost parameters: $a_6$ & 3.92 &[$0.949,12.707$] \\
Group 6 cost parameters: $b_6$ & 6.233 &[$2.171,21.789$] \\
Parameter of the technology shock ($W$): $\tilde{a}_1$ & $8.654\times 10^{-8}$ &[$1\times10^{-8},2.02\times 10^{-7}$] \\
Parameter of the technology shock ($W$): $\tilde{a}_2$ & $7.055\times 10^{-8}$&[$1\times10^{-8},1.436\times10^{-6}$]\\
$\overline{w}$ & 42.975 &[$31.427,105.377$] \\
\bottomrule  
\end{tabular}}
\caption*{\footnotesize {\bf Notes.} The table displays maximum likelihood estimates of the parameters, with the group membership defined in Figure \ref{fig:detrendQ}, using BLS's PPI for oil and gas drilling (PCU213111213111P) as the price deflator. The third column displays the 95\% confidence interval estimated using the subsampling method.}
\end{center}
\end{table}

\begin{figure}[ht!]
\begin{center}
\caption {\bf Estimated PDFs of Costs and Demand Shocks}
\includegraphics[scale=0.35]{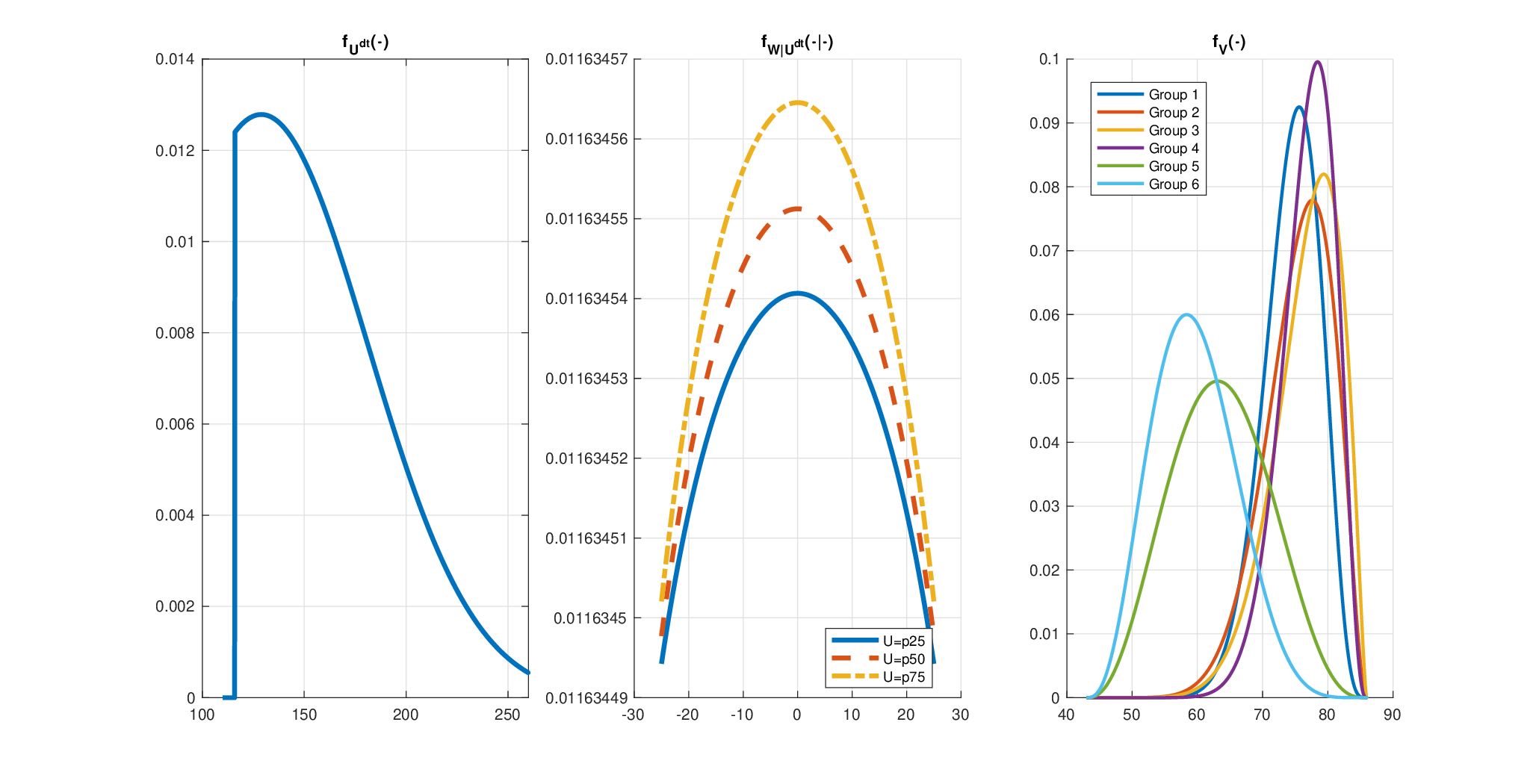}\label{fig:estimationsBLS}
\caption*{\footnotesize {\bf Notes.} These figures display (left to right) the estimated density of (a) (de-trended) demand shock $U^{\mathrm{dt}}$; (b) technology shock $W$ given $U^{\mathrm{dt}}\in\{p25, p50, p75\}$; and (c) private costs, by group, with means and variances $(74.434,
75.413,
76.732,
76.849,
63.477,
59.57)$ and $(18.209,
26.422,
25.36,
16.558,
53.657,
39.249)$, respectively.}
 \end{center}
\end{figure}

%\clearpage
\pdfbookmark[1]{References}{References}

\bibliographystyle{econometrica}

\bibliography{bib_cournot}

%\clearpage 

\setcounter{section}{0}
\setcounter{equation}{0}
\setcounter{assumption}{0}
\setcounter{lemma}{0}
\setcounter{theorem}{0}
\setcounter{ex}{0}

\renewcommand{\thesection}{S}
\renewcommand{\thesubsection}{S.\arabic{subsection}}
\renewcommand{\theequation}{S.\arabic{equation}}
\renewcommand{\theassumption}{S.\arabic{assumption}}

\renewcommand{\thelemma}{S.\arabic{lemma}}
\renewcommand{\thetheorem}{S.\arabic{theorem}}

\renewcommand{\theex}{S.\arabic{ex}}

%\appendix

%\begin{center}
%{\Large\bf S. Supplementary Appendix: Extensions}
%\end{center}

\section{Supplementary Appendix: Extensions}

%\section*{{\bf S}. Extensions\label{section:extensions}}
In this section, we consider four extensions and discuss their identification. First, we study differentiated Cournot competition. Second, we study the possibility that firms do not play Bayesian Cournot-Nash equilibrium by allowing them to have different conduct parameters. Third, we consider a nonlinear demand. Fourth, we consider Cournot competition with a selective entry. Throughout this section, we maintain Assumption \ref{assump:distributions} and further assume that $U_t$ is supported on a compact interval $[\underline u, \bar u]$ with $\underline u < \bar u <\infty$, instead of $[\underline u, \infty)$. We also introduce new assumptions as needed.

\subsection*{Differentiated Products}

In this section, we extend our model to allow for differentiated products. For that purpose, we assume that each firm $i$ faces an inverse demand function of the form \begin{equation}
\mathfrak p_i^{\mathrm{df}} (Q_{it} , {\mathbf Q}_{-it}, U_t) = U_t -\beta_i Q_{it} - \sum_{j \neq i } \beta_j Q_{j t}, 
\label{eq:demanddif}
\end{equation}where $\beta_i>0$ for all $ i \in \mathscr{I}$ and $Q_{-it}$ is a $(\mathcal{I}-1)\times1$ vector of quantities produced by $i$'s competitors in market $t$. As in Section \ref{section:model}, we continue to assume that firm $i$'s total variable cost function $\mathfrak{v c}_i(\cdot;\cdot,\cdot)$ is defined in (\ref{eq:vc}), and that Assumption \ref{assump:distributions} holds. As before, $i$'s variable cost is known only by firm $i$, but we assume that firms commonly know the degree of product differentiation, which is constant. 

Next, we generalize Assumption \ref{assump:distributions2} to allow for product differentiation. 
Let $\mathds{M}_2$ be a $\mathcal{I}\times\mathcal{I}$ matrix whose $(i,j)$-element is given by \begin{equation*}
 \mathds{M}_{2, (i,j)} = \left\{ \begin{array}{cc}
0 & \text{if } i=j , \\
-\beta_j/ (\lambda + 2\beta_i) & \text{if } i \neq j, 
\end{array}\right. 
\end{equation*}
and let $\boldsymbol{\beta} = (\beta_1, \dots, \beta_\mathcal{I})$ and $ \mathbf{m}_{2} = ((\lambda + 2\beta_1 )^{-1},\dots, (\lambda + 2\beta_{\mathcal{I}} )^{-1} ) $ be a $\mathcal{I}\times 1$ vectors.

\begin{assumption}\label{assump:difgood}
The following conditions hold: 
\begin{enumerate}[label=(\roman*)]
\item $\beta_i > 0$ for all $i \in \mathscr I$ and $\lambda >0$.
\item $\min \{ ( \mathds{I}_\mathcal{I} - \mathds{M}_2 )^{-1} \mathbf{m}_{2} \} > 0$.
\item $
\ \underline{u} - \boldsymbol{\beta}^\prime \{ ( \mathds{I}_\mathcal{I} - \mathds{M}_2 )^{-1} \left[ (\underline{u} - \underline{w} ) \mathbf{m}_{2} - \mathds{M}_2 \mathrm{diag}( \mathbf{m}_{2} ) \mu_{\mathbf{V}} \right] - \mathrm{diag}( \mathbf{m}_{2} ) \underline{\mathbf{v}} \} \geq 0.$
\item $\min \left \{ ( \mathds{I}_\mathcal{I} - \mathds{M}_2 )^{-1} \left[ (\underline{u} - \bar{w} ) \mathbf{m}_{2} - \mathds{M}_2 \mathrm{diag}( \mathbf{m}_{2} ) \mu_{\mathbf{V}} \right] - \mathrm{diag}( \mathbf{m}_{2} ) \bar{\mathbf{v}} \right\} \geq 0 $.
\end{enumerate}
\end{assumption}

Several remarks about this assumption are noteworthy. Assumption \ref{assump:difgood}-(i) is a standard extension from the homogenous good case. For Assumption \ref{assump:difgood}-(ii), we show in Lemma \ref{lem:difgood} below that $ (\mathds{I}_\mathcal{I} - \mathds{M}_2)$ is invertible, and furthermore, it is automatically satisfied when $\beta_1 = \dots = \beta_{\mathcal{I}}$, as in Section \ref{section:model}. It guarantees that the equilibrium strategy is strictly monotonic in costs. Assumption \ref{assump:difgood}-(iii) and (iv) ensure that the equilibrium price and quantities are nonnegative.  

Thus, Assumption \ref{assump:difgood} nests Assumption \ref{assump:distributions2} for homogenous products as a special case. Furthermore, as with Assumption \ref{assump:distributions2}, Assumption \ref{assump:difgood} is also satisfied when the smallest demand intercept, $\underline u $, is sufficiently large in comparison with the other parameters. 

These assumptions are sufficient to guarantee the existence and uniqueness of Bayesian Nash equilibrium in strictly monotonic strategies. We formalize this result below.

\begin{lemma}	\label{lem:difgood}
If Assumptions \ref{assump:distributions} and \ref{assump:difgood} hold, there exists a unique Bayesian Cournot-Nash equilibrium. Specifically, each equilibrium strategy $\mathfrak \q_i^{\mathrm{df}} (v_i , w, u)$ is given by $i$th element of the vector $ ( \mathds{I}_\mathcal{I} - \mathds{M}_2 )^{-1} \left[ ({u} - {w} ) \mathbf{m}_{2} - \mathds{M}_2 \mathrm{diag}( \mathbf{m}_{2} ) \mu_{\mathbf{V}} \right] - \mathrm{diag}( \mathbf{m}_{2} ) {\mathbf{v}}$. It is strictly increasing in $u$ and strictly decreasing in the other arguments, $v_i$ and $w$.
\end{lemma}

\begin{proof}\label{proof:difgood}
By similar arguments to the ones in Section \ref{section:model}, here the first-order conditions are \begin{equation*}
\q_i^{\mathrm{df}} ( v , w , u ) = \frac{u - \sum_{j \neq i} \beta_j \E[ {\q}_j^{\mathrm{df}} (\mathbf{V}_{j,t},w,u) ] - w - v}{\lambda + 2\beta_i} \quad \forall \ i \in \mathscr{I} .
\end{equation*}Existence of the equilibrium strategies follows immediately by checking that $\{ \q_i^{\mathrm{df}} : i \in \mathscr I \}$ satisfy these first-order conditions, as well as the second-order conditions, which is trivial because $-2\beta_i - \lambda <0$ for all $i$. Observe that such strategies are nonnegative due to Assumption \ref{assump:difgood} and are also strict-monotonic as specified in the lemma.

To establish uniqueness, we follow similar steps to the ones in Appendix \ref{section:proof}. Let $\{\tilde{\mathfrak q}_i : i \in \mathscr I \}$ be equilibrium strategies and consider any fixed $(w,u)$. Note that they must satisfy \begin{equation*}
\tilde{\q} ( v , w , u ) =  \tilde{\mathfrak g}_{1}^{\mathrm{df}} ( w,u ) -  \mathrm{diag}( \mathbf{m}_{2} ) \mathbf v
\end{equation*}where $\tilde{\mathfrak g}_{1}^{\mathrm{df}} ( w,u )$ is a $\mathcal{I} \times 1$ vector whose $i^\text{th}$ component is given by \begin{equation*}
\tilde{\mathfrak g}_{i,1} ( w,u ) = \left\{ u - w - \sum_{j \neq i} \beta_j \E[ \tilde{\q}_j(\mathbf{V}_{j,t},w,u) ] \right\} / (\lambda + 2\beta_i ) .
\end{equation*}
Then, establishing the existence and uniqueness of an equilibrium reduces to establishing the existence and uniqueness of a vector-valued function $\tilde{\mathfrak g}_{1}$ that satisfies \begin{equation*}
\tilde{\mathfrak g}_{1} ( w,u ) - \mathrm{diag}( \mathbf{m}_{2} ) \mathbf v =  \mathds M_2 \left[ \tilde{\mathfrak g}_{1} ( w,u ) -  \mathrm{diag}( \mathbf{m}_{2} ) \mu_{\mathbf V} \right] + (u - w ) \mathbf{m}_{2}  - \mathrm{diag}( \mathbf{m}_{2} ) \mathbf v \  \forall  (w,u , \mathbf v) .
\end{equation*}
But this follows if and only if $\mathds I_{\mathcal{I}} - \mathds M_2$ is nonsingular. To check this result, note that we can write $\mathds{M}_2 = \mathrm{diag}( \beta_1 (\lambda + 2\beta_1 )^{-1},\dots, \beta_\mathcal{I} (\lambda + 2\beta_\mathcal{I} )^{-1}  ) - \mathbf{m}_{2} \boldsymbol{\beta}^\prime $ and \begin{equation*}
\mathds I_{\mathcal{I}} - \mathds M_2 = \mathrm{diag}((\lambda + \beta_1)(\lambda + 2\beta_1 )^{-1},\dots,(\lambda + \beta_\mathcal{I}) (\lambda + 2\beta_\mathcal{I} )^{-1}  ) + \mathbf{m}_{2} \boldsymbol{\beta}^\prime .
\end{equation*}Then, by Sherman-Morrison formula, $\mathds I_{\mathcal{I}} - \mathds M_2$ is invertible if and only if \begin{equation*}
1 + \boldsymbol{\beta}^\prime \mathrm{diag}((\lambda + \beta_1)^{-1}(\lambda + 2\beta_1 ),\dots,(\lambda + \beta_\mathcal{I})^{-1} (\lambda + 2\beta_\mathcal{I} )  )  \mathbf{m}_{2} \neq 0 . 
\end{equation*}Finally, the desired result follows because the LHS is equal to $1 + \sum_{i \in \mathscr{I}} \beta_i/ (\lambda + \beta_i)$.
\end{proof}

Next, we briefly discuss how the previous identification strategy applies to this model. 
In particular, we show how we can exploit the change in the conditional quantiles of market-clearing price with respect to the change in the conditional quantiles of $i$'s to identify $\beta_i$. Once we identify slope parameters, identifying other model parameters is almost identical to when products are homogenous. 

In particular, for firm $i \in \mathscr{I}$ and two distinct quantiles $\alpha, \alpha^{\prime}\in[0,1]$, applying similar steps as we did to get (\ref{eq:idb}) to the inverse demand function (\ref{eq:demanddif}), conditional on $i$'s competitors producing at their minimum, i.e., ${\bf Q}_{-i}=\underline{\bf q}_{-i}$, we obtain 
\begin{eqnarray*}
F^{-1}_{P | \mathbf{Q}_{-i} } (\alpha| \underline{\mathbf q}_{-i} ) &=& \underline u - \beta_i F^{-1}_{Q_{i} | \mathbf{Q}_{-i} } \left(1- \alpha| \underline{\mathbf q}_{-i} \right) - \sum_{j \neq i} \beta_j \underline q_j , \\
F^{-1}_{P | \mathbf{Q}_{-i} } (\alpha^\prime | \underline{\mathbf q}_{-i} ) &=& \underline u - \beta_i F^{-1}_{Q_{i} | \mathbf{Q}_{-i} } \left(1- \alpha^\prime | \underline{\mathbf q}_{-i} \right) - \sum_{j \neq i} \beta_j \underline q_j .
\end{eqnarray*}
Here $\alpha\neq \alpha^{\prime}$ implies that $F^{-1}_{Q_{i} | \mathbf{Q}_{-i} } \left(1- \alpha| \underline{\mathbf q}_{-i} \right) \neq F^{-1}_{Q_{i} | \mathbf{Q}_{-i} } \left(1- \alpha^\prime | \underline{\mathbf q}_{-i} \right)$, so subtracting the first equation from the second identifies $\beta_i$ as
\begin{equation}
\beta_i = \frac{F^{-1}_{P | \mathbf{Q}_{-i} } (\alpha^\prime | \underline{\mathbf q}_{-i} ) - F^{-1}_{P | \mathbf{Q}_{-i} } (\alpha| \underline{\mathbf q}_{-i} ) }{ F^{-1}_{Q_{i} | \mathbf{Q}_{-i} } \left(1- \alpha| \underline{\mathbf q}_{-i} \right) - F^{-1}_{Q_{i} | \mathbf{Q}_{-i} } \left(1- \alpha^\prime | \underline{\mathbf q}_{-i} \right) } .
\label{eq:idbdif}
\end{equation}
The choice of the quantiles was arbitrary, suggesting that $\beta_i$ is over-identified here. Once $\{ \beta_i : i \in \mathscr{I} \}$ are identified, we can recover the demand shock as $U =P + \sum_{i \in \mathscr{I}} \beta_i Q_i$ and identify its CDF as $F_U(u) =F_{ P + \sum_{i } \beta_i Q_i } (u)$ for $u\in\mathbb{R}$. Then the other parameters can also be identified in the same way. 

\subsection*{Conduct Parameters\label{section:CV}}

In this section, following closely \cite{bre89} and \cite{GenesoveMullin1998}, we extend our model to include conduct parameters. 
As before, firm $i\in \mathscr I$ observes $(V_i,W, U)$ and chooses quantity to maximize its interim expected profit (\ref{eq:interprofit}). However, now, in the first-order conditions, we introduce new parameters $\{ \vartheta_i \geq 0: i \in \mathscr I\}$, where $\vartheta_i$ is $i$'s conjecture about the effect of changing its output on the industry output. To be specific, for each $i \in \mathscr I$ and for given strategies $\mathfrak{\q}^{\mathrm{cp}}_{-i} = ( \mathfrak{\q}^{\mathrm{cp}}_{1},\dots,\mathfrak{\q}^{\mathrm{cp}}_{i-1}\mathfrak{\q}^{\mathrm{cp}}_{i+1}, \dots, \mathfrak{\q}^{\mathrm{cp}}_{\mathcal I} )$ such that $\mathfrak{\q}^{\mathrm{cp}+}_{-i} = \sum_{j\neq i} \mathfrak{\q}^{\mathrm{cp}}_j$, after observing $(V_i,W,U) = (v_i , w ,u)$, where $V_i \perp ( \mathbf V_{-i} , W, U )$, firm $i$ output solves 
\begin{eqnarray*}
 & & u - \beta \left\{q_i + \E \left[ \mathfrak{\q}^{\mathrm{cp}+}_{-i}(\mathbf V_{-i} , W, U) | W = w, U = u \right] \right\} \\
 & & \qquad \quad - \quad q_i \times \beta\times \underbrace{\frac{\partial \left\{q_i+ \E \left[ \mathfrak{\q}^{\mathrm{cp}+}_{-i}(\mathbf V_{-i} , W , U) | W = w, U = u \right] \right\}}{\partial q_i}}_{=\vartheta_i} = v_i+ w + \lambda q_i.
\end{eqnarray*}

Now suppose that each firm $i$ correctly believes that the other firms respond the same way to its choice, i.e., $\kappa_i:=\partial \E [ \mathfrak{\q}^{\mathrm{cp}}_j ( \mathbf V_{-i} , w ,u ) ] /\partial q_i \leq 0$ for all $(w,u)$, so that $\vartheta_i = 1+ (\mathcal I-1) \kappa_i$. Then, extending the arguments of Appendix \ref{section:proof}, we can show that there exists a unique vector of functions $(\mathfrak{\q}^{\mathrm{cp}}_1,\dots,\mathfrak{\q}^{\mathrm{cp}}_{\mathcal I} )$ that satisfies \begin{equation}
 \q_i^\mathrm{cp} ( v_i , w ,u ) = \frac{u -\beta \E[ \mathfrak{\q}^{\mathrm{cp}+}_{-i}(\mathbf{V}_{-i},W,U) | W = w, U = u] - w - v_i}{\lambda + \beta (\mathcal I-1)\kappa_i + 2\beta} \quad \forall \ i \in \mathscr I .
\label{eq:CV}
\end{equation}
As before, we note that we can ensure nonnegative quantities and price by taking $\underline u$ sufficiently large enough relative to other parameters.
Moreover, for every $i \in \mathscr I$, we can show that $\mathfrak q_i^\mathrm{cp} (v_i , w, u)$ is linear in $v_i$ and strictly monotonic in $\{v_i, w, u\}$, i.e., 
\begin{eqnarray*}
\frac{\partial \mathfrak q_i^\mathrm{cp}}{\partial v_i} (v_i , w, u) < 0 , \ \frac{\partial \mathfrak q_i^\mathrm{cp}}{\partial w} (v_i , w, u) < 0, \ \text{and}\ \frac{\partial \mathfrak q_i^\mathrm{cp}}{\partial u} (v_i , w, u) >0 .
\end{eqnarray*}

Thus, as before we can use the joint distribution $F_{P^\mathrm{cp}, \mathbf Q^\mathrm{cp}}$, to identify the demand parameter $\beta$ by replacing $(P,Q_i, Q_{-i}^+)$ with $(P^\mathrm{cp},Q_i^\mathrm{cp}, Q_{-i}^{\mathrm{cp}+})$ in (\ref{eq:idb}), where $P =\mathfrak p (Q^{+\mathrm{cp}},U) $, $Q^{+\mathrm{cp}} = \sum_{i \in \mathscr I} Q_{i}^\mathrm{cp}$, $Q_{i}^\mathrm{cp} = \mathfrak q^\mathrm{cp}_i (V_{i} , W,U)$, and $\mathbf Q^\mathrm{cp}= (Q_{1}^\mathrm{cp}, \dots, Q_{\mathcal I }^\mathrm{cp})$. Then the intercept $U$ and its distribution $F_U$ can be identified too.

Before identifying the cost parameter $\lambda$, we first consider the problem of identifying $\tilde\lambda_i : =  \lambda + ( \mathcal{I} -1 )\beta \kappa_i$ for $i \in \mathscr{I}$. From the first-order conditions (\ref{eq:CV}), we have that $
Q_i^\mathrm{cp} = \frac{U -\beta \E \left( Q_{-i}^{\mathrm{cp}+} | W , U \right) - W - V_i}{\tilde\lambda_i  + 2\beta}  \quad \forall \  i \in \mathscr{I}. $ 
Then it follows from the law of iterated expectation that for  $\forall \  i \in \mathscr{I}$, 
\begin{equation*}
\mu_{Q_i^\mathrm{cp} | U} (u) = \frac{u -\beta \mu_{ Q_{-i}^{\mathrm{cp}+} | U }(u) - \mu_{V_i}}{\tilde\lambda_i + 2\beta}    \quad \forall \   u \in [\underline u, \bar u]   .
\end{equation*}
Then, choosing any $u^\prime \neq u$, we can identify $
\tilde\lambda_i= \frac{\beta \left[ \mu_{ Q_{-i}^{\mathrm{cp}+}|U} (u) - \mu_{ Q_{-i}^{\mathrm{cp}+}|U} (u^\prime) \right] + ( u - u^\prime )}{\mu_{Q_i^\mathrm{cp} | U} (u^\prime) - \mu_{Q_i^\mathrm{cp} | U} (u)}  - 2 \beta  \quad \forall \  i \in \mathscr{I}. $
Even though $\tilde\lambda_i$ is over-identified,  to identify $\lambda$ and consequently $\kappa_i$, we need additional restrictions on the conduct parameters.
For instance, we can set $\kappa_1 = -1/(\mathcal I-1)$ if we assume that firm 1 is a price taker, or $\kappa_1 = 0$ if firm 1 is playing Cournot; in such cases, we have that $\lambda = \tilde\lambda_1 + \beta$ or $\lambda   = \tilde\lambda_1$. Then, using $\lambda$, we can identify 
\begin{equation*}
\kappa_i = \frac{1}{ \beta (\mathcal I-1)} \left\{ \frac{\beta \left[ \mu_{ Q_{-i}^{\mathrm{cp}+}|U} (u) - \mu_{ Q_{-i}^{\mathrm{cp}+}|U} (u^\prime) \right] + ( u - u^\prime )}{\mu_{Q_i^\mathrm{cp} | U} (u^\prime) - \mu_{Q_i^\mathrm{cp} | U} (u)} - \lambda - 2 \beta \right\}. 
\end{equation*}
Finally, by applying arguments similar to the ones in Section \ref{section:identification}, we can also identify the means $\{ \mu_{V_i} : i \in \mathscr I \}$, as well as the distributions $F_{V_i}$ and $F_{W|U}$.

\subsection*{Nonlinear Demand} \label{sec:nonlinear}

In this subsection, we consider an inverse market demand given by \begin{equation}
{\mathfrak p}^{\mathrm{nl}} (c, u ; \beta ) , 
\label{eq:nonlineardemand}
\end{equation}where ${\mathfrak p}^\mathrm{nl}$ is a continuous function defined on $ \mathbb R_{+} \times \mathscr U \times \mathscr B $, $\mathscr U $ and $\mathscr B$ are known compact intervals of $\mathbb R_{++}$ such that $[ \underline u, \bar u ] \subset \mathscr U$, and $ \beta $ is the demand parameter. For instance, we may take $\mathfrak{p}^\mathrm{nl} (c, u ; \beta ) = \exp \left( u - \beta c \right)$, which produces the log-linear inverse demand $\log[\mathfrak{p}^\mathrm{nl} ( c , u ; \beta) ] = u - \beta c $.

\begin{assumption}	\label{assump:nonlinear-demand}

The inverse demand function $\mathfrak{p}^\mathrm{nl}$ is known up to the parameter $ \beta $ and 
\begin{enumerate}[label=(\roman*)]

\item For all $(c,u ,b) \in \mathbb R_{+} \times \mathscr U \times \mathscr B$, we have ${\mathfrak p}^\mathrm{nl} (c, u ; b ) > 0 $ and $\lim_{c \rightarrow \infty} {\mathfrak p}^\mathrm{nl} (c, u ; b ) = 0 $ .

\item It admits two continuous partial derivatives on $ \mathbb R_{+} \times \mathscr U \times \mathscr B$ such that \begin{equation*}
\frac{\partial{\mathfrak p}^\mathrm{nl} }{\partial c} ( c , u ; b ) < 0 \ \ \text{and} \ \ \frac{\partial {\mathfrak p}^\mathrm{nl} }{\partial u} ( c, u ; b ) > 0, \qquad \forall \ (c, u,b) \in \mathbb R_{+} \times \mathscr U \times \mathscr B .
\end{equation*}

\item \label{assump:nonlinearid} The following implication holds for all $(u,u^\prime ) \in \left[\underline u,\bar u \right]^2$ and $(b,b^\prime) \in \mathscr B^2$, and for any pair $(c , c^\prime) \in \mathbb R_{+}^2$ such that $c \neq c^\prime: 
\mathfrak{p}^\mathrm{nl} (c , u; b ) = \mathfrak{p}^\mathrm{nl}( c, u^\prime; b^\prime ) \ \text{and} \ \mathfrak{p}^\mathrm{nl} (c^\prime, u; b ) = \mathfrak{p}^\mathrm{nl} (c^\prime,u^\prime; b^\prime ) \ \rightarrow \ ( u ,b ) = \left( u ^\prime,b^\prime \right)$.

\end{enumerate}

\end{assumption}

The first part of this assumption is standard. The second establishes that the inverse demand is strictly decreasing in the total consumption and strictly increasing in the demand shock. Finally, the third part is a technical assumption that restricts the shape of the demand function. To understand this condition, let us consider an example.

\begin{ex} \label{ex:nonlinear1}

Consider $\mathfrak{p}^\mathrm{nl} ( c , u ; \beta ) = \exp \left( u - \beta c \right)$ and pick any $c \neq c^\prime$. Then 
\begin{equation*}
 \exp \left( u - b \times c \right)= \exp \left( u^\prime - b^\prime \times c \right) \ \text{and} \ \exp \left( u - b \times c^\prime \right)= \exp \left( u^\prime - b^\prime \times c^\prime \right)
\end{equation*}
imply $( u-u^\prime) + (b^\prime - b ) c = 0$ and $( u - u^\prime) + (b^\prime - b ) c^\prime = 0 $. This is a system of two linear equations with two unknowns, $( u - u^\prime) $ and $(b^\prime - b )$. Since $c \neq c^\prime$, this system has a unique solution at zero: $ u - u^\prime = 0$ and $b^\prime - b = 0$. Thus, $(u,b ) = \left(u^\prime,b^\prime \right) $.

\end{ex}

Before we proceed to identify this model, we establish the existence of nondecreasing equilibrium strategies $\{ \mathfrak q_i^\mathrm{nl} (\cdot, w, u): [\underline{v}_i,\bar{v}_i] \rightarrow \mathbb R_{+} :i \in \mathscr I\}$ that satisfy \begin{align*}
\mathfrak q_i^\mathrm{nl} (v_i , w , u )  =  \underset{q_{i}\in \mathbb R_+}{\arg\max} \ q_{i}\times \E\left\{ \mathfrak p^\mathrm{nl} \left[ q_{i} + {\q}^{\mathrm{nl}+}_{-i}(\mathbf{V}_{-i} , W,U) , U ; \beta\right] \Big| V_{i} = v_i , W = w, U = u \right\} \\
 \ - \ {\left[ (v_{i} + w) q_{i}+ \frac{\lambda}{2} q_{i}^2 \right]},
\end{align*}for all $i \in \mathscr I$ and for each $(w,u) \in [ \underline w, \bar w ] \times [ \underline u , \bar u]$. Since we have that \begin{align*}
\E\left\{\mathfrak p^\mathrm{nl} \left[ q_{i} + {\q}^{\mathrm{nl}+}_{-i}(\mathbf{V}_{-i} , W,U) , U ; \beta\right] \Big| V_{i} = v_i , W = w, U = u \right\} =\E\left\{ \mathfrak p^\mathrm{nl} \left[ q_{i} + {\q}^{\mathrm{nl}+}_{-i}(\mathbf{V}_{-i} , w,u) , u ; \beta\right] \right\} ,
\end{align*}
and this expression does not depend on $v_i$, existence of non-increasing equilibrium strategies follows from \citet[Corollary 2.1]{Athey2001}. Besides showing the existence of equilibrium strategies, strict monotonicity and continuity are required to develop an empirical framework. To our best knowledge, the theoretical literature does not provide conditions of the primitives to guarantee these properties. Given the empirical focus of our article, we impose these properties in the next assumption to derive the identification results.

\begin{assumption}	\label{assump:nonlinear-equilibrium}
The following statements are satisfied for every firm $i \in \mathscr I$.\begin{enumerate}[label=(\roman*)]

\item The equilibrium strategy $\mathfrak q_i^\mathrm{nl}$ continuous on $ [\underline{v}_i,\bar{v}_i] \times [\underline w, \bar w] \times [\underline{u},\bar{u}] $.

\item $ \mathfrak q_i^\mathrm{nl} (\cdot, w , u )$ and $ \mathfrak q_i^\mathrm{nl} (v_i, \cdot,u)$ are strictly decreasing for all $(w , u) $ and $(v_i, u)$, respectively.

\item $ \mathfrak q_i^\mathrm{nl} (v_i, w, \cdot ) $ is strictly increasing for every $ (v_i, w)$.

\end{enumerate}

\end{assumption}

Before proceeding, we briefly discuss how the conditions of this assumption can be obtained or verified in certain cases. For instance, consider the log-linear case and take $\underline{u} > 0$ to be sufficiently large so that all firms always produce positive quantities. Then, a vector  equilibrium strategies $({\q}_1^{\mathrm{nl}} , \dots, {\q}_{\mathcal{I}}^{\mathrm{nl}} )$ can be implicitly characterized by the next first-order conditions: for all $i  \in \mathscr{I}$ and $(v_i ,  w , u )  \in [\underline{v}_i,\bar{v}_i] \times [\underline w, \bar w] \times [\underline{u},\bar{u}]$, \begin{multline*}
\exp [u - \beta {\q}_i^{\mathrm{nl}} ( v_i , w,u) ]  \times  \E\left\{ \exp\left[ -\beta {\q}^{\mathrm{nl}+}_{-i}(\mathbf{V}_{-i} , w,u) \right]   \right\}   - (v_{i} + w) -  \lambda {\q}_i^{\mathrm{nl}} ( v_i , w,u)   \\
   -  \  \beta  {\q}_i^{\mathrm{nl}} ( v_i , w,u)    \times      \exp [u - \beta {\q}_i^{\mathrm{nl}} ( v_i , w,u) ]  \times  \E\left\{ \exp\left[ -\beta {\q}^{\mathrm{nl}+}_{-i}(\mathbf{V}_{-i} , w,u) \right]   \right\}    \ =  \ 0 .
\end{multline*}So, Schauder fixed-point theorem implies that there exists a vector of equilibrium strategies $({\q}_1^{\mathrm{nl}} , \dots, {\q}_{\mathcal{I}}^{\mathrm{nl}} )$ such that each $ {\q}_i^{\mathrm{nl}}$ is twice continuously differentiable on $ [\underline{v}_i,\bar{v}_i] \times [\underline w, \bar w] \times [\underline{u},\bar{u} ]$; hence, Assumption \ref{assump:nonlinear-equilibrium}-(i) follows immediately. Verifying the second and third conditions is more involved as there is no closed-form expression for $({\q}_1^{\mathrm{nl}} , \dots, {\q}_{\mathcal{I}}^{\mathrm{nl}} )$. However, for a given value of $(\beta , \lambda)$, this task can still be performed by computational methods for approximating fixed points or, more specifically, Bayesian Nash equilibrium strategies \cite[see, e.g.,][]{arman08approx}.

Now let $P^\mathrm{nl} = {\mathfrak p}^\mathrm{nl} (Q^{\mathrm{nl} +},U) $ and $\mathbf Q^\mathrm{nl} = ( Q_{1}^\mathrm{nl} ,\dots, Q_{{\mathcal I}}^\mathrm{nl})$ be the equilibrium prices and quantities produced, respectively, under the inverse demand function (\ref{eq:nonlineardemand}) and Assumptions \ref{assump:nonlinear-demand} and \ref{assump:nonlinear-equilibrium}, i.e., $Q^\mathrm{nl}_i = \mathfrak q_i^\mathrm{nl} (V_i , W , U )$ for $i \in \mathscr I$. For identification purposes, we suppose that the joint distribution $F_{P^\mathrm{nl}, \mathbf Q^\mathrm{nl}}$ is known by the researcher. When the sample size increases to infinity, we can consistently estimate this joint CDF from a random sample of prices and quantities generated from the same equilibrium strategy.

Starting with the identification of the demand parameter $\beta$, choose any firm $i\in\mathscr I$. Using arguments similar to the ones in Lemma \ref{lem:testableimplications}, we can show that the conditional quantile function $ F^{-1}_{Q_{-i}^\mathrm{nl} | Q_i^\mathrm{nl}} (\cdot| \underline q_i^\mathrm{nl})$ is strictly increasing, where $\underline q_i^\mathrm{nl} : = \mathfrak q_i^\mathrm{nl} (\bar v _i , \bar w , \underline u )$ can be identified as $\underline q_i^\mathrm{nl} = \inf\{q \in \mathbb R_{+}: F_{Q_i^\mathrm{nl}} (q) > 0 \}$. Then, for any $(\alpha, \alpha^\prime) \in [0,1]^2$ such that $\alpha \neq \alpha^\prime$, we have \begin{eqnarray*}
F^{-1}_{P^\mathrm{nl} | Q_i^\mathrm{nl}} \left(\alpha \middle| \underline q_i^\mathrm{nl} \right) &=& \mathfrak{p}^\mathrm{nl} \left[ \underline q_i^\mathrm{nl} + F^{-1}_{Q_{-i}^\mathrm{nl} | Q_i^\mathrm{nl}} (1-\alpha| \underline q_i^\mathrm{nl}) , \underline u ; \beta \right] , \nonumber \\
F^{-1}_{P^\mathrm{nl} | Q_i^\mathrm{nl}} \left(\alpha' \middle| \underline q_i^\mathrm{nl} \right) &=& \mathfrak{p}^\mathrm{nl} \left[ \underline q_i^\mathrm{nl} + F^{-1}_{Q_{-i}^\mathrm{nl} | Q_i^\mathrm{nl}} (1-\alpha'| \underline q_i^\mathrm{nl}) , \underline u ; \beta \right] .
\end{eqnarray*}

Hence, by Assumption \ref{assump:nonlinear-demand}-(iii) and since $ F^{-1}_{Q_{-i}^\mathrm{nl} | Q_i^\mathrm{nl}} (1-\alpha| \underline q_i^\mathrm{nl}) \neq F^{-1}_{Q_{-i}^\mathrm{nl} | Q_i^\mathrm{nl}} (1-\alpha^\prime| \underline q_i^\mathrm{nl})$, $(\underline u , \beta)$ can be identified as the unique solution of the system of equations \begin{eqnarray*}
F^{-1}_{P^\mathrm{nl} | Q_i^\mathrm{nl}} \left(\alpha \middle| \underline q_i^\mathrm{nl} \right) &=& \mathfrak{p}^\mathrm{nl} \left[ \underline q_i^\mathrm{nl} + F^{-1}_{Q_{-i}^\mathrm{nl} | Q_i^\mathrm{nl}} (1-\alpha| \underline q_i^\mathrm{nl}) , u ; b \right] , \nonumber \\
F^{-1}_{P^\mathrm{nl} | Q_i^\mathrm{nl}} \left(\alpha' \middle| \underline q_i^\mathrm{nl} \right) &=& \mathfrak{p}^\mathrm{nl} \left[ \underline q_i^\mathrm{nl} + F^{-1}_{Q_{-i}^\mathrm{nl} | Q_i^\mathrm{nl}} (1-\alpha'| \underline q_i^\mathrm{nl}) , u ; b \right] ,
\end{eqnarray*}
with respect to $(u,b) \in \mathscr U \times \mathscr B$. Although there is no warranty that a closed-form expression for the solution exists, we show in Example \ref{ex:nonlinear1-id} below that a closed-form solution exists for the log-linear case. The choice of $i$ and the pair of quantiles $(\alpha, \alpha^\prime)$ were arbitrary, which means $\beta$ is over-identified as in the linear case.

Now that $\beta$ has been identified, we can recover the demand shock as $
U = \mathfrak{p}^{\mathrm{nl},-1} ( Q^{\mathrm{nl} +} , P^\mathrm{nl} ; \beta )$,
where $ \mathfrak{p}^{\mathrm{nl},-1} (q , \cdot ; \beta )$ denotes the functional inverse of $ \mathfrak{p}^{\mathrm{nl}} ( q , \cdot ; \beta )$; such an inverse exists by Assumption \ref{assump:nonlinear-demand}. Then, the CDF of $U$ can be identified as $F_U (u) = F_{\mathfrak{p}^{\mathrm{nl},-1} (Q^{\mathrm{nl} +}, P^\mathrm{nl} ;\beta )} (u)$. As an illustration, we apply the precedent identification strategies to Example \ref{ex:nonlinear1}.

\begin{ex}	 \label{ex:nonlinear1-id}

Considering the log-linear case of Example \ref{ex:nonlinear1}, we can identify 
\begin{eqnarray*}
\beta & = & \frac{\log \left[ F^{-1}_{P^\mathrm{nl} | Q_i^\mathrm{nl}} (\alpha^\prime | \underline q_i^\mathrm{nl})\right] - \log \left[ F^{-1}_{P^\mathrm{nl} | Q_i^\mathrm{nl}} (\alpha| \underline q_i^\mathrm{nl}) \right]}{ F^{-1}_{Q^+_{-i} | Q_i^\mathrm{nl}} (1 - \alpha| \underline q_i^\mathrm{nl} ) - F^{-1}_{Q^{^\mathrm{nl}+}_{-i} | Q_i^\mathrm{nl}} (1 - \alpha^\prime| \underline q_i^\mathrm{nl} ) } , \\
\underline u & = & \log \left[ F^{-1}_{P^\mathrm{nl} | Q_i^\mathrm{nl}} (\alpha| \underline q_i^\mathrm{nl}) \right] + \beta \left[ \underline q_i^\mathrm{nl} + F^{-1}_{Q_{-i}^\mathrm{nl} | Q_i^\mathrm{nl}} (1-\alpha^\prime| \underline q_i^\mathrm{nl}) \right].
\end{eqnarray*}
Since $ \mathfrak{p}^{\mathrm{nl},-1} ( q , p ;\beta ) = \log(p) + \beta q$, we can recover the demand shock as $U = \log(P^\mathrm{nl}) + \beta Q^{\mathrm{nl} +}$.
\end{ex}

Next, we consider identifying the cost parameter $\lambda>0$. Because of the nonlinearity of the demand function and the nonparametric distributions, we need to make a location assumption about the conditional mean $\mu_{W|U}$.

\begin{assumption}	\label{assump:nonlinear-condmeanind}
We have that $\mu_{W|U}(u) =0$ for every $u \in [\underline u,\bar u]$. 
\end{assumption}
 
%[{\color{red} R1 says: Like before, explain what sort of correlation this assumption allows.}]

Note that this assumption implies that $W$ and $U$ must be uncorrelated, but they do not need to be independent. For instance, the volatility of $W$ can still depend on the value of $U$, i.e., $E (W^2 | U=u)$ can depend on $u$.

 For a generic function $\psi$ and $j \in \mathbb N$, let $\mathfrak D_j \psi (x)$ denote the derivative of $\psi$ with respect to the $j$th argument evaluated at $x$. Equilibrium strategies satisfy the first-order conditions \begin{eqnarray}
& & \E\left[ \mathfrak q_i^\mathrm{nl} (V_i , W , U ) \times \mathfrak D_1 \mathfrak p^\mathrm{nl} \left[\mathfrak q_i^\mathrm{nl} (V_i , W , U ) + {\q}^{\mathrm{nl}+}_{-i}(\mathbf{V}_{-i} , W,U) , U ; \beta\right] | V_{i} = v_i , W = w , U = u \right] \nonumber \\
& & \qquad + \ \E\left\{ \mathfrak p^\mathrm{nl} \left[ \mathfrak q_i^\mathrm{nl} (V_i , W , U ) + {\q}^{\mathrm{nl}+}_{-i}(\mathbf{V}_{-i} , W,U\right) , U ; \beta] | V_{i} = v_i , W = w , U = u \right\} \nonumber \\
& & \qquad - \ \left[ (w + v_{i}) + {\lambda} \mathfrak q_i^\mathrm{nl} (v_i , w , u ) \right] = 0, \quad \forall i, \forall (v_i,w,u).
\label{eq:focnonlinear}
\end{eqnarray}
Then, taking the conditional expectation given $U = u$ and using Assumption \ref{assump:nonlinear-condmeanind} yield
 \begin{eqnarray*}
& & \E\left\{ \mathfrak q_i^\mathrm{nl} (V_i , W , U ) \times \mathfrak D_1 \mathfrak p^\mathrm{nl} [\mathfrak q_i^\mathrm{nl} (V_i , W , U ) + {\q}^{\mathrm{nl}+}_{-i}(\mathbf{V}_{-i} , W,U) , U ; \beta] \middle| U = u \right\} \\
& & \qquad + \ \E\left\{ \mathfrak p^\mathrm{nl} [ \mathfrak q_i^\mathrm{nl} (V_i , W , U ) + {\q}^{\mathrm{nl}+}_{-i}(\mathbf{V}_{-i} , W,U) , U ; \beta] \middle| U = u \right\}  -  \mu_{V_{i}} - {\lambda} \E\left[ \mathfrak q_i^\mathrm{nl} (V_i , W , U ) \middle| U = u \right] = 0 .
\end{eqnarray*}Observe that this condition can be written in terms of the standard expected marginal revenue and expected marginal cost:
 \begin{eqnarray}
 \underbrace{\E\left[Q_i^\mathrm{nl} \times \mathfrak D_1 \mathfrak p^\mathrm{nl} \left( Q^{\mathrm{nl}+} , U ; \beta \right) + \mathfrak p^\mathrm{nl} \left( Q^{\mathrm{nl}+} , U ; \beta \right)\middle| U = u \right]}_{\texttt{expected marginal revenue}} = \underbrace{ \mu_{V_{i}} + {\lambda}\times \E(Q_i^\mathrm{nl}|U = u) }_{\texttt{expected marginal cost}} .
 \label{eq:nonlinearidlambda}
\end{eqnarray}Except for the mean $ \mu_{V_{i}}$, the other variables in this expression are identified objects. Thus, we can subtract away $ \mu_{V_{i}}$ to identify the cost parameter $\lambda$. Specifically, after evaluating the expected marginal revenue in (\ref{eq:nonlinearidlambda}) at $U=u$ and $U=u^\prime$, being $u\neq u^\prime$, we obtain \begin{eqnarray*}
\lambda & = & \frac{\E\left[Q_i^\mathrm{nl} \times \mathfrak D_1 \mathfrak p^\mathrm{nl} \left( Q^{\mathrm{nl}+} , U; \beta \right) + \mathfrak p^\mathrm{nl} \left( Q^{\mathrm{nl}+} , U ; \beta \right) \middle| U = u^\prime \right] } { \mu_{Q_i^\mathrm{nl} | U } \left( u^\prime \right) - \mu_{Q_i^\mathrm{nl} | U } \left( u \right)} \\
 & & \ \ - \ \frac{\E\left[Q_i^\mathrm{nl} \times \mathfrak D_1 \mathfrak p^\mathrm{nl} \left( Q^{\mathrm{nl}+} , U; \beta \right) + \mathfrak p^\mathrm{nl} \left( Q^{\mathrm{nl}+} , U ; \beta \right) \middle| U = u \right] } { \mu_{Q_i^\mathrm{nl} | U } \left( u^\prime \right) - \mu_{Q_i^\mathrm{nl} | U } \left( u \right)} .
\end{eqnarray*}
Strict monotonicity of the equilibrium strategies guarantees that $\mu_{Q_i^\mathrm{nl} | U } \left( u^\prime \right) - \mu_{Q_i^\mathrm{nl} | U } ( u)\neq 0$, thereby identifying $\lambda$. 
In fact, since our choices of $i \in \mathcal I$ and $u^\prime \neq u$ were arbitrary, $\lambda$ is also over-identified. Now that $\lambda$ has been identified, the unconditional means $\{ \mu_{V_1},\dots, \mu_{V_{\mathcal I}}\}$ can be recovered by noting that $
\mu_{V_i} = \E\left[Q_i^\mathrm{nl} \times \mathfrak D_1 \mathfrak p^\mathrm{nl} \left( Q^{\mathrm{nl}+} , U ; \beta \right) + \mathfrak p^\mathrm{nl} \left( Q^{\mathrm{nl}+} , U ; \beta \right) \right]  - {\lambda} \times \E ( Q_i^\mathrm{nl} ), \forall \ i \in \mathscr I
$
and the first term on the right-hand side is an identified object.

Next, we establish the nonparametric identification of the distributions $\{F_{V_i} : i \in \mathscr I \}$. To do so, we impose an additional assumption on the support of $F_W$.

\begin{assumption}	\label{assump:nonlinear-condmeanind-3}
The support of $F_{W}$ is symmetric around zero, i.e., $\bar w = - \underline w$ .
\end{assumption}

%[{\color{red} R1 says: Give one liner intuition for this assumption.}]

Heuristically, this is a technical assumption that states that negative cost shocks can be as large as positive ones. It treats positive shock to technology, which reduces cost, the same way as a negative shock. More specifically, for given $\alpha \in (0,1)$ and $i \in \mathscr{I}$, \ref{assump:nonlinear-condmeanind-3} allows us to identify $( F_{V_i}^{-1} (\alpha) ,  \underline w , \bar w  ) $ from equations (\ref{eq:idFthetanonlinear1}) and (\ref{eq:idFthetanonlinear2}) in Appendix A.

\begin{theorem} \label{thm:nonlinear-identification}

Suppose that $F_{P^\mathrm{nl}, \mathbf{Q}^\mathrm{nl}}$ is known and that Assumptions \ref{assump:distributions}, \ref{assump:nonlinear-demand}, \ref{assump:nonlinear-equilibrium}, \ref{assump:nonlinear-condmeanind}, and \ref{assump:nonlinear-condmeanind-3} hold. Then the distributions $\{F_{V_i} : i \in \mathscr I \}$ are nonparametrically identified.

\end{theorem}

\begin{proof}
Pick any $i \in \mathscr I$ and $\alpha \in [0,1]$. In what follows, we show that $F_{V_i}^{-1}(\alpha)$ is identified. After evaluating the first-order condition (\ref{eq:focnonlinear}) at $(v_i, w, u) = \left( F_{V_i}^{-1}(\alpha ) ,  \bar w , \underline u  \right)$ and rearranging, it follows that $F_{V_i | W,U}^{-1}(\alpha| \bar w , \underline u )=F_{V_i }^{-1} \left(\alpha  \right)$ is equal to 
\begin{eqnarray}
F_{V_i }^{-1} \left(\alpha  \right)&=& \mathfrak q_i^\mathrm{nl} \left[ F_{V_i }^{-1} \left(\alpha\right) , \bar w , \underline u \right]\notag\\
 &&\times \E\left\{ \mathfrak D_1 \mathfrak p^\mathrm{nl} \left[ \mathfrak q_i^\mathrm{nl} \left(V_i , W , U \right) + {\q}^{\mathrm{nl}+}_{-i}(\mathbf{V}_{-i} , W, U) , U ; \beta \right] \middle| V_i = F_{V_i}^{-1}(\alpha) , W = \bar w, U = \underline u \right\} \nonumber \nonumber \\
& &  + \  \E\left\{ \mathfrak p^\mathrm{nl} \left[ \mathfrak q_i^\mathrm{nl} (V_i , W , U ) + {\q}^{\mathrm{nl}+}_{-i}(\mathbf{V}_{-i} , W,U) , U ; \beta \right] \middle|  V_i = F_{V_i}^{-1}(\alpha) , W = \bar w, U = \underline u \right\} \nonumber \\
& &  - \  \bar w \ - \  {\lambda} \mathfrak q_i^\mathrm{nl} \left[ F_{V_i}^{-1}(\alpha ) , \bar w , \underline\upsilon \right]. \label{eq:idFthetanonlinear}
\end{eqnarray}

We can write the first term in expression (\ref{eq:idFthetanonlinear}), for any $j\neq i$, in terms of observables as \begin{align*}
 \mathfrak q_i^\mathrm{nl} \left[ F_{V_i }^{-1} \left(\alpha  \right), \bar w , \underline u \right] = F^{-1}_{Q_i^\mathrm{nl} |W , U } \left(1-\alpha \middle|  \bar w , \underline u \right)  =  F^{-1}_{Q_i^\mathrm{nl} |V_j, W , U } \left(1 - \alpha \middle| \bar v_j , \bar w , \underline u \right) = F^{-1}_{Q_i^\mathrm{nl} |Q_j^\mathrm{nl}} \left(1 - \alpha \middle| \underline q_j^\mathrm{nl} \right).
\end{align*}
The first equality follows from the strict monotonicity of the equilibrium strategies, the second equality follows from the conditional independence $Q_i^\mathrm{nl}\perp V _j | (W,U)$, and the third from the fact that the events $Q_j^\mathrm{nl} = \underline q_j^\mathrm{nl}$ and $( V_j,W,U ) = (\bar v_j, \underline w , \underline u)$ are equivalent. 

Now considering the second term in (\ref{eq:idFthetanonlinear}), we write
 \begin{eqnarray*}
 & & \E\left\{ \mathfrak D_1 \mathfrak p^\mathrm{nl} \left[ \mathfrak q_i^\mathrm{nl} \left( V_i , W , U \right) + {\q}^{\mathrm{nl}+}_{-i}(\mathbf{V}_{-i} , W, U) , U ; \beta \right] \middle|  V_i = F_{V_i}^{-1}(\alpha) , W = \bar w, U = \underline u \right\}  \\
 & & \ \ = \ \E\left[ \mathfrak D_1 \mathfrak p^\mathrm{nl} \left\{ \mathfrak q_i^\mathrm{nl} \left[ F_{V_i}^{-1}(\alpha) , \bar w , \underline u \right] + {\q}^{\mathrm{nl}+}_{-i}(\mathbf{V}_{-i} , W , U ) , \underline u ; \beta \right) \middle| V_i = F_{V_i}^{-1}(\alpha) , W = \bar w, U = \underline u \right] \\
  & & \ \ = \ \E\left[\mathfrak D_1 \mathfrak p^\mathrm{nl} \left\{ \mathfrak q_i^\mathrm{nl} \left[ F_{V_i}^{-1}(\alpha) , \bar w , \underline u \right] + {\q}^{\mathrm{nl}+}_{-i}(\mathbf{V}_{-i} , W, U ) , \underline u ; \beta \right\} \middle| V_i = \bar v_i , W = \bar w, U = \underline u \right]  \\
  & & \ \ = \ \E\left\{ \mathfrak D_1 \mathfrak p^\mathrm{nl} \left[  F^{-1}_{Q_i^\mathrm{nl} |Q_j^\mathrm{nl}} \left(1 - \alpha \middle| \underline q_j ^\mathrm{nl} \right) + Q_{-i}^{\mathrm{nl} +} , \underline u ; \beta \right] \middle| Q_i^\mathrm{nl} = \underline q_i^\mathrm{nl} \right\}.
  \end{eqnarray*}The first equality follows by the fact that, since we are conditioning on $V_i = F_{V_i}^{-1}(\alpha) $, $V_i$ can be treated as a constant inside the conditional expectation. The second follows from the conditional independence $V_i\perp\mathbf{V}_{-i} | (W, U) $, which implies $F_{\mathbf{V}_{-i} | V_i, W, U} \left[ \cdot |  F_{V_i}^{-1}(\alpha) , \bar w , \underline u \right] = F_{\mathbf{V}_{-i} | V_i, W, U} \left( \cdot |  \bar v_i , \bar w , \underline u \right) = F_{\mathbf{V}_{-i} | W, U} \left( \cdot |  \bar w , \underline u \right).$ And finally, the third equality follows by the fact that $Q_i^\mathrm{nl} = \underline q_i^\mathrm{nl}$ and $( V_i,W,U ) = (\bar v _i, \bar w , \underline u)$ are equivalent events. Similarly, the third term in (\ref{eq:idFthetanonlinear}) can be written as \begin{align*}
 \E\left\{ \mathfrak p^\mathrm{nl} \left[ \mathfrak q_i^\mathrm{nl} (V_i , W , U ) + {\q}^{\mathrm{nl}+}_{-i}(\mathbf{V}_{-i} , W,U) , U ; \beta \right] \middle|  V_i = F_{V_i}^{-1}(\alpha) , W = \bar w, U = \underline u \right\} \\
 = \E\left\{ \mathfrak p^\mathrm{nl} \left[  F^{-1}_{Q_i^{\mathrm{nl}} |Q_j^{\mathrm{nl}}} \left(1 - \alpha | \underline q_j ^\mathrm{nl} \right) + Q_{-i}^{\mathrm{nl}} , \underline u; \beta \right] \middle| Q_i^{\mathrm{nl}} = \underline q_i^\mathrm{nl} \right\} . % \label{eq:expnolinear2}
\end{align*}
Combining the precedent results with (\ref{eq:idFthetanonlinear}), the quantile $F_{V_i}^{-1}(\alpha ) $ can be rewritten as 
\begin{eqnarray}
F_{V_i}^{-1}(\alpha )  &=&  F^{-1}_{Q_i^\mathrm{nl} |Q_j^\mathrm{nl}} (1- \alpha | \underline q_j^\mathrm{nl} ) \times \E\left\{ \mathfrak D_1 \mathfrak p^\mathrm{nl} \left[  F^{-1}_{Q_i^\mathrm{nl} |Q_j^\mathrm{nl}} \left( 1 - \alpha \middle| \underline q_j^\mathrm{nl} \right) + Q_{-i}^{\mathrm{nl} +} , \underline u ; \beta \right] \middle| Q_i^\mathrm{nl} = \underline q_i^\mathrm{nl} \right\} \notag \\
 &&\!\! \!\!\!\!\!\!\!\!\!\!+   \E\left\{ \mathfrak p^\mathrm{nl} \left[ F^{-1}_{Q_i^\mathrm{nl} |Q_j^\mathrm{nl}} \left(1 - \alpha \middle| \underline q_j \right) + Q_{-i}^{\mathrm{nl} +} , \underline u ; \beta^{\mathrm{nl}} \right] \middle| Q_i^\mathrm{nl} = \underline q_i^\mathrm{nl} \right\}- \bar w - \lambda F^{-1}_{Q_i^\mathrm{nl} |Q_j^\mathrm{nl}}(1 - \alpha| \underline{q}_j^\mathrm{nl}).\qquad\quad \label{eq:idFthetanonlinear1}
\end{eqnarray}
By symmetric arguments, we can replace $(\bar w , \underline q_i)$ with $(\underline w , \bar q_i)$ in (\ref{eq:idFthetanonlinear1}), and we get  
\begin{eqnarray}
F_{V_i}^{-1}(\alpha )  &=&  F^{-1}_{Q_i^\mathrm{nl} |Q_j^\mathrm{nl}} \left(1-\alpha \middle| \bar q_j^\mathrm{nl} \right) \times \E\left\{ \mathfrak D_1 \mathfrak p^\mathrm{nl} \left[  F^{-1}_{Q_i^\mathrm{nl} |Q_j^\mathrm{nl}} \left(1 - \alpha \middle| \bar q_j^\mathrm{nl} \right) + Q_{-i}^{\mathrm{nl} +} , \bar u ; \beta \right] \middle| Q_i^\mathrm{nl} = \bar q_i^\mathrm{nl} \right\}\notag  \\
 &&\!\! \!\!\!\!\!\!\!\!\!\!+  \E\left\{ \mathfrak p^\mathrm{nl} \left[  F^{-1}_{Q_i^\mathrm{nl} |Q_j^\mathrm{nl}} \left(1  - \alpha \middle| \bar q_j \right) + Q_{-i}^{\mathrm{nl} +} , \bar u; \beta^{\mathrm{nl}} \right] \middle| Q_i^\mathrm{nl} = \bar q_i^\mathrm{nl} \right\} - \underline w - \lambda F^{-1}_{Q_i^\mathrm{nl} |Q_j^\mathrm{nl}} \left( 1 - \alpha \middle| \bar q_j ^\mathrm{nl} \right).\qquad\quad\label{eq:idFthetanonlinear2}  
\end{eqnarray}

To identify $F_{V_i}^{-1} (\alpha)$, $\underline w$ and $\bar w$, we observe that (\ref{eq:idFthetanonlinear1}) and (\ref{eq:idFthetanonlinear2}) together with the restriction $\bar w = - \underline w$ (Assumption \ref{assump:nonlinear-condmeanind-3}) form a system of three linear equations with three unknowns ($F_{V_i}^{-1} (\alpha),\underline w,\bar w)$ with a unique solution. As $( i , \alpha)$ was arbitrary, $F_{V_i}$ is identified $\forall i \in \mathscr I$.  
\end{proof}

We complete this subsection with a discussion about the identification of $F_{W|U}$. In Section \ref{section:identification}, this conditional CDF has been identified from the characteristic function, which was expressed as a function of identified objects. This expression was obtained, essentially, by exploiting the linearity of the equilibrium strategies. However, in this nonlinear setting, the arguments of Section \ref{section:identification} cannot be applied, even if we assume that $\{q_1^\mathrm{nl}, \ldots, q_\mathcal{I}^\mathrm{nl}\}$ are the unique equilibrium strategies. To see this point, note that under this extra assumption the functional forms of the equilibrium strategies $\{ \mathfrak{q}_i^\mathrm{nl}(\cdot,w,u) : i \in \mathscr I\}$ can be identified as the unique fixed point of the mapping $\mathfrak T (\mathfrak{q}_1,\dots, \mathfrak{q}_\mathcal{I} ) = \left(\mathfrak T_1 (\mathfrak{q}_1,\dots, \mathfrak{q}_\mathcal{I} ) ,\dots, \mathfrak T_\mathcal{I} (\mathfrak{q}_1,\dots, \mathfrak{q}_\mathcal{I} ) \right),$ defined on the space of functions satisfying the conditions of Assumption \ref{assump:nonlinear-equilibrium}, and given by \begin{equation*}
\mathfrak T_i (\mathfrak{q}_1,\dots, \mathfrak{q}_\mathcal{I} )[v_i] =  \underset{q_{i}\in \mathbb R_+}{\arg\max} \ q_{i}\times \E\left\{ \mathfrak p^\mathrm{nl} \left[ q_{i} + {\q}_{-i}(\mathbf{V}_{-i} , w, u) , u ; \beta\right] \right\} - {\left[ (v_i + w) q_{i}+ \frac{\lambda}{2} q_{i}^2 \right]} 
\end{equation*}for each $i \in \mathscr I$. Observe that $\mathfrak T$ is an identified object as both $F_\mathbf{V}$ and $\lambda$, as well as $\beta$, have already been identified. Even though the equilibrium strategies can be identified, a closed-form expression may not be available. Without that, we do not know whether the conditional characteristic function of $W$ given $U$ can be obtained from $F_{P,\mathbf Q}$.

\subsection*{Selective Entry} \label{sec:secentry}

So far, we have assumed that the number of firms in the market is exogenous.  
However, in some cases, a firm may decide not to serve a market due to a high entry cost. 
If this decision is based on an estimate of the variable cost of production, then an entry cost leads to selective entry: only those firms that expect their cost to be sufficiently low enter the market. 
Thus, the subset of firms active in the market is not a representative sample. 

Next, we consider selective entry by symmetric firms with linear demand. 
We begin by specifying the timing of the game. First, each potential entrant $i \in \mathscr I$, privately observes a signal $S_{i}$ of its private (but unknown) cost $V_{i}$. Second, all potential entrants observe the entry cost of $C$, which may vary across markets, and then simultaneously decide whether to enter the market and pay $C$. Third, upon entry, firms observe the number of entrants, their private cost, and the demand and technology shock $(U, W)$ and quantities.

Let $\mathbf S = (S_1,\dots, S_{\mathcal I})$ denote the vector of signals. 
For notational simplicity, we assume that the set of potential entrants ${\mathscr I}$ is fixed. We make the following assumption about $C$, $(U, W)$, and the relationship between $V_i$ and the signal $S_i$. 

\begin{assumption}	\label{assump:entry}

The following conditions hold.

\begin{enumerate}[label=(\roman*)]

\item $\{(\mathbf S , \mathbf V) , C , ( W, U)\}$ are mutually independent.

\item $\{ (S_i,V_i) : i =1,\dots, {\mathcal I} \}$ are IID as $F_{S,V}$ with support $[0 , 1] \times [\underline v , \bar v]$. Further, the marginal distribution of signal $F_{S}$ is normalized to be uniform on $[0,1]$.
 
\item The conditional distribution of $V_i$ given $S_i = s_i$ is stochastically ordered in signals: $s^\prime \geq s$ implies $F_{V|S}(v|s^\prime) \leq F_{V|S}(v|s)$ for all $(v,s) \in [\underline v, \bar v] \times [0,1]$. 

\item For every $s \in [0,1]$, $F_{V|S}(\cdot|s)$ admits density $f_{V|S}(\cdot|s)$ that is strictly positive and continuously differentiable on $(\underline{v},\bar{v})$ and expectation $\mu_{V|S}(s)$ continuous on $[0,1]$.

\item The entry cost $C$ has support $[\underline c , \bar c]$ and its CDF admits a PDF that is strictly positive and continuously differentiable on $(\underline c , \bar c)$.

\end{enumerate}

\end{assumption}

This assumption is similar to Assumptions 1 and 2 of \cite{GentryLi2014} in auctions with selective entry. 
Hereafter, we assume a symmetric equilibrium exists and firm $i$ enters a market if and only if its signal $S_i$ is less than some threshold. This threshold signal is the highest signal that makes a firm indifferent between entering or not entering. Let $E_i$ be a random variable that equals one if firm $i$ enters, and zero otherwise. 
Then, we can write $E_i=0$ if $S_i > \mathfrak{s}(C)$ and $E_i=1$ if $S_i \leq \mathfrak{s}(C)$, for some nonincreasing (threshold) function $\mathfrak{s}: [ \underline c , \bar c] \rightarrow [0,1]$ that satisfies $0 \leq \mathfrak{s} ( \bar c) < \mathfrak{s} (\underline c) $. So every firm enters with positive probability, and $\{ E_i : i \in {\mathscr I}\}$ are identically distributed. 

In the second stage, the strategies can be represented as follows. Let $\mathbf e = (e_1,\dots,e_{\mathcal I}) \in \{0,1\}^{\mathcal I}$ is a vector such that $e_i = 1 $ if firm $i$ entered the market and zero otherwise, and denote $e^+ = \sum_{i \in {\mathscr I}} e_i$. Then 
$
 \mathfrak{\q}^{\mathrm{se}}_i (v_i , w, u , c , \mathbf e ) = \left\{ \begin{array}{ll}
0 & \text{if} \ e_i = 0 , \\
 \frac{1}{\lambda + (e^+ +1)\beta} \left[ u - w - \tilde\mu_{V}( c) \right] - \frac{v_i - \tilde\mu_{V}( c ) }{\lambda + 2 \beta} & \text{if} \ e_i = 1 ,
 \end{array} \right.
$ where we note that, by Assumption \ref{assump:entry}-(ii), $\tilde\mu_{V}( c ) := \E [ V_i | S_i \leq \mathfrak{s}(c) ] $ does not depend on $i$ and is nonincreasing in $c$.

Let $P^\mathrm{se} $ and $\mathbf{Q}^\mathrm{se} $ be the equilibrium price and vector of quantities under selective entry, respectively, and let $\mathbf E = (E_1,\dots,E_{\mathcal I})$ denote the vector of entry decisions, such that ${\bf E}={\bf e}$ is the realization of entry decisions. Selective entry does not affect the identification of the demand slope $\beta$ and parameter of the cost function $\lambda$, so the previous identification arguments apply here. In particular, suppose from the data we observe the conditional joint CDF $F_{P^{\mathrm{se}},\mathbf{Q}^{\mathrm{se}} | \mathbf{E}}$. Then using the fact that conditional on $\mathbf E = (1,1,0,\dots,0)$, the events $\{Q_1 = \underline q_1 \}$ and $\{(V_1, W, U , \tilde\mu_V (C)) = (\bar v_1, \bar w, \underline u, \tilde\mu_V (\bar c) )\}$ are equivalent, we can apply the identification results from Section \ref{section:identification} to the conditional distribution $F_{P^{\mathrm{se}},\mathbf{Q}^{\mathrm{se}} | \mathbf{E}} (\cdot,\cdot| 1,1,0,\dots,0)$ and identify $\{ \beta, F_U\}$. Then, using Assumption \ref{assump:nonlinear-condmeanind}, we can also identify $\lambda$ from the conditional expectation of $Q_1^\mathrm{se}$ given $P^\mathrm{se} - \beta (Q^\mathrm{se}_1 + Q^\mathrm{se}_2) = u $ and $\mathbf E = (1,1,0,\dots,0)$, as before.

However, if we follow the identification strategy of Section \ref{section:identification}, we can identify only the truncated cost distributions $ F_{V|E} ( v | 1)=\mathbb{P} (V_i \leq v |S_i\leq \mathfrak{s}(C))$, which can identify $F_{W|U}$. Next, we determine conditions under which we can identify $F_{V|S}$.  
Suppose we observe a continuous variable $Z$ with support $[\underline z , \bar z]$ that affects the entry cost, i.e.\ $C =\mathfrak c(Z)$ for some unknown function $\mathfrak c$, but does not affect $(V_i,S_i)$.

\begin{assumption}	\label{assump:entry2}
Suppose the joint CDF is such that $F_{V,S|Z}(v,s|z)=F_{V,S}(v,s)$. 
\end{assumption}

Observe that we can identify the entry frequency by $\mathfrak{s}^\ast (z):=\mathfrak{s}[\mathfrak c(z)]=\E ( E_{i}|Z=z )/{\mathcal I}$ for any $z \in [\underline z , \bar z]$. Moreover, by following the above identification strategy under the conditional $Z=z$, we can also recover the truncated conditional distribution $F_{V|S}^* (v ; s) : = \mathbb P ( V_i \leq v | S_i \leq s)$ for any $s = \mathfrak{s}^\ast (z)$ with $z \in [\underline z , \bar z]$. From the threshold-crossing equilibrium condition, note that we can write $
F_{V|S}^*(v; s ) = \frac{1}{{s}} \int_0^{{s}} F_{V|S}(v|\xi)d\xi,$ where we have suppressed the dependence of the threshold on the entry cost $C$. Differentiating this expression with respect to $s$ yields $F_{V| S}( v | s )=\frac{\partial}{\partial s}[{s}\times F_{V|S}^*(v;{s})]$. So now, for each $z$ in the support of $Z$ and under Assumption \ref{assump:entry2}, we can identify $\mathfrak{s}^\ast(z)$, and hence $F_{V|S} [V|\mathfrak{s}^\ast(z)]$.

\end{document}